\documentclass[twocolumn]{aastex631}

\usepackage{gensymb}
\usepackage{amsmath}
\usepackage{csvsimple}
\usepackage{graphicx}
\usepackage{moresize}
\usepackage{xspace}
\newcommand{\Like}{\mathcal{L}}
\newcommand{\LogLike}[1]{\log{\left(\Like_{#1}\right)}}
\begin{document}
\title{The Fourth HAWC Catalog of Very-High-Energy Gamma-Ray Sources}

\correspondingauthor{Samuel Groetsch}
\email{groetsch@icecube.wisc.edu}

\author{R.~Alfaro}

\affiliation{Instituto de F\'{i}sica, Universidad Nacional Autónoma de México, Ciudad de Mexico, Mexico}

\author{C.~Alvarez}

\affiliation{Universidad Autónoma de Chiapas, Tuxtla Gutiérrez, Chiapas, México}

\author{E.~Anita-Rangel}

\affiliation{Instituto de Astronom\'{i}a, Universidad Nacional Autónoma de México, Ciudad de Mexico, Mexico }

\author{M.~Araya}

\affiliation{Universidad de Costa Rica, San José 2060, Costa Rica}

\author{J.C.~Arteaga-Velázquez}

\affiliation{Universidad Michoacana de San Nicolás de Hidalgo, Morelia, Mexico }

\author{D.~Avila Rojas}

\affiliation{Instituto de Astronom\'{i}a, Universidad Nacional Autónoma de México, Ciudad de Mexico, Mexico }

\author{H.A.~Ayala Solares}

\affiliation{Temple University, Department of Physics, 1925 N. 12th Street, Philadelphia, PA 19122, USA}

\author{P.~Bangale}

\affiliation{Temple University, Department of Physics, 1925 N. 12th Street, Philadelphia, PA 19122, USA}

\author{E.~Belmont-Moreno}

\affiliation{Instituto de F\'{i}sica, Universidad Nacional Autónoma de México, Ciudad de Mexico, Mexico}

\author{A.~Bernal}

\affiliation{Instituto de Astronom\'{i}a, Universidad Nacional Autónoma de México, Ciudad de Mexico, Mexico }

\author{K.S.~Caballero-Mora}

\affiliation{Universidad Autónoma de Chiapas, Tuxtla Gutiérrez, Chiapas, México}

\author{T.~Capistrán}

\affiliation{Instituto de Astronom\'{i}a, Universidad Nacional Autónoma de México, Ciudad de Mexico, Mexico }

\author{A.~Carramiñana}

\affiliation{Instituto Nacional de Astrof\'{i}sica, Óptica y Electrónica, Puebla, Mexico}

\author{F.~Carreón}

\affiliation{Instituto de Astronom\'{i}a, Universidad Nacional Autónoma de México, Ciudad de Mexico, Mexico }

\author{S.~Casanova}

\affiliation{Institute of Nuclear Physics Polish Academy of Sciences, PL-31342 IFJ-PAN, Krakow, Poland }

\author{U.~Cotti}

\affiliation{Universidad Michoacana de San Nicolás de Hidalgo, Morelia, Mexico }

\author{J.~Cotzomi}

\affiliation{Facultad de Ciencias F\'{i}sico Matemáticas, Benemérita Universidad Autónoma de Puebla, Puebla, Mexico }

\author{S.~Coutiño de León}

\affiliation{Instituto de Física Corpuscular, CSIC, Universitat de València, E-46980, Paterna, Valencia, Spain}

\author{E.~De la Fuente}

\affiliation{Departamento de F\'{i}sica, Centro Universitario de Ciencias Exactase Ingenierias, Universidad de Guadalajara, Guadalajara, Mexico }

\author{P.~Desiati}

\affiliation{Dept. of Physics and Wisconsin IceCube Particle Astrophysics Center, University of Wisconsin{\textemdash}Madison, Madison, WI, USA}

\author{N.~Di Lalla}

\affiliation{Department of Physics, Stanford University: Stanford, CA 94305–4060, USA}

\author{R.~Diaz Hernandez}

\affiliation{Instituto Nacional de Astrof\'{i}sica, Óptica y Electrónica, Puebla, Mexico}

\author{B.L.~Dingus}

\affiliation{Los Alamos National Laboratory, Los Alamos, NM, USA }

\author{M.A.~DuVernois}

\affiliation{Dept. of Physics and Wisconsin IceCube Particle Astrophysics Center, University of Wisconsin{\textemdash}Madison, Madison, WI, USA}

\author{J.C.~Díaz-Vélez}

\affiliation{Dept. of Physics and Wisconsin IceCube Particle Astrophysics Center, University of Wisconsin{\textemdash}Madison, Madison, WI, USA}

\author{K.~Engel}

\affiliation{Department of Physics, University of Maryland, College Park, MD, USA }

\author{T.~Ergin}

\affiliation{Department of Physics and Astronomy, Michigan State University, East Lansing, MI, USA }

\author{C.~Espinoza}

\affiliation{Instituto de F\'{i}sica, Universidad Nacional Autónoma de México, Ciudad de Mexico, Mexico}

\author{K.~Fang}

\affiliation{Dept. of Physics and Wisconsin IceCube Particle Astrophysics Center, University of Wisconsin{\textemdash}Madison, Madison, WI, USA}

\author{N.~Fraija}

\affiliation{Instituto de Astronom\'{i}a, Universidad Nacional Autónoma de México, Ciudad de Mexico, Mexico }

\author{S.~Fraija}

\affiliation{Instituto de Astronom\'{i}a, Universidad Nacional Autónoma de México, Ciudad de Mexico, Mexico }

\author{J.A.~García-González}

\affiliation{Tecnologico de Monterrey, Escuela de Ingenier\'{i}a y Ciencias, Ave. Eugenio Garza Sada 2501, Monterrey, N.L., Mexico, 64849}

\author{F.~Garfias}

\affiliation{Instituto de Astronom\'{i}a, Universidad Nacional Autónoma de México, Ciudad de Mexico, Mexico }

\author{N.~Ghosh}

\affiliation{Department of Physics, Michigan Technological University, Houghton, MI, USA }

\author{M.M.~González}

\affiliation{Instituto de Astronom\'{i}a, Universidad Nacional Autónoma de México, Ciudad de Mexico, Mexico }

\author{J.A.~González}

\affiliation{Universidad Michoacana de San Nicolás de Hidalgo, Morelia, Mexico }

\author{J.A.~Goodman}

\affiliation{Department of Physics, University of Maryland, College Park, MD, USA }

\author{S.~Groetsch}

\affiliation{Dept. of Physics and Wisconsin IceCube Particle Astrophysics Center, University of Wisconsin{\textemdash}Madison, Madison, WI, USA}

\affiliation{Department of Physics, Michigan Technological University, Houghton, MI, USA }

\author{D.~Guevel}

\affiliation{Department of Physics, Michigan Technological University, Houghton, MI, USA }

\author{S.~Hernández-Cadena}

\affiliation{Tsung-Dao Lee Institute \& School of Physics and Astronomy, Shanghai Jiao Tong University, 800 Dongchuan Rd, Shanghai, SH 200240, China}

\author{F.~Hueyotl-Zahuantitla}

\affiliation{Universidad Autónoma de Chiapas, Tuxtla Gutiérrez, Chiapas, México}

\author{P.~Hüntemeyer}

\affiliation{Department of Physics, Michigan Technological University, Houghton, MI, USA }

\author{A.~Iriarte}

\affiliation{Instituto de Astronom\'{i}a, Universidad Nacional Autónoma de México, Ciudad de Mexico, Mexico }

\author{S.~Kaufmann}

\affiliation{Universidad Politecnica de Pachuca, Pachuca, Hgo, Mexico }

\author{D.~Kieda}

\affiliation{Department of Physics and Astronomy, University of Utah, Salt Lake City, UT, USA }

\author{A.~Lara}

\affiliation{Instituto de Geof\'{i}sica, Universidad Nacional Autónoma de México, Ciudad de Mexico, Mexico }

\author{K.~Leavitt}

\affiliation{Department of Physics, Michigan Technological University, Houghton, MI, USA }

\author{T.~Lewis}

\affiliation{Department of Physics, Michigan Technological University, Houghton, MI, USA }

\author{H.~León Vargas}

\affiliation{Instituto de F\'{i}sica, Universidad Nacional Autónoma de México, Ciudad de Mexico, Mexico}

\author{J.T.~Linnemann}

\affiliation{Department of Physics and Astronomy, Michigan State University, East Lansing, MI, USA }

\author{A.L.~Longinotti}

\affiliation{Instituto de Astronom\'{i}a, Universidad Nacional Autónoma de México, Ciudad de Mexico, Mexico }

\author{G.~Luis-Raya}

\affiliation{Universidad Politecnica de Pachuca, Pachuca, Hgo, Mexico }

\author{O.~Martinez}

\affiliation{Facultad de Ciencias F\'{i}sico Matemáticas, Benemérita Universidad Autónoma de Puebla, Puebla, Mexico }

\author{J.~Martínez-Castro}

\affiliation{Centro de Investigaci\'on en Computaci\'on, Instituto Polit\'ecnico Nacional, M\'exico City, M\'exico.}

\author{H.~Martínez-Huerta}

\affiliation{Universidad de Monterrey, San Pedro Garza Garc´ıa, Nuevo Le´on, M´exico}

\author{J.A.~Matthews}

\affiliation{Dept of Physics and Astronomy, University of New Mexico, Albuquerque, NM, USA }

\author{P.~Miranda-Romagnoli}

\affiliation{Universidad Autónoma del Estado de Hidalgo, Pachuca, Mexico }

\author{E.~Moreno}

\affiliation{Facultad de Ciencias F\'{i}sico Matemáticas, Benemérita Universidad Autónoma de Puebla, Puebla, Mexico }

\author{M.~Mostafá}

\affiliation{Temple University, Department of Physics, 1925 N. 12th Street, Philadelphia, PA 19122, USA}

\author{M.~Najafi}

\affiliation{Department of Physics, Michigan Technological University, Houghton, MI, USA }

\author{A.~Nayerhoda}

\affiliation{Institute of Nuclear Physics Polish Academy of Sciences, PL-31342 IFJ-PAN, Krakow, Poland }

\author{L.~Nellen}

\affiliation{Instituto de Ciencias Nucleares, Universidad Nacional Autónoma de Mexico, Ciudad de Mexico, Mexico }

\author{R.~Noriega-Papaqui}

\affiliation{Universidad Autónoma del Estado de Hidalgo, Pachuca, Mexico }

\author{N.~Omodei}

\affiliation{Department of Physics, Stanford University: Stanford, CA 94305–4060, USA}

\author{M. Osorio-Archila}

\affiliation{Instituto de F\'{i}sica, Universidad Nacional Autónoma de México, Ciudad de Mexico, Mexico}

\author{E.~Ponce}

\affiliation{Facultad de Ciencias F\'{i}sico Matemáticas, Benemérita Universidad Autónoma de Puebla, Puebla, Mexico }

\author{Y.~Pérez Araujo}

\affiliation{Instituto de F\'{i}sica, Universidad Nacional Autónoma de México, Ciudad de Mexico, Mexico}

\author{E.G.~Pérez-Pérez}

\affiliation{Universidad Politecnica de Pachuca, Pachuca, Hgo, Mexico }

\author{Q.~Remy}

\affiliation{Max-Planck Institute for Nuclear Physics, 69117 Heidelberg, Germany}

\author{A.~Rodriguez Parra}

\affiliation{Universidad Michoacana de San Nicolás de Hidalgo, Morelia, Mexico }

\author{D.~Rosa-González}

\affiliation{Instituto Nacional de Astrof\'{i}sica, Óptica y Electrónica, Puebla, Mexico}

\author{M.~Roth}

\affiliation{Los Alamos National Laboratory, Los Alamos, NM, USA }

\author{H.~Salazar}

\affiliation{Facultad de Ciencias F\'{i}sico Matemáticas, Benemérita Universidad Autónoma de Puebla, Puebla, Mexico }

\author{A.~Sandoval}

\affiliation{Instituto de F\'{i}sica, Universidad Nacional Autónoma de México, Ciudad de Mexico, Mexico}

\author{J.~Serna-Franco}

\affiliation{Instituto de F\'{i}sica, Universidad Nacional Autónoma de México, Ciudad de Mexico, Mexico}

\author{Y.~Son}

\affiliation{University of Seoul, Seoul, Rep. of Korea}

\author{R.W.~Springer}

\affiliation{Department of Physics and Astronomy, University of Utah, Salt Lake City, UT, USA }

\author{O.~Tibolla}

\affiliation{Universidad Politecnica de Pachuca, Pachuca, Hgo, Mexico }

\author{K.~Tollefson}

\affiliation{Department of Physics and Astronomy, Michigan State University, East Lansing, MI, USA }

\author{I.~Torres}

\affiliation{Instituto Nacional de Astrof\'{i}sica, Óptica y Electrónica, Puebla, Mexico}

\author{R.~Torres-Escobedo}

\affiliation{Tsung-Dao Lee Institute \& School of Physics and Astronomy, Shanghai Jiao Tong University, 800 Dongchuan Rd, Shanghai, SH 200240, China}

\author{R.~Turner}

\affiliation{Department of Physics, Michigan Technological University, Houghton, MI, USA }

\author{E.~Varela}

\affiliation{Facultad de Ciencias F\'{i}sico Matemáticas, Benemérita Universidad Autónoma de Puebla, Puebla, Mexico }

\author{L.~Villaseñor}

\affiliation{Facultad de Ciencias F\'{i}sico Matemáticas, Benemérita Universidad Autónoma de Puebla, Puebla, Mexico }

\author{I.J.~Watson}

\affiliation{University of Seoul, Seoul, Rep. of Korea}

\author{H.~Wu}

\affiliation{Dept. of Physics and Wisconsin IceCube Particle Astrophysics Center, University of Wisconsin{\textemdash}Madison, Madison, WI, USA}

\author{S.~Yu}

\affiliation{Department of Physics, Pennsylvania State University, University Park, PA, USA }

\author{X.~Zhang}

\affiliation{Institute of Nuclear Physics Polish Academy of Sciences, PL-31342 IFJ-PAN, Krakow, Poland }

\author{H.~Zhou}

\affiliation{Tsung-Dao Lee Institute \& School of Physics and Astronomy, Shanghai Jiao Tong University, 800 Dongchuan Rd, Shanghai, SH 200240, China}

\author{C.~de León}

\affiliation{Universidad Michoacana de San Nicolás de Hidalgo, Morelia, Mexico }




\vspace{1cm}
\begin{abstract}
We present an updated catalog of TeV gamma-ray sources based on the fifth pass of data from the High-Altitude Water Cherenkov (HAWC) Observatory. This release benefits from improved event reconstruction and nearly three additional years of observations. It also incorporates a systematic multi-source fitting framework, enabling more flexible and accurate modeling of the gamma-ray sky. This fitting procedure was modeled after the manual approach used in HAWC analyses of individual sources and regions as well as other gamma-ray catalogs, like the 4FGL. In addition to more varied modeling of source morphology and spectral parameters compared to previous HAWC catalogs, this catalog uses a robust modeling of Galactic diffuse TeV emission. The fitting procedure uses both point-like and symmetric Gaussian spatial templates to model the source morphology. The spectral shape of the emission is modeled with either a simple power-law or log-parabola to explore curvature in the spectral energy distribution. We report 85 sources at the 4$\sigma$ level, including 11 sources not associated with any TeVCat source using a distance-based association criterion. Distance-based association with the 1LHAASO catalog results in 22 4HWC sources without a counterpart. Additionally, there are 12 sources not associated with any physical counterpart in the Low- or High-Mass X-Ray Binary, the ATNF, or \textit{Fermi} Pulsar, or SNR catalogs of sources. Five of the aforementioned sources have no counterpart in any of the catalogs searched and represent an opportunity for follow-up observations.

\end{abstract}



\section{Introduction} \label{sec:intro}
Accurate and sensitive large-scale surveys of the gamma-ray sky are invaluable tools both in their ability to directly study and analyze possible gamma-ray-emitting objects, and to direct other instruments and scientists towards areas of interest. Close study of the variety of very-high-energy (VHE; $>$100~GeV) and ultra-high-energy (UHE; $>$100~TeV) gamma-ray sources provides insights into many of the pressing questions in modern astrophysics, including, but certainly not limited to, the study of Dark Matter, the mechanisms behind the acceleration of cosmic rays, and the constraint on any new physics beyond the standard model at these very-high and ultra-high energies. 

The most precise instruments in the VHE gamma-ray energy range are Imaging Atmospheric Cherenkov Telescopes (IACTs), like the High-Energy Stereoscopic System (H.E.S.S. \citep{HESS}), the Major Atmospheric Gamma Imaging Cherenkov (MAGIC) telescopes \citep{Magic}, the Very-Energetic Radiation Imaging Telescope Array System (VERITAS \citep{Veritas}), and the upcoming Cherenkov Telescope Array Observatory (CTAO \citep{CTA}), which are capable of extremely accurate measurements of both source position and spectral properties. However, these very same properties also make large-scale surveys of the sky difficult for such instruments. While IACTs typically have good angular resolution ($\le 0.1\degree$), they also typically have small fields of view (FoVs)--- in the single digits of degrees. This, in addition to the low duty cycle necessitated by the requirement for dark and cloudless nights, makes a comprehensive survey of the visible sky from an IACT instrument a challenging endeavor. As such, the production of a survey of the TeV gamma-ray sky is more easily undertaken by Extensive-Air-Shower (EAS) arrays like the Large High-Altitude Air Shower Observatory (LHAASO) \citep{1LHAASO}, Tibet-AS$\gamma$ \citep{Tibet}, the Astrophysical Radiation with Ground-based Observatory at YangBaJing (ARGO-YBJ) \citep{ARGO}, and the High-Altitude Water Cherenkov (HAWC) Observatory. While each of these instruments are unique and complementary, they generally have very-high duty cycles and very large FoVs, which make them preeminent survey instruments for the TeV gamma-ray sky. 

\begin{figure*}[t]
    \centering
    \includegraphics[width=0.999\textwidth]{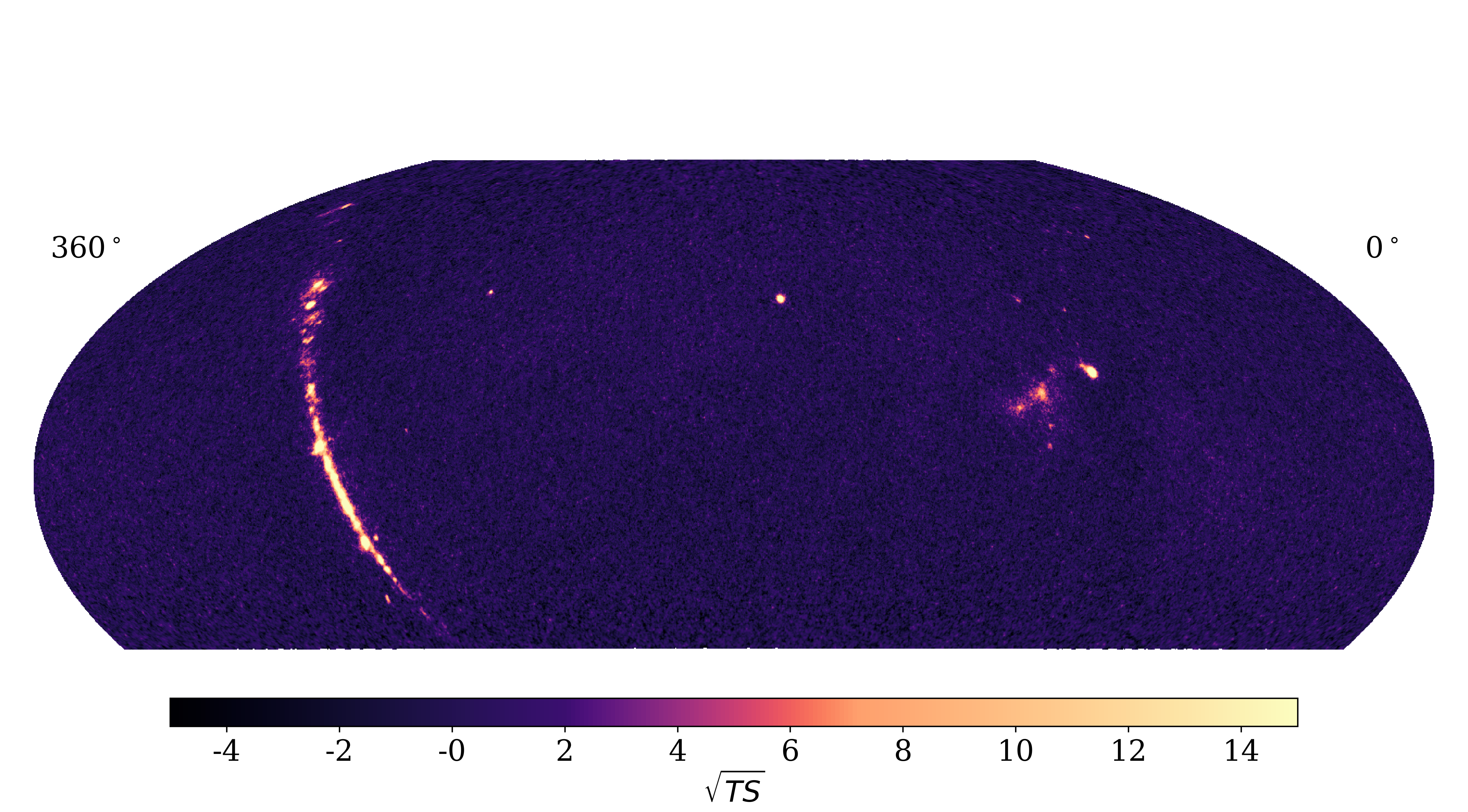}
    \caption{The full all-sky significance map from the HAWC map used to create the 4HWC catalog. This map is in celestial coordinates, and the significance shown is calculated using a point-source assumption with a fixed $-2.6$ power-law index.}
    \label{fig:AllSkyMap}
\end{figure*}

The HAWC Observatory has been operating for nearly 10 years, observing the Northern gamma-ray sky at energies from hundreds of GeV to hundreds of TeV. We present an automated multi-source-fit analysis of the latest Pass~5 \citep{Pass5Perf} all-sky data (Figure~\ref{fig:AllSkyMap}) from the HAWC Observatory. This new method performs a likelihood-ratio-based fitting procedure of subsections of the HAWC dataset. These fits do not use any results from previous HAWC analyses or other TeV astronomy results. As a result, the catalog may not contain all sources from the previous HAWC catalogs, unlike, e.g., the 4FGL catalog. However, this reduces the possible biases from using previous HAWC data reconstructions or results that other instruments may introduce. A systematic and comprehensive survey provides useful results on new and old source populations like microquasars, TeV halos, and other astrophysical objects. All-sky surveys also serve to provide valuable context and motivation for pointing instruments to guide their limited observation time to produce in-depth studies of the morphology and spectra of these objects. While a survey cannot provide the same level of detail and comprehensive results that a dedicated study of a small region can, it does provide a powerful starting point for others to utilize. 

\section{Instrument and Data Selection and Analysis}

\subsection{The HAWC Detector}

The HAWC main array is made up of 300 large steel tanks containing approximately 200,000 liters of water each in internal water- and light-tight bladders. Each bladder has four upward facing PhotoMultiplier Tubes (PMTs) mounted at the bottom: one larger 10-inch PMT at the center of the tank and three smaller 8-inch PMTs forming an equilateral triangle with the larger PMT at its center. 


The reconstructed gamma-ray dataset covers nearly two thirds of the gamma-ray sky (declination $-26$\degree\, to $64$\degree). The relevant technical details of both the physical array and the newest Pass~5 air-shower reconstruction methods can be found in two papers: \citet{Pass5Perf} and \citet{NimPaper}. 

Two major improvements characterize the 4HWC catalog: a new automated analysis method and the improvement of the data quality and quantity. The Pass~5 dataset has many dramatic improvements over the Pass~4 dataset used to construct the 3HWC catalog \citep{3HWC}. Some of the most impactful improvements relate to low-energy and/or high-zenith-angle events. At the lowest energies, the background rejection has improved significantly due to improved noise-rejection techniques. Greatly improved gamma/hadron separation algorithms improve the quality of data; different optimal gamma/hadron separation cuts were introduced for events that land inside or outside the footprint of the HAWC main array. Further optimized cuts were developed for different zenith angles, drastically improving performance at higher zeniths ($>37$\degree). In older HAWC analyses using the previous Pass~4 datasets  the smallest-footprint events were excluded due to high noise and poor fitting. Using a new shower-plane fitting method, the quality of these events was dramatically improved, allowing two additional bins with median gamma-ray energy $<$500~GeV. These additional bins provide a significant improvement at the lower end of the HAWC energy range, closing the gap between HAWC and other instruments in the GeV--TeV band like the \textit{Fermi}-LAT. 

Due to these improvements, as well as others described in \citet{Pass5Perf}, the effective area of HAWC was improved by up to a factor of five at low energies and, at the highest energies, the effective area nearly matches the total physical area of the detector. At the largest zenith angles, the Pass~5 dataset has a nearly three times better angular resolution compared to Pass~4 and approximately four times better gamma/hadron separation, measured using the quality factor defined in \citet{Pass5Perf}. 

The primary remaining background, even after applying the improved gamma/hadron separation, is from cosmic rays. This background is directly estimated using two different methods described in detail in Section~4.3 of \citet{HAWCCrab2019}. The first method, direct integration, is used for bins with sufficient statistics during a two-hour period. This method is the standard HAWC method for estimating the isotropic cosmic-ray background and was originally developed for use by the Milagro Observatory \citep{MilagroDirectIntegration}. The expected background can be modeled as: \\
\begin{equation}
\begin{split}
    N(R.A.,Dec.) =  \iint E&(HA, R.A.) R(t) \\ &*\epsilon(R.A.,HA,t) dtd\Omega\,,
\end{split}
\end{equation}
where $E(HA, R.A.)$ represents the probability of an event comes from the differential solid angle $d\Omega$ at a given Right Ascension (R.A.) and hour angle (HA), $R(t)$ is the count rate of the detector at time t, and $\epsilon (R.A.,HA,t)$ is 1 if the event falls within the correct R.A. and Declination (Dec.) range for the spatial bin at which you are looking and 0 otherwise. The details and more complete definitions of this direct-counting method are given in Section~5 of \citet{MilagroDirectIntegration}. The second background-estimation method is used in the highest energy bins, which don't have large-enough statistics to use the direct-integration method. This method uses an all-sky estimate of the isotropic cosmic-ray background and draws a random location in the sky using local coordinates, converting them to the R.A. and Dec. of the event. This process is repeated 10,000 times for each event to generate a background map, which is then normalized by the number of events in the data. This provides a smoothed background map in the highest energy bins. These two methods provide a robust estimate of the background of cosmic rays that survive the gamma/hadron cuts.

In total, the dataset used in this work has a live time of approximately 2,726~days. The previous 3HWC catalog had an uptime of 1,523~days. The dataset uses the 1-D fHit binning scheme. This scheme divides the air-shower events into 11 bins by the fraction of PMTs hit by the air-shower. The bins are defined such that the number of events in each bin reduces by approximately half. These bins are then subdivided into events which central axis or core lands inside or outside the footprint of the main HAWC array. All 22 bins in total are used in this analysis. This additional data, alongside the discussed improvements in data reconstruction quality, improves the sensitive energy range of this catalog. At nearly overhead zenith angles ($<10\degree$) for a point-like source with a power-law index of $-2.65$ and flux approximately 5\% of the Crab, the fit spectral parameters are valid from around 0.11~TeV to 232~TeV. For high zenith angles ($>35\degree$) using the same source, the spectral parameters should be considered valid from 0.94~TeV to 196~TeV. This estimate can change depending on the exact flux level, spectral shape, and declination of each individual source.

\section{Source Search Methodology}
\vspace{0.5cm}
\subsection{The Automated Likelihood Pipeline Search} \label{subsection:Fitmethod}
\begin{figure*}[]
    \centering
    \includegraphics[width=1.00\textwidth]{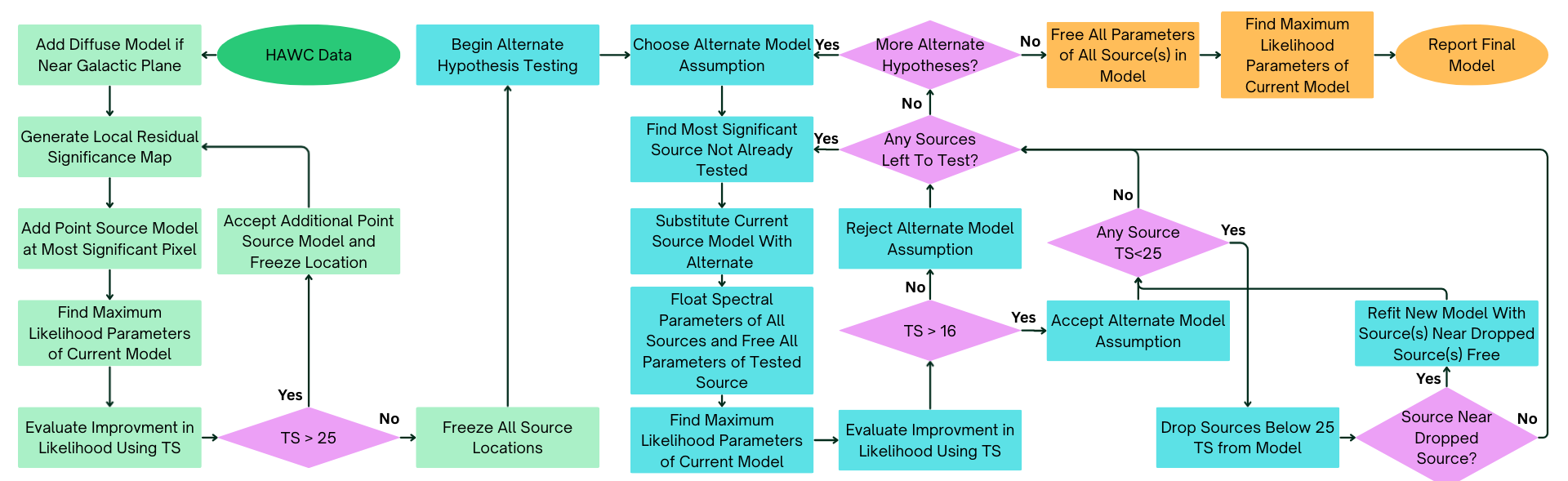}
    \caption{Flowchart of the ALPS method. Beginning with the dark green ``HAWC Data" oval node, the green sections correspond to the Point-Source-Adding phase. The cyan sections show the flow of the Alternate-Hypothesis-Testing phase. The orange sections show the final refitting section that precedes the final output model. Each diamond-shaped purple node shows a branching point in the algorithm.}
    \label{fig:FlowChart}
\end{figure*}

\vspace{0.5cm}
The Automated Likelihood Pipeline Search (ALPS) method works to create a reproducible, automated, and unbiased pipeline for finding and fitting gamma-ray sources in any region of the sky in HAWC data. A visual flowchart of the ALPS method is shown in Figure~\ref{fig:FlowChart}. The basis of this pipeline is the Maximum Likelihood Estimation (MLE) of gamma-ray model parameters and the likelihood ratio test to evaluate the statistical improvement of different models based on the work in the \textit{Fermi}-LAT extended source catalog \citep{FermiExt}. The full likelihood function evaluated for each set of model parameters is:
\begin{equation}
    \log{\mathcal{L}\left(\theta\right)} = \sum_{b=0}^{10}\sum_{p=0}^N \log\left(\frac{E(x_{b,p}\lvert\theta)^{x_{b,p}}\times e^{-E(x_{b,p}\lvert\theta)}}{x_{b,p}!}\right)
    \label{eq:Likelihood}\,,
\end{equation}

\noindent where $x_{b,p}$ is the observed gamma-ray counts and $E(x_{b,p}\lvert\theta)$ is the expected photon counts in a given pixel, $p$, in a given data bin, $b$, using a forward-folded gamma-ray model plus background with parameters $\theta$. Using this likelihood, we can maximize the value of $\log{\mathcal{L}\left(\theta\right)}$ to find the MLE parameters, $\theta$. Then, to compare different models, we construct a Test Statistic (TS) based on Wilks' theorem \citep{Wilks} under the condition that the parameters form a nested pair of models. The TS is calculated using a likelihood ratio as follows:
\begin{equation}
    TS = -2*\left(\LogLike{Null} - \LogLike{Alternate} \right)\, ,
\end{equation}
where the null model may represent a pure background assumption or a less complex model and the alternate model is the model with additional parameters not present in the null model. The distribution of this TS is a $\chi^2$ with degrees of freedom equal to the number of additional parameters in the alternate model compared to the null. The validity of assuming Wilks' theorem and the $\chi^2$ distribution of this TS was discussed in the 2HWC catalog \citep{2HWC}. For areas away from possible sources, the background significance distribution ($\sqrt{TS}$) follows the predicted Gaussian distribution well, confirming that Wilks' theorem is valid.

\subsubsection{On- and Off-Plane Region-Of-Interest Selection} \label{subsection:ROIselect}

The process of computing the MLE parameters is very computationally intensive and the processing time grows highly non-linearly with the amount of significant gamma-ray emission in a region and the number of free parameters. For each additional source, between four to six additional degrees of freedom are added to the model parameter space. Each degree of freedom expands the number of times this likelihood must be calculated to perform the maximization of the log likelihood.

To complete an all-sky survey in a feasible time frame, the sky must be subdivided into many smaller Regions Of Interest (ROIs), which can be run independently of each other in parallel. For the Galactic Plane, we used many small sliding ROIs to fit the sky $b \le |10|\degree$. These ROIs are $6\degree$ in longitude and $20\degree$ in latitude and are centered starting at $(l=-1\degree,b=0\degree)$ and move $3\degree$ until $(l=245\degree,b=0\degree)$. Thus, all portions of the sky $b \le |10|\degree$ and $-1\degree \le l \le 245\degree$ will be contained in a minimum of two ROIs. Any sources that are on the edge of an ROI will be close to the center of another. This provides a cross-check for any sources found in these regions, but also presents a challenge for automatically removing duplicate sources in these overlapping regions when combining the results from all ROIs. 

For regions away from the Galactic Plane, ROIs were defined as $10\degree \times 10\degree$ squares around any pixel with a $\sqrt{TS} > 5$ and a location more than $5\degree$ away from any previously identified pixel. The only exception to this was the region near the known TeV Halo sources Geminga and Monogem. A special extended rectangular ROI contains both sources since using the smaller, standard off-plane ROIs here would cause contamination issues for both sources. This method of subdividing the sky leads to a total of 128 ROIs, which are each analyzed using the method described in the following subsections.

\subsubsection{Point-Source-Adding Phase}
The analysis pipeline begins by selecting a given ROI from those detailed in the previous section. The pixel with the largest significance value is chosen as the first seed location for a new source model to be added. 

If an ROI does not contain the Galactic Plane (ROI centered at $\lvert b \rvert >10\degree$ in Galactic coordinates), the pipeline moves immediately to adding a source model at this maximum significance pixel.  

For regions that do contain the Galactic Plane, a Galactic diffuse model is added before any fitting begins. This model is generated using the HERMES framework \citep{HERMES} to calculate the predicted gamma-ray emission from Galactic cosmic rays. Our HERMES model uses the default cosmic-ray distribution assumed by DRAGON \citep{DRAGON1,DRAGON2}, the matter distribution from the DAME CO survey \citep{dame2003largeextensioncfagalactic} to trace molecular gas, and the HI4PI survey \citep{HI4PI} to trace atomic gases. The HERMES code is capable of computing many radiative processes, but we limit our modeling to two of the most significant at HAWC energies--- Inverse Compton (IC) and pion decay \citep{HERMES}. Once the predicted emission is computed from these processes, it is added as a template with a fixed normalization and shape to the model and the fitting proceeds to generate a new residual significance map of the ROI with the emission from the HERMES template model taken into account. The most significant pixel in this residual emission significance map is now taken as the new seed location for adding a point source model. The point-like source has free location parameters and a power-law spectral assumption given by:
\begin{equation}
    \frac{dN}{dE} = N_0\left(\frac{E}{E_0}\right)^{-\alpha}\enspace ,
\end{equation}
 where $N_0$, $E_0$, and $\alpha$ are the flux normalization, pivot energy, and spectral index, respectively. The pivot energy is fixed at 2~TeV and is the value that minimizes the correlation between the flux normalization and spectral index for a Crab-like source at 0\degree \space zenith. This model represents the simplest assumptions in terms of both spatial and spectral modeling. The location of the source is allowed to vary over a range of 3\degree \space and both the normalization and index of the source spectrum are free. The MLE values for these four parameters are fit using the maximum likelihood framework described in \citet{3ml}. After the best-fit parameters are found, the location of the source is fixed and a new local residual significance map is made which subtracts the emission from all models, including the HERMES template if present. A new source is placed at the location of the most significant pixel in this residual map and the process repeats.
 
 This point source fitting process is repeated until the TS value calculated using the models before and after adding the source falls below 25. This TS threshold of 25 corresponds to a p-value of $5.03 \times 10^{-5}$, or a z-score of approximately 3.89, when we look at adding an additional point source with four free parameters. All TS thresholds for accepting new source models and accepting alternate source modeling hypotheses are chosen to correspond with a similar level of significance. 
 
 \subsubsection{Morphology Hypothesis Testing Phase}

 We then test whether each source is morphologically extended. We characterize an extended source as a symmetric two-dimensional Gaussian with an extension parameter represented by its standard deviation, $\sigma$. The pipeline begins with the source with the highest individual TS value found by calculating the TS of the model with and without each source one by one. The reduction in likelihood found by removing each source corresponds to their individual TS. The TS threshold for this phase is set at 16 (p-value of $3.17 \times 10^{-5}$) as only one additional extension parameter is added when altering the model. 

 If the tested source, which has had its location freed, is best-fit as extended, any other sources that now have an individual TS below 25 are removed from the model. The entire model is then refit, allowing any source near a dropped source to fully refit in its absence. Both location and spectral parameters are free for nearby sources, while locations far away from any dropped sources remain fixed, but their spectrum is still free. The distance at which a source is considered nearby is 0.3\degree \space plus the sum of the extension parameters, if any, of the two sources.

  \subsubsection{Spectral Curvature Testing Phase}
  
 After the morphology testing phase of the pipeline is completed, we test for evidence of spectral curvature in each of the remaining sources. The chosen model to test for this curvature is a log parabola spectral model:
 \begin{equation}
     \frac{dN}{dE} = N_0\left(\frac{E}{E_0}\right)^{-\alpha-\beta\log{\left(\frac{E}{E_0}\right)}}\enspace ,
 \end{equation}
 which collapses to the power-law spectral assumption when $\beta=0$. 

 After the spectral curvature test is completed the lest remaining step is a final refit of the model. The final refit allows all source locations, extensions if applicable, and spectral parameters to be free and performs one simultaneous fit for all parameters to ensure that the true maximum likelihood parameters for the finalized model has been found. This step also calculates the statistical errors for each parameter by sampling the likelihood curves.

\subsection{Systematic Uncertainties}
\label{sec:systematic}

Systematic uncertainties for the catalog as a whole were estimated by taking a sample of ROIs in different regions of the HAWC sky. These ROIs include the regions containing the Crab, Cygnus Cocoon, Galactic Center, Mrk~421, Mrk~501, and M87. These regions were chosen to include low- and high- zenith areas of the sky as well as sources of likely Galactic and extragalactic origin. They are also contain a mix of point-like and extended sources as well as a mix of log-parabola and power-law spectral models.

To estimate the pointing systematics for Right Ascension (R.A.) and Dec., the locations of the point source models fit to the Crab, Mrk~501, and M87 were compared with IACT observations. The deviations were added in quadrature to compute an estimated pointing systematic uncertainty of 0.04\degree \space in R.A. and 0.02\degree \space in Dec. 

The source model for Mrk~421 was excluded from this pointing study as the best-fit model is an extended source and represents systematic uncertainty in the extension estimation. Mrk~421 is an extragalactic source, and is expected to appear as a point source; however, this model has a best-fit extension of 0.09\degree \space or about two pixel widths in the HEALPix maps used to spatially bin the HAWC skymaps \citep{HealPix}. This best-fit extension of 0.09\degree \space represents a systematic uncertainty in the estimation of the extension of its emission and is quoted as the systematic uncertainty for all sources in the catalog.  

The systematic uncertainty of the spectral parameters were fit using alternate detector response functions. These functions simulate systematic differences in the performance of HAWC compared to our current modeling. The details of all the different effects tested by these alternate response functions are described in Section~4.5 of \citet{HAWCCrab2019}. However, for this analysis, the bin-selection uncertainty described in Section~4.5.6 was not used as the 2D binning scheme for the energy estimator maps in HAWC is not present in the fHit binning scheme. Additionally, the alternate detector response corresponding to the partial array used for earlier HAWC analyses was removed as this data is now a negligible portion of the data and does not contribute to the systematic uncertainties. Adding the effects of each of the applicable categories of systematics in quadrature and averaging over the sources contained in the six ROIs sampled results in overall spectral systematic uncertainties for the flux measurement of +4.71\% and $-10.73$\%. For the spectral index, we find an overall systematic of +4.01\% and $-4.85$\%. For the spectral curvature parameter $\beta$ of any log-parabola source spectrum models, we find an overall systematic of +15.38\% and $-8.11$\%. 
\vspace{-0.1cm}

\subsection{Source Detection Rate Estimation}

To estimate the efficiency and false-positive rate of the ALPS methodology on HAWC data, two different sets of fake sky maps were made to test the ability of the pipeline to find above-threshold sources and its propensity to detect false sources. 

\subsubsection{Source Recovery Estimation}
For the first set of maps, we selected 125 random 10\degree $\times$10\degree \space ROI within the sky viewable by HAWC.  We then injected 1--4 sources with randomized locations within the ROI with randomized morphology and power-law spectra. Randomized parameters were chosen using triangular distributions. Spectral index values ranged from 1.7 to 3.5 centered on 2.7, Differential Flux ($\frac{1}{\mathrm{TeV} \mathrm{cm}^{2} \mathrm{s}}$) ranged from $6\times 10^{-15}$ to $6\times 10^{-13}$  centered on $1.2\times 10^{-13}$. Sources were equally likely to be extended or point-like. The extension parameter was randomly chosen between 0.2\degree \space and 2.5\degree \space centered at 0.5\degree. We Poisson fluctuated the expected background (for that region of the sky) and the injected source emission. Of these 125 generated maps, 97 were able to recover the exact injected morphology within $2\sigma$ statistical uncertainty for all injected sources. Out of a total of 239 injected sources, the method recovered 190. An average of 10\% of injected sources were injected with fluxes below the expected sensitivity for a point-like source and are likely true negatives. Some of the incorrect modeling included overlapping extended sources with similar spectra being fit as a single extended source, low-flux-density extended sources unable to fit a seed point source nearby, and high-flux-density extended sources close to near-threshold point sources causing the significance of the weaker point source to drop below detection. 

Some of these issues represent statistical or physical problems with modeling complex regions. However, an important result of the test is that, of the 28 regions that did not match the exact models within $2\sigma$, none reported more sources than injected or sources without a corresponding injected source. Zero false sources were reported during this test--- that is, the method only underestimated the number of sources. The ALPS method was able to recover approximately 80\% of the total injected sources, of which 90\% are expected to be above the detection threshold.   

\subsubsection{False-Positive Rate Estimation}
To directly estimate the false-positive rate, 100 more maps were made, using the same randomized locations in the HAWC sky, with no significant or detectable sources injected, simulating pure background and undetectable source contributions ($\le 10\%$ sensitive flux). 
This test found zero sources above the 25~TS threshold for inclusion in the catalog and 22 sources above the 16~TS. Using Wilks' theorem, the expected TS distribution should follow a $\chi^2$ with degrees of freedom equal to four. The tail probability of a TS greater than 16 would be a pre-trials p-value of 0.003, corresponding to only 0.3 expected sources in 100 runs. However, within the 10\degree \space X 10\degree \space ROI, we estimate that there are approximately 88 independent regions using the average point-spread function (PSF) of HAWC. Adjusting the pre-trials significance using 88 trials, as the source is located at the most significant pixel in the map so we need only consider independent trials, results in a corrected p-value of 0.265. Using this adjusted p-value, we estimate an expected 23.32 sources passing the 16 TS threshold. Using this same process to correct the 25 TS threshold results in an expected 0.387 sources passing the 25 TS threshold. This result agrees with the observed outcomes of the 88 completed trials observing zero sources. While a total of 128 ROIs were fit using the ALPS method, due to the overlapping construction of the Galactic Plane ROIs, fewer than 90 independent ROIs were fit. Using the results above and extrapolating to the number of independent ROIs fit during the 4HWC catalog allows us to put an upper limit of 0.5 false positives across the catalog.

\subsection{Disambiguating Sources in Overlapping Regions} \label{subsection:Assoc}

Section \ref{subsection:ROIselect} stated that regions of the sky near the Galactic Plane are contained in two different ROIs that are analyzed separately. This leads to the result that nearly all sources in this very source-rich region are double-counted if the source lists from each ROI are naively appended to each other to create the catalog. To avoid biasing the list, we adopted an automated method of disambiguating sources in overlapping regions. A simple way to automatically disambiguate these sources from each other is to assume they are circles with a radius equal to their $68\%$ containment radius. For point sources, this means their radius will be equal to the PSF at their declination and, for extended sources, they will have a radius equal to their extension parameter and the PSF added in quadrature. Once the sources have been simplified to circles, they can be associated with each other based upon their overlapping area compared to their total area when combined, called a Surface Fraction (SF).   

\begin{equation}
    SF = \frac{Source_1\cap Source_2}{Source_1\cup Source_2}
    \label{fig:AsssocCriterion}
\end{equation}

This geometric interpretation for SF, used in analysis of simulated H.E.S.S. data in \citep{OverlapHESS} and shown in Equation \ref{fig:AsssocCriterion}, ranges from 0 for completely non-overlapping to 1 if their position and size are identical. 
A threshold of 0.25 was used to disambiguate the final list given in this work. This number was determined empirically by applying a range of SF thresholds and picking the smallest value that did not combine multiple sources in the same ROI. One example is 4HWC J1906+0613 and 4HWC 1908+0616, which are found in the same ROI by the pipeline but would combine for lower thresholds of SF. This threshold of 0.25 was independently found in \citep{OverlapCTA} using a series of mock catalogs to optimize the false positive and negative rates. When two or more sources pass the association threshold, the source closest to the center of the ROI in which it was fit is chosen to remain, to minimize possible edge effects.

\section{4HWC Source List}

We present the 4HWC pipeline results here. This is first given in summary in the form of Table~\ref{table:sourceList}, which details all 85 sources, and also by drawing the location and extension of each source on top of the HAWC map (Figures~\ref{fig:Galplot1}--\ref{fig:Offplane7}). We compare this new HAWC catalog with several catalogs described in each subsection. Additionally, those sources new to TeV catalogs will be discussed individually. 

\subsection{Visualization of Results}

\begin{figure*}[]
    \centering
    \includegraphics[width=0.75\textwidth]{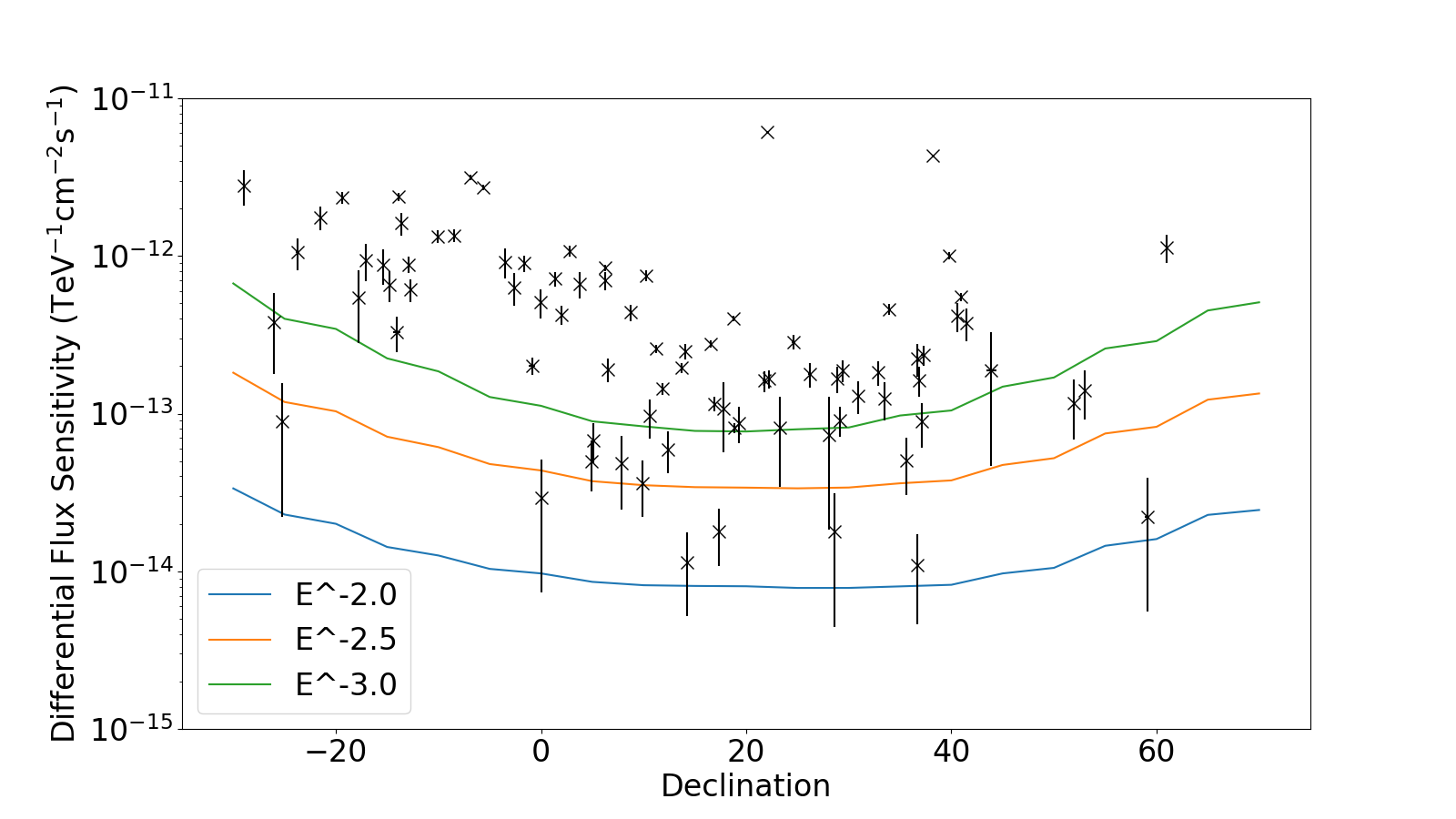}
    \caption{Sensitivity Curves for the various spectral assumptions (power-law indices $-2.0$, $-2.5$, and $-3.0$) with the flux of the 4HWC sources plotted for reference.}
    \label{fig:Sensitivity}
\end{figure*}

Figure~\ref{fig:Sensitivity} highlights several sensitivity curves for different power-law assumptions and the fluxes of the 85 sources in the 4HWC catalog. The sensitivity curve is formed by computing the required flux of a point-like source with the given simple-power-law index to be detected above $5\sigma$ at least 50\% of the time. The flux normalization is reported at 2~TeV.

\begin{figure*}[]
    \centering
    \includegraphics[width=0.75\textwidth]{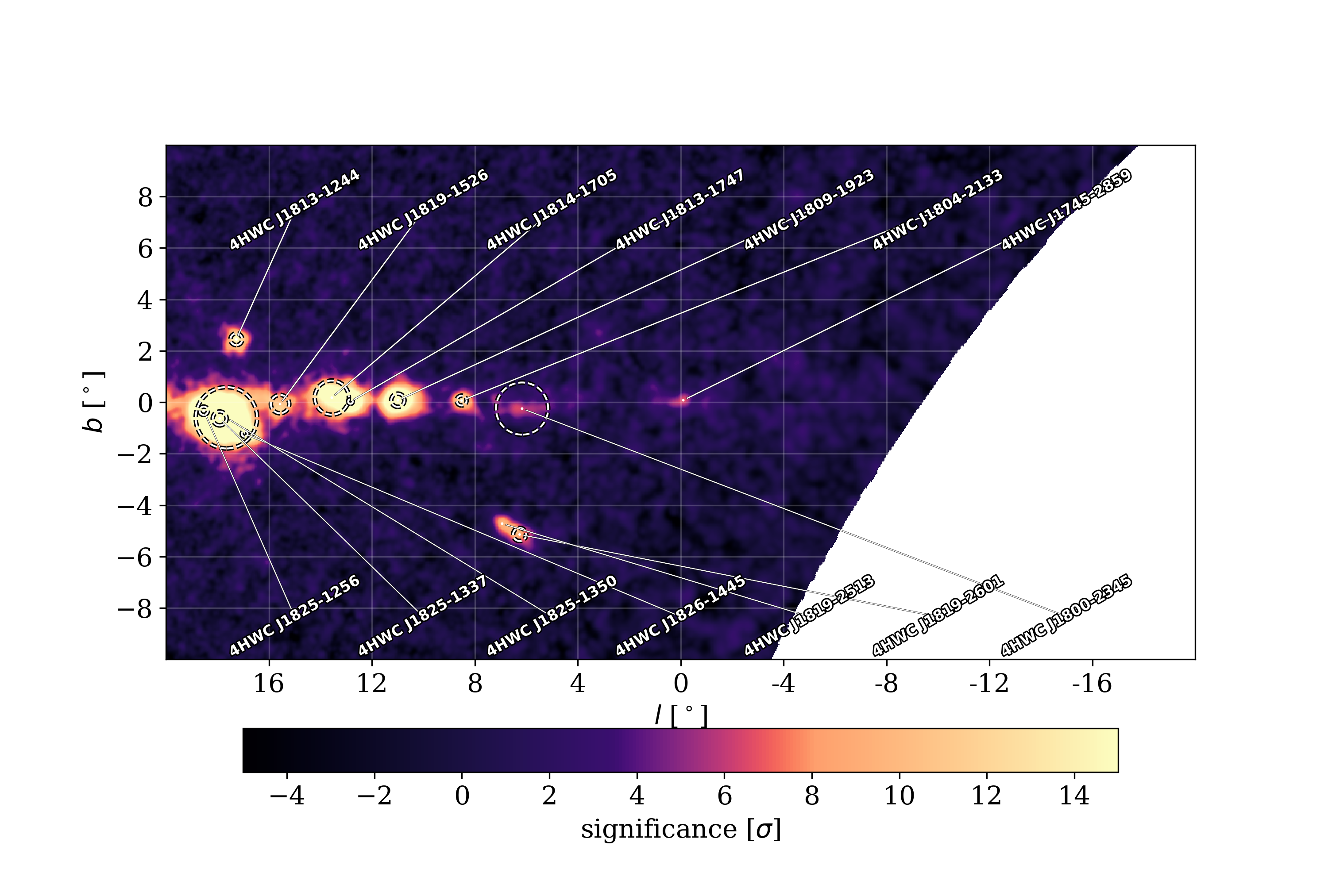}
    \caption{HAWC significance map of region $l$ = $-20$\degree \space to $l$ = 20\degree, $\lvert b \rvert <$ 10\degree \space with the 4HWC sources overlaid as dashed circles with radius equal to the extension parameter. A small dot represents the center of the source model emission. Map generated using a point-source assumption with a $-2.6$ power-law index. The white or transparent portion of the map represents empty portions of the sky beyond the zenith cut of the data set.}
    \label{fig:Galplot1}
\end{figure*}

\begin{figure*}[]
    \centering
    \includegraphics[width=0.75\textwidth]{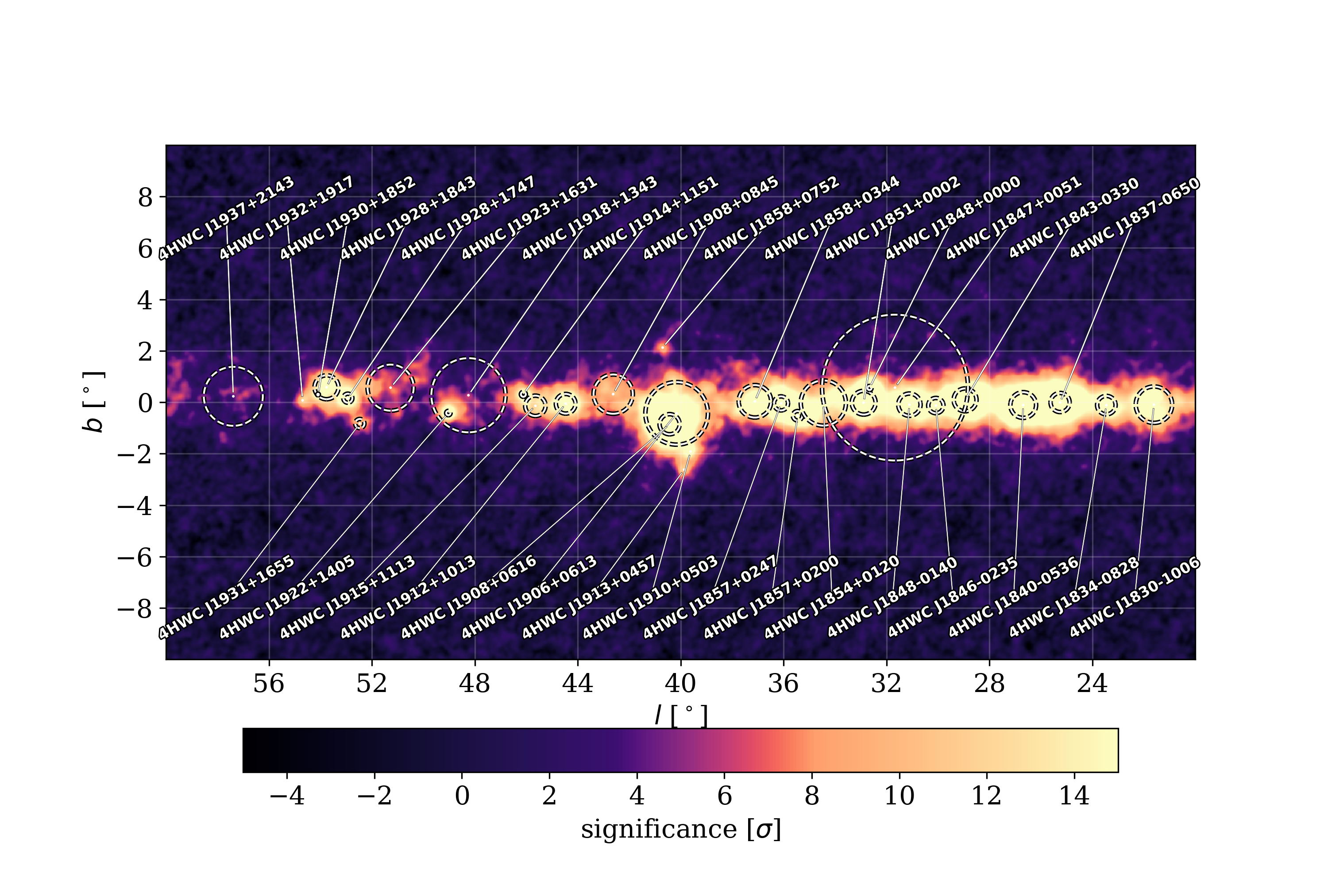}
    \caption{HAWC significance map of region $l$ = 20\degree \space to $l$ = 60\degree, $\lvert b \rvert <$ 10\degree \space with the 4HWC sources overlaid as dashed circles with radii equal to the extension parameter. A small dot represents the center of the source model emission. Map generated using a point-source assumption with a $-2.6$ power-law index.}
    \label{fig:Galplot2}
\end{figure*}

\begin{figure*}[]
    \centering
    \includegraphics[width=0.75\textwidth]{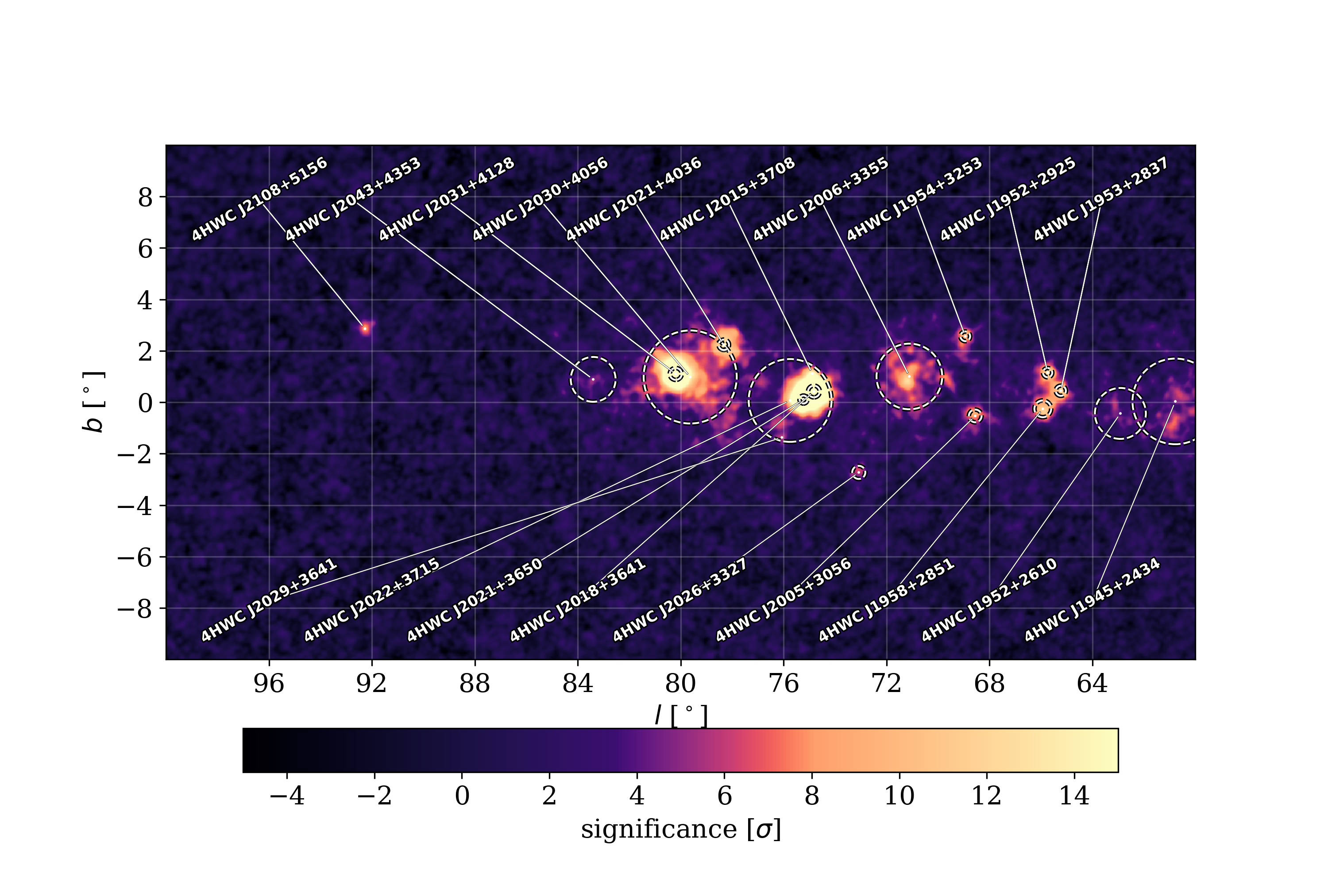}
    \caption{HAWC significance map of region $l$ = 60\degree \space to $l$ = 100\degree, $\lvert b \rvert <$ 10\degree \space with the 4HWC sources overlaid as dashed circles with radii equal to the extension parameter. A small dot represents the center of the source model emission. Map generated using a point-source assumption with a $-2.6$ power-law index.}
    \label{fig:Galplot3}
\end{figure*}

\begin{figure*}[]
    \centering
    \includegraphics[width=0.75\textwidth]{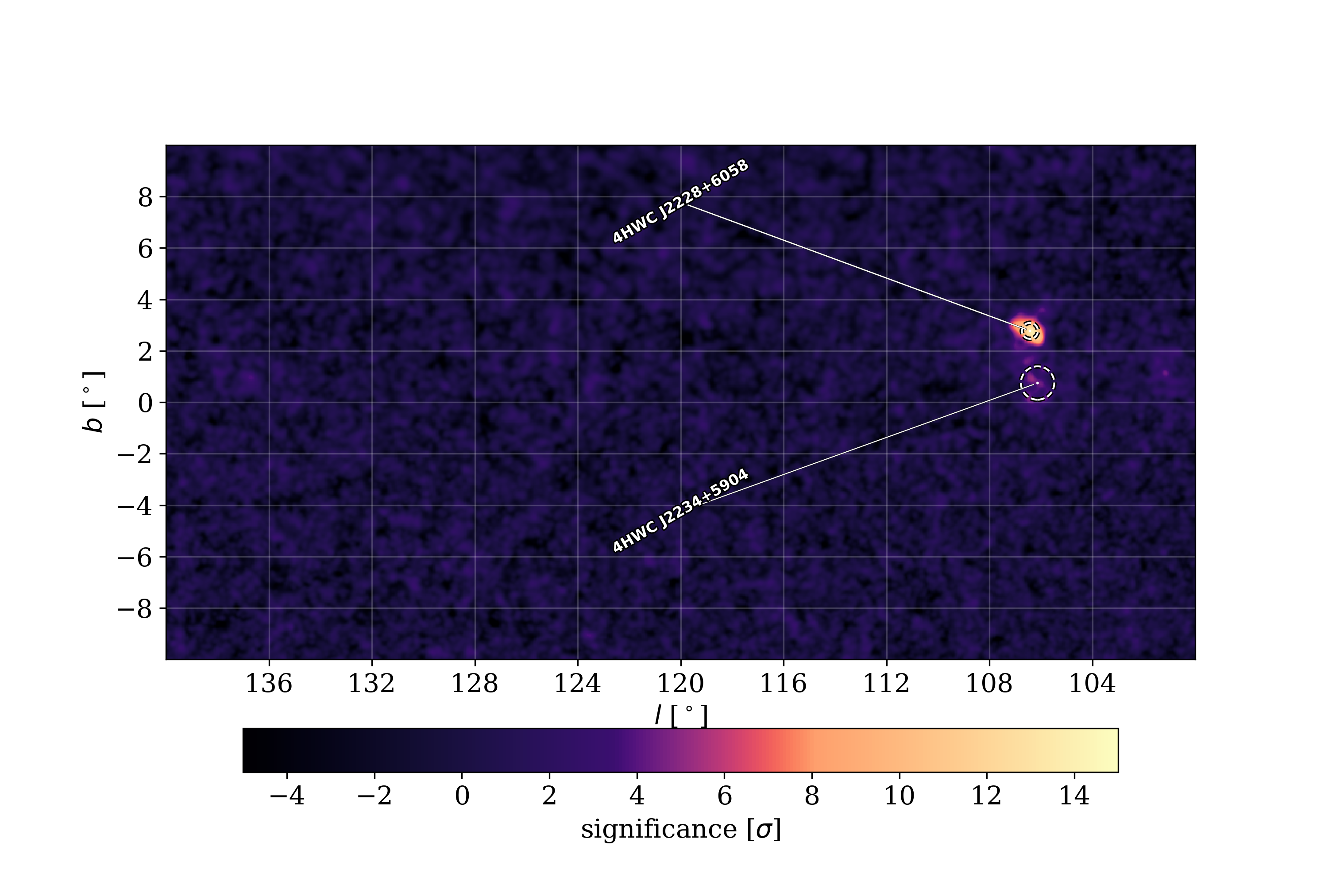}
    \caption{HAWC significance map of region $l$ = 100\degree \space to $l$ = 140\degree, $\lvert b \rvert <$ 10\degree \space with the 4HWC sources overlaid as dashed circles with radii equal to the extension parameter. A small dot represents the center of the source model emission. Map generated using a point-source assumption with a $-2.6$ power-law index.}
    \label{fig:Galplot4}
\end{figure*}

\begin{figure*}[]
    \centering
    \includegraphics[width=0.75\textwidth]{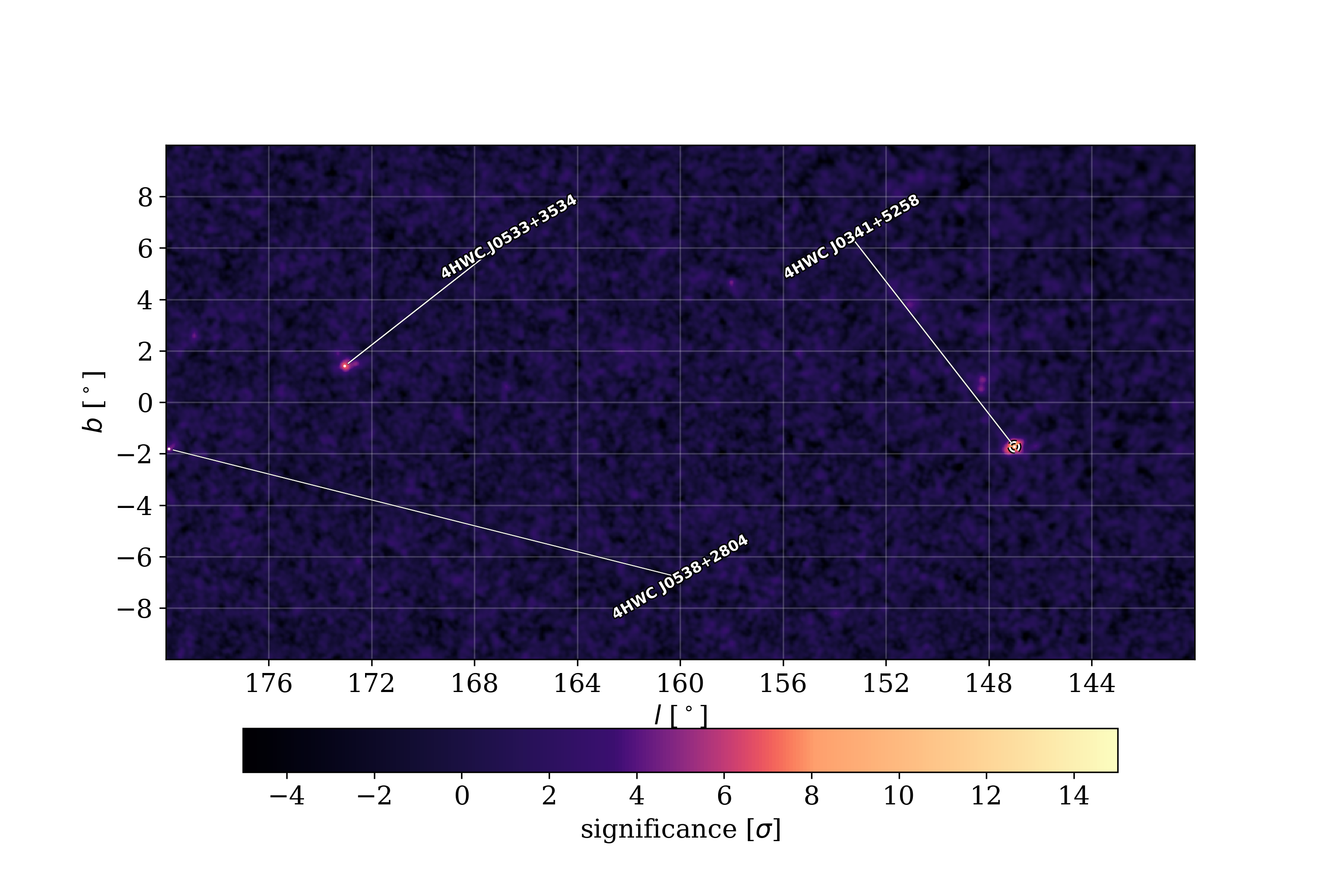}
    \caption{HAWC significance map of region $l$ = 140\degree \space to $l$ = 180\degree, $\lvert b \rvert <$ 10\degree \space with the 4HWC sources overlaid as dashed circles with radii equal to the extension parameter. A small dot represents the center of the source model emission. Map generated using a point-source assumption with a $-2.6$ power-law index.}
    \label{fig:Galplot5}
\end{figure*}

\begin{figure*}[]
    \centering
    \includegraphics[width=0.75\textwidth]{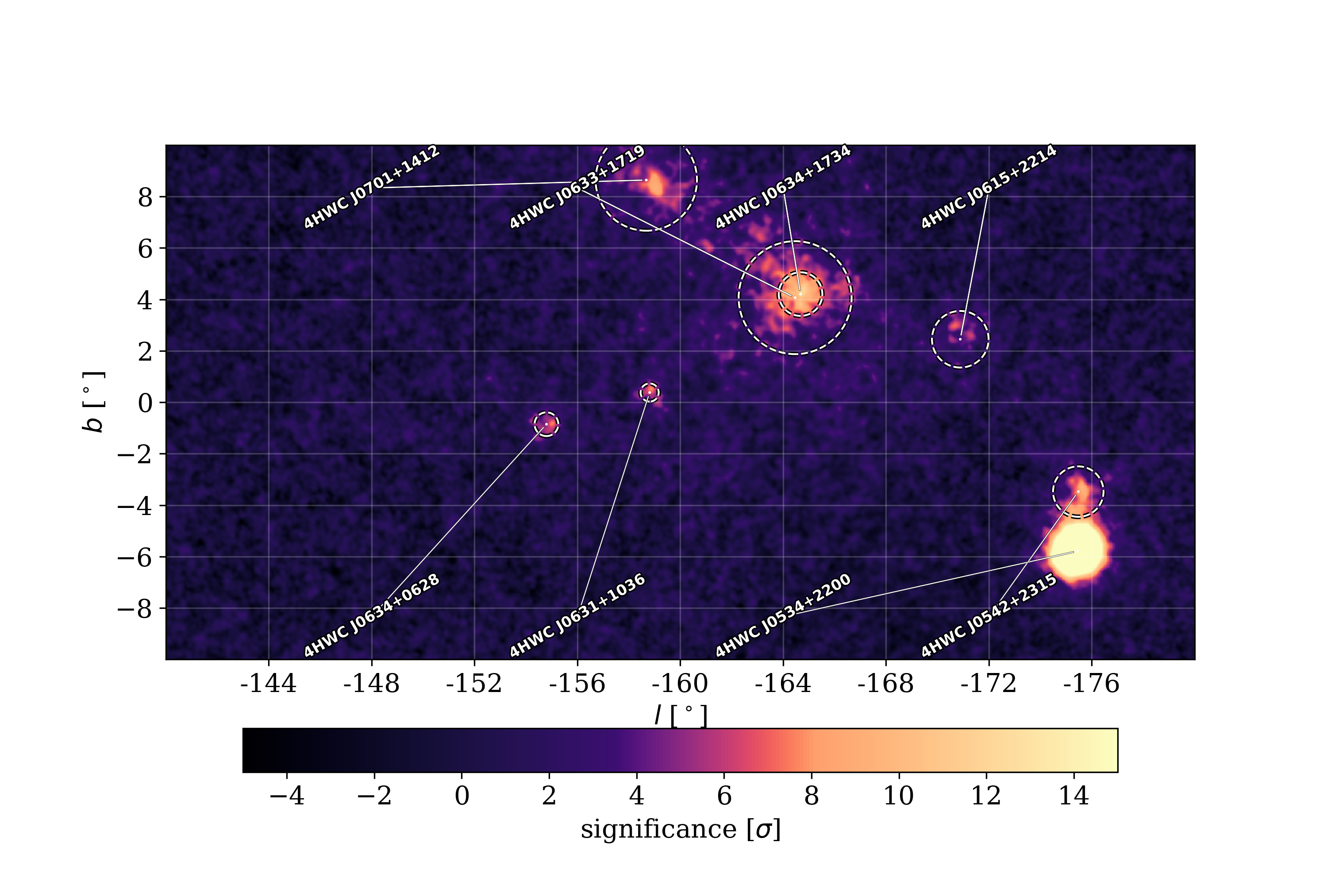}
    \caption{HAWC significance map of region $l$ = $-180$\degree \space to $l$ = $-140$\degree , $\lvert b \rvert <$ 10\degree \space with the 4HWC sources overlaid as dashed circles with radii equal to the extension parameter. A small dot represents the center of the source model emission. Map generated using a point-source assumption with a $-2.6$ power-law index.}
    \label{fig:Galplot6}
\end{figure*}

\begin{figure}[]
    \centering
    \includegraphics[trim={0 1.5cm 0 0},width=0.40\textwidth]{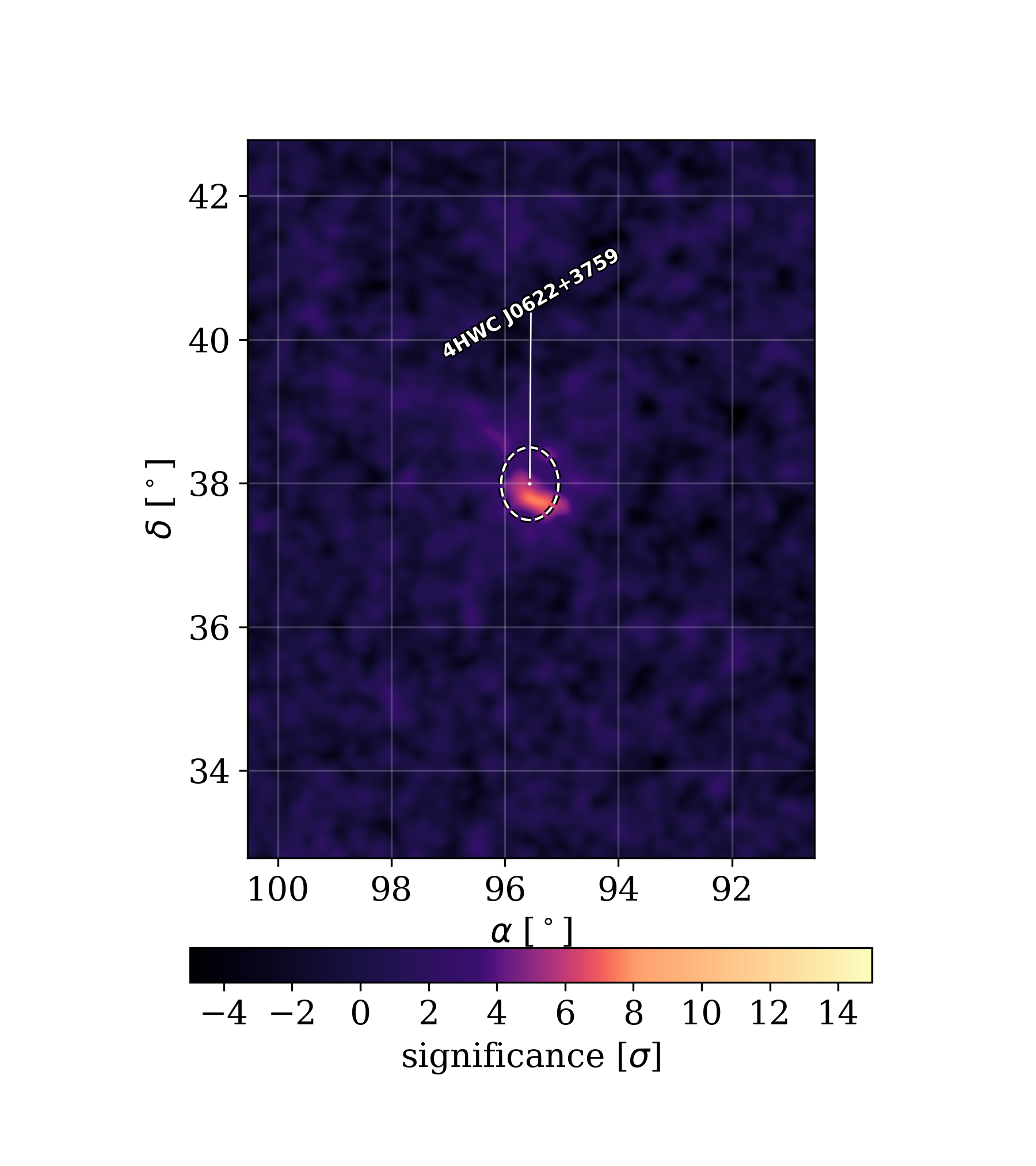}
    \caption{HAWC significance map of region R.A.~=~95.54\degree \space and Dec. = 37.78\degree \space with the 4HWC source overlaid as a dashed circle with radius equal to the extension parameter. A small dot represents the center of the source model emission. Map generated using a point-source assumption with a $-2.6$ power-law index.}
    \label{fig:Offplane1}
\end{figure}

\begin{figure}[]
    \centering
    \includegraphics[trim={0 1.5cm 0 0},width=0.40\textwidth]{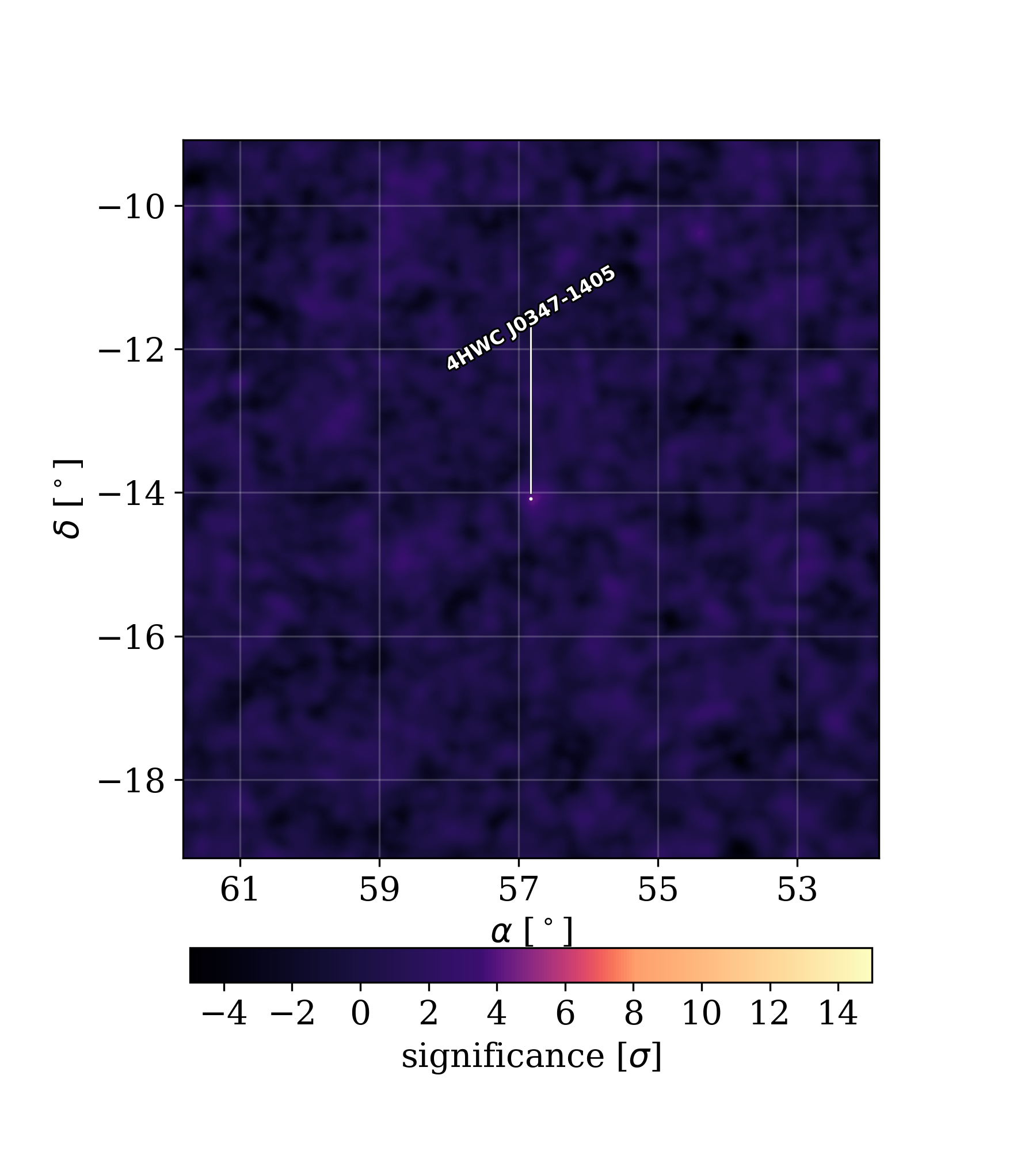}
    \caption{HAWC significance map of region R.A.~=~56.82\degree \space and Dec. = $-14.09$\degree \space with the 4HWC source overlaid as a dashed circle with radius equal to the extension parameter. A small dot represents the center of the source model emission. Map generated using a point-source assumption with a $-2.6$ power-law index.}
    \label{fig:Offplane2}
\end{figure}

\begin{figure}[]
    \centering
    \includegraphics[trim={0 1.5cm 0 0},width=0.40\textwidth]{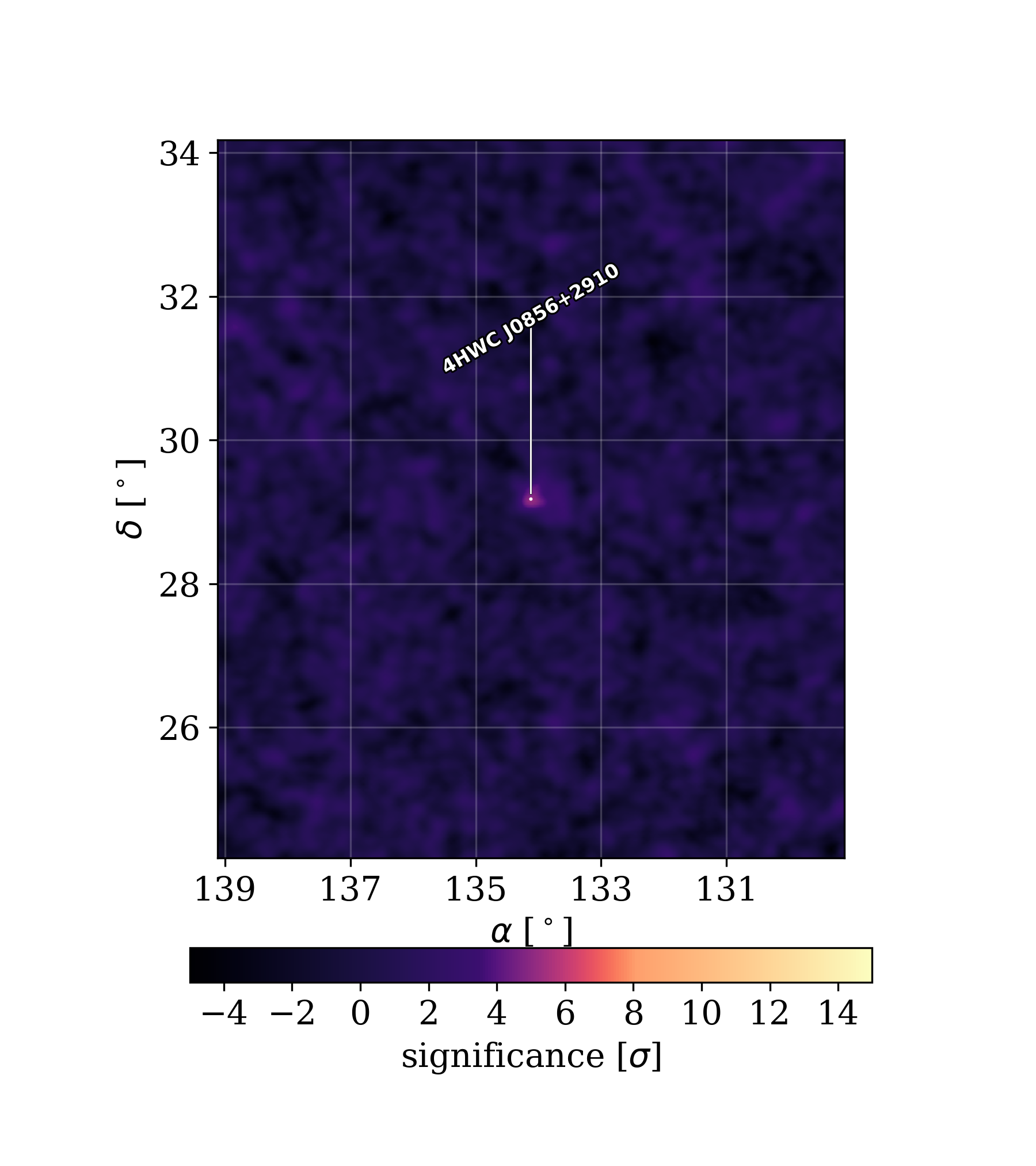}
    \caption{HAWC significance map of region R.A.~=~134.12\degree \space and Dec. = 29.18\degree \space with the 4HWC source overlaid as a dashed circle with radius equal to the extension parameter. A small dot represents the center of the source model emission. Map generated using a point-source assumption with a $-2.6$ power-law index.}
    \label{fig:Offplane4}
\end{figure}

\begin{figure}[]
    \centering
    \includegraphics[trim={0 1.5cm 0 0},width=0.40\textwidth]{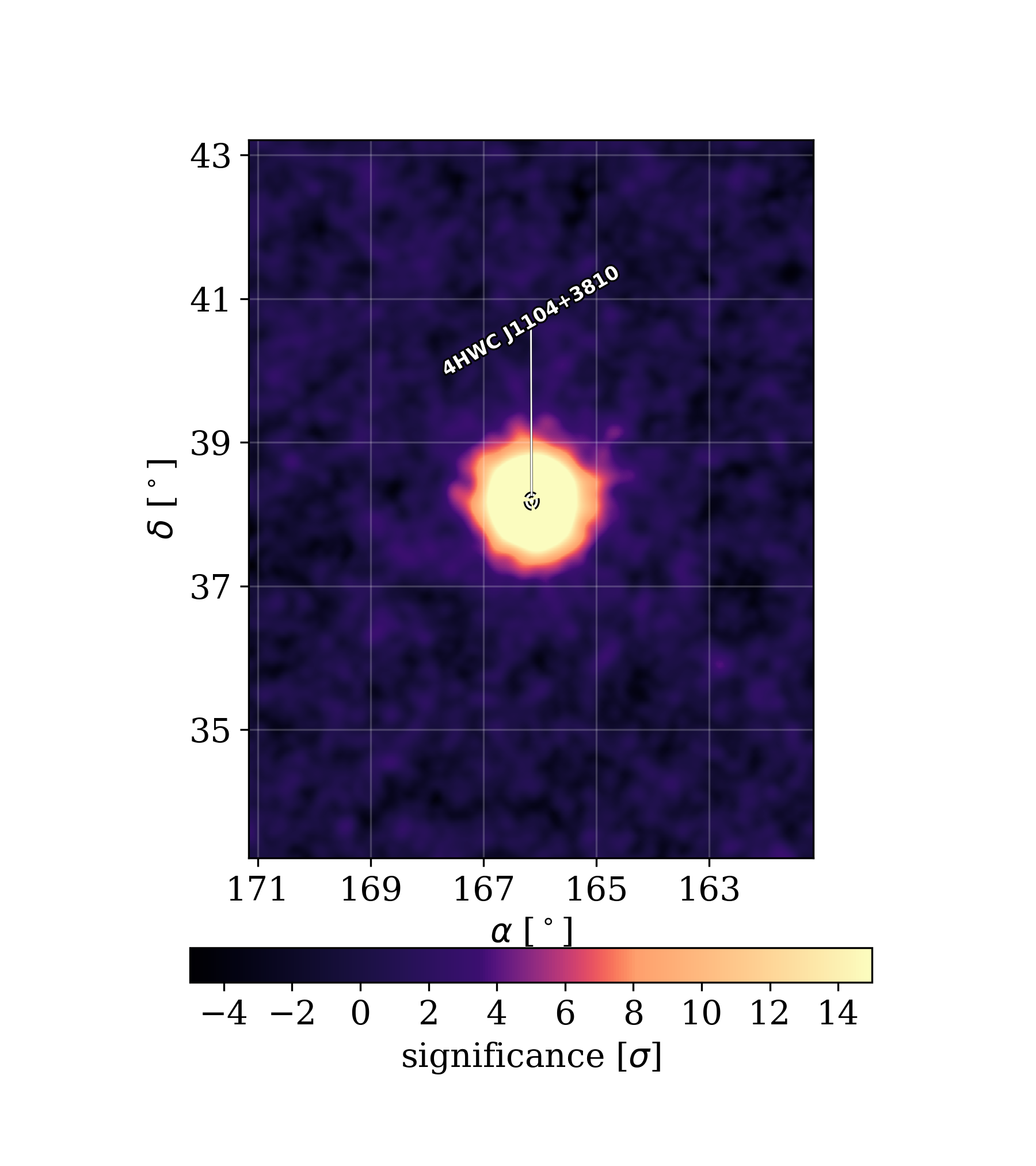}
    \caption{HAWC significance map of region R.A.~=~166.16\degree \space and Dec. = 38.21\degree \space with the 4HWC source overlaid as a dashed circle with radius equal to the extension parameter. A small dot represents the center of the source model emission. Map generated using a point-source assumption with a $-2.6$ power-law index.}
    \label{fig:Offplane5}
\end{figure}

\begin{figure}[]
    \centering
    \includegraphics[trim={0 1.5cm 0 0},width=0.45\textwidth]{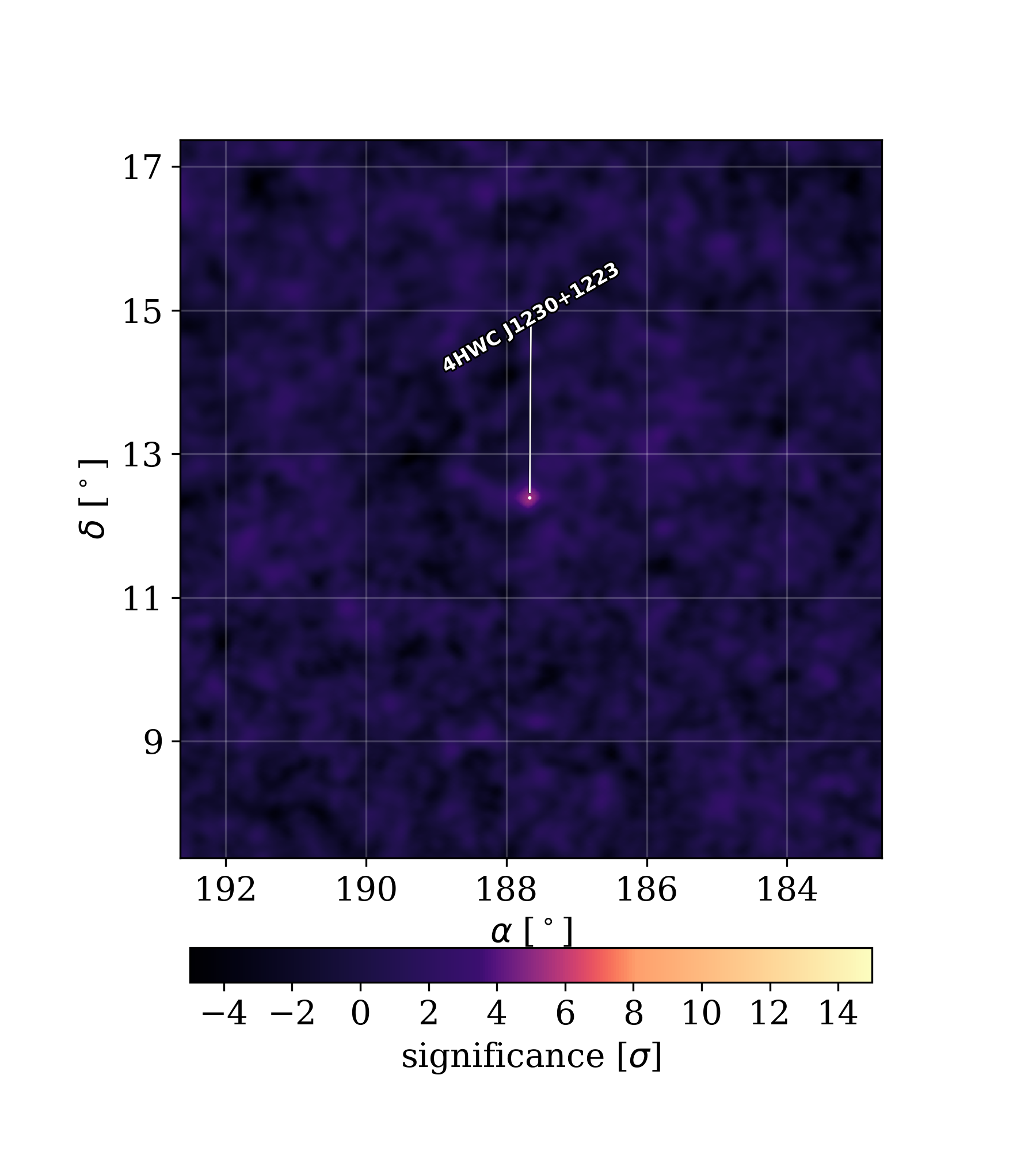}
    \caption{HAWC significance map of region R.A.~=~187.65\degree \space and Dec. = 12.37\degree \space with the 4HWC source overlaid as a dashed circle with radius equal to the extension parameter. A small dot represents the center of the source model emission. Map generated using a point-source assumption with a $-2.6$ power-law index.}
    \label{fig:Offplane6}
\end{figure}

\begin{figure}[]
    \centering
    \includegraphics[trim={0 1.5cm 0 0},width=0.45\textwidth]{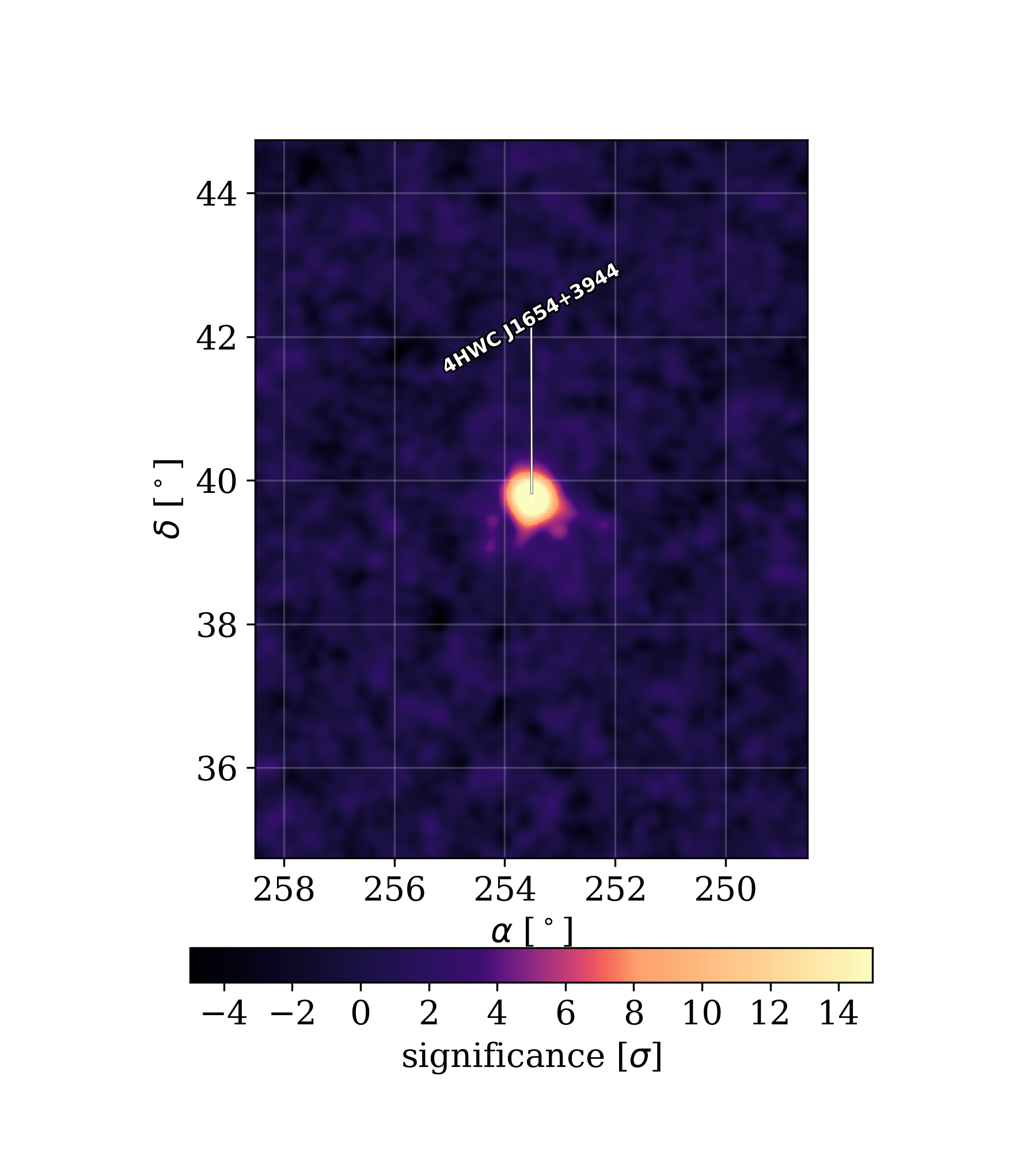}
    \caption{HAWC significance map of region R.A.~=~253.52\degree \space and Dec. = 39.74\degree \space with the 4HWC source overlaid as a dashed circle with radius equal to the extension parameter. A small dot represents the center of the source model emission. Map generated using a point-source assumption with a $-2.6$ power-law index.}
    \label{fig:Offplane7}
\end{figure}

Figures~\ref{fig:Galplot1} through \ref{fig:Offplane7} provide a visual look at the morphology data in Table~\ref{table:sourceList}. The significance is computed by adding a point-like source with a fixed index of $-2.6$ with a free flux normalization. The reported significance is the $\sqrt{TS}$ and, due to the point-like assumption, under-predicts the significance of extended emission. Some regions may appear to be of low significance while the corresponding extended source is much more significant.

Figures~\ref{fig:Galplot1} through \ref{fig:Galplot6} highlight the sources found by the sliding ROIs along the Galactic Plane, while Figures~\ref{fig:Offplane1} through \ref{fig:Offplane7} show the sources identified by the higher-latitude off-Plane ROIs. The majority of sources identified by the ALPS method are in the former category, due largely in part to the significant suppression of gamma-ray emission at TeV energies of extragalactic sources.

\startlongtable
\begin{deluxetable*}{c||c|c|c||c|c|c||c||c}
    \centering
    \tabletypesize{\tiny}
    \tablecaption{4HWC Source List, ordered by R.A., with location and spectral parameters of each source as well as the significance and nearest association in TeVCat. All uncertainties are purely statistical and do not include the systematics. Seven sources marked with a $\dagger$~were~accepted as curved, but the MLE was unable to properly fit a log-parabola model within physically reasonable boundaries. These sources likely have significantly curved spectra, but within the ALPS fitting process, they were not able to be determined as such and dedicated analyses are needed to confirm these morphologies.}
    \label{table:sourceList}
    \tablehead{
        \multicolumn{1}{c||}{} & \multicolumn{3}{c||}{Location and Extension} & \multicolumn{3}{c||}{Spectral Parameters} & \multicolumn{1}{c||}{} & \colhead{} \\
        \multicolumn{1}{c||}{Name} & \colhead{R.A.} & \colhead{Dec.} & \multicolumn{1}{c||}{Extension} & \colhead{Flux Norm} & \colhead{Index} & \multicolumn{1}{c||}{Beta} & \multicolumn{1}{c||}{TS} & \colhead{TeVCat Assoc. (distance)} \\
        \multicolumn{1}{c||}{} & \colhead{[\degree]} & \colhead{[\degree]} & \multicolumn{1}{c||}{[\degree]} & \colhead{$10^{-13}\frac{1}{\mathrm{TeV} \mathrm{cm}^{2} \mathrm{s}}$} & \colhead{} & \multicolumn{1}{c||}{} & \multicolumn{1}{c||}{} & \colhead{[\degree]}
    }
    \startdata
      4HWC J0341+5258 & $55.49^{+0.06}_{-0.06}$ & $52.98^{+0.04}_{-0.04}$ & $0.18^{+0.03}_{-0.03}$ & $1.53^{+0.57}_{-0.47}$ & $-2.29^{+0.11}_{-0.1}$ & - & 122 & LHAASO J0341+5258 (0.148) \\
4HWC J0347-1405 & $56.82^{+0.35}_{-0.35}$ & $-14.09^{+0.08}_{-0.07}$ & - & $3.28^{+0.92}_{-0.73}$ & $-3.5^{+0.15}_{-0.06}$ & - & 30 & - \\
4HWC J0533+3534 & $83.44^{+0.02}_{-0.02}$ & $35.57^{+0.02}_{-0.03}$ & - & $0.5^{+0.22}_{-0.18}$ & $-2.38^{+0.15}_{-0.14}$ & - & 61 & 1LHAASO J0534+3533 (0.091) \\
4HWC J0534+2200 & $83.63^{+0.001}_{-0.001}$ & $22.01^{+0.001}_{-0.002}$ & - & $61.0^{+0.39}_{-0.39}$ & $-2.5^{+0.01}_{-0.01}$ & $0.09^{+0.01}_{-0.01}$ & 82677 & Crab (0.003) \\
4HWC J0538+2804 & $84.58^{+0.03}_{-0.09}$ & $28.08^{+0.05}_{-0.04}$ & - & $0.73^{+0.45}_{-0.31}$ & $-2.87^{+0.11}_{-1.09}$ & - & 27 & - \\
4HWC J0542+2315 & $85.74^{+0.09}_{-0.1}$ & $23.26^{+0.07}_{-0.08}$ & $0.98^{+0.07}_{-0.07}$ & $2.97^{+1.88}_{-1.25}$ & $-1.15^{+0.65}_{-0.45}$ & $0.3^{+0.13}_{-0.09}$ & 411 & HAWC J0543+233 (0.141) \\
4HWC J0615+2214 & $93.79^{+0.13}_{-0.14}$ & $22.24^{+0.13}_{-0.14}$ & $1.1^{+0.12}_{-0.11}$ & $7.25^{+0.97}_{-0.91}$ & $-2.54^{+0.03}_{-0.04}$ & - & 181 & IC 443 (0.496) \\
4HWC J0622+3759$^{\dagger}$ & $95.56^{+0.08}_{-0.08}$ & $37.99^{+0.08}_{-0.07}$ & $0.5^{+0.09}_{-0.05}$ & $1.87^{+0.34}_{-0.171}$ & $-2.32^{+0.07}_{-0.06}$ & - & 144 & LHAASO J0621+3755 (0.116) \\
4HWC J0631+1036 & $97.79^{+0.06}_{-0.06}$ & $10.61^{+0.05}_{-0.06}$ & $0.35^{+0.05}_{-0.05}$ & $1.29^{+0.39}_{-0.34}$ & $-2.29^{+0.11}_{-0.08}$ & - & 91 & 3HWC J0631+107 (0.123) \\
4HWC J0633+1719$^{\dagger}$ & $98.48^{+0.08}_{-0.08}$ & $17.32^{+0.07}_{-0.07}$ & $2.2^{+0.07}_{-0.06}$ & $29.58^{+1.14}_{-1.27}$ & $2.314^{+0.016}_{-0.015}$ & - & 2061 & Geminga (0.368) \\
4HWC J0634+1734$^{\dagger}$ & $98.51^{+0.06}_{-0.06}$ & $17.57^{+0.06}_{-0.06}$ & $0.85^{+0.06}_{-0.05}$ & $5.5^{+0.27}_{-0.29}$ & $-2.26^{+0.031}_{-0.029}$ & - & 417 & Geminga Pulsar (0.199) \\
4HWC J0634+0628 & $98.56^{+0.06}_{-0.06}$ & $6.48^{+0.06}_{-0.06}$ & $0.46^{+0.1}_{-0.05}$ & $3.05^{+0.55}_{-0.51}$ & $-2.45^{+0.06}_{-0.06}$ & - & 150 & HAWC J0635+070 (0.541) \\
4HWC J0701+1412$^{\dagger}$ & $105.4^{+0.1}_{-0.1}$ & $14.22^{+0.08}_{-0.08}$ & $1.97^{+0.08}_{-0.08}$ & $19.8^{+0.98}_{-1.02}$ & $-2.339^{+0.021}_{-0.019}$ & - & 940 & 2HWC J0700+143 (0.3) \\
4HWC J0856+2910 & $134.12^{+0.04}_{-0.07}$ & $29.18^{+0.04}_{-0.04}$ & - & $0.91^{+0.19}_{-0.2}$ & $-2.95^{+0.14}_{-1.04}$ & - & 31 & - \\
4HWC J1104+3810 & $166.15^{+0.004}_{-0.004}$ & $38.18^{+0.004}_{-0.004}$ & $0.09^{+0.005}_{-0.005}$ & $44.2^{+0.83}_{-0.81}$ & $-2.95^{+0.02}_{-0.02}$ & $0.28^{+0.02}_{-0.02}$ & 10782 & Markarian 421 (0.07) \\
4HWC J1230+1223 & $187.67^{+0.04}_{-0.03}$ & $12.39^{+0.04}_{-0.03}$ & - & $0.59^{+0.18}_{-0.17}$ & $-2.63^{+0.13}_{-0.59}$ & - & 31 & M87 (0.032) \\
4HWC J1654+3944 & $253.51^{+0.01}_{-0.01}$ & $39.74^{+0.01}_{-0.01}$ & - & $10.0^{+0.54}_{-0.52}$ & $-2.75^{+0.11}_{-0.09}$ & $0.28^{+0.06}_{-0.05}$ & 782 & Markarian 501 (0.046) \\
4HWC J1740+0950 & $265.04^{+0.04}_{-0.02}$ & $9.85^{+0.02}_{-0.02}$ & - & $0.36^{+0.15}_{-0.13}$ & $-2.2^{+0.14}_{-0.13}$ & - & 89 & 3HWC J1739+099 (0.099) \\
4HWC J1745-2859 & $266.26^{+0.07}_{-0.06}$ & $-28.98^{+0.04}_{-0.04}$ & - & $28.0^{+17.1}_{-8.87}$ & $-2.95^{+0.14}_{-0.87}$ & - & 48 & Galactic Centre Ridge (0.152) \\
4HWC J1800-2345 & $270.14^{+0.18}_{-0.19}$ & $-23.75^{+0.18}_{-0.18}$ & $1.01^{+0.12}_{-0.11}$ & $40.6^{+9.94}_{-8.99}$ & $-2.69^{+0.08}_{-0.07}$ & - & 106 & HESS J1800-240B (0.286) \\
4HWC J1804-2133 & $271.1^{+0.04}_{-0.04}$ & $-21.56^{+0.04}_{-0.04}$ & $0.19^{+0.04}_{-0.04}$ & $19.4^{+3.4}_{-3.42}$ & $-2.88^{+0.08}_{-0.07}$ & - & 164 & HESS J1804-216 (0.146) \\
4HWC J1809-1923 & $272.38^{+0.02}_{-0.02}$ & $-19.39^{+0.02}_{-0.02}$ & $0.26^{+0.02}_{-0.02}$ & $27.9^{+2.48}_{-2.38}$ & $-2.56^{+0.03}_{-0.03}$ & - & 1371 & HESS J1809-193 (0.266) \\
4HWC J1813-1244 & $273.36^{+0.03}_{-0.03}$ & $-12.74^{+0.03}_{-0.03}$ & $0.23^{+0.03}_{-0.02}$ & $7.01^{+1.19}_{-1.14}$ & $-2.59^{+0.06}_{-0.06}$ & - & 267 & HESS J1813-126 (0.06) \\
4HWC J1813-1747 & $273.36^{+0.01}_{-0.01}$ & $-17.8^{+0.01}_{-0.01}$ & $0.09^{+0.02}_{-0.02}$ & $5.6^{+2.65}_{-2.35}$ & $-1.45^{+0.5}_{-0.49}$ & $0.26^{+0.1}_{-0.1}$ & 614 & HESS J1813-178 (0.059) \\
4HWC J1814-1705 & $273.58^{+0.04}_{-0.04}$ & $-17.09^{+0.05}_{-0.05}$ & $0.66^{+0.04}_{-0.04}$ & $20.8^{+4.86}_{-6.2}$ & $-1.78^{+0.32}_{-0.24}$ & $0.2^{+0.07}_{-0.05}$ & 717 & 2HWC J1814-173 (0.232) \\
4HWC J1819-1526 & $274.81^{+0.06}_{-0.06}$ & $-15.44^{+0.07}_{-0.06}$ & $0.36^{+0.07}_{-0.06}$ & $11.9^{+3.25}_{-2.78}$ & $-2.73^{+0.07}_{-0.07}$ & - & 149 & SNR G015.4+00.1 (0.291) \\
4HWC J1819-2513 & $274.85^{+0.03}_{-0.04}$ & $-25.23^{+0.05}_{-0.03}$ & - & $0.89^{+0.54}_{-0.37}$ & $-1.99^{+0.22}_{-0.42}$ & - & 84 & V4641 Sgr (0.178) \\
4HWC J1819-2601 & $274.92^{+0.04}_{-0.04}$ & $-26.03^{+0.06}_{-0.06}$ & $0.26^{+0.04}_{-0.04}$ & $4.5^{+2.83}_{-1.89}$ & $-2.23^{+0.15}_{-0.14}$ & - & 123 & V4641 Sgr (0.626) \\
4HWC J1825-1350 & $276.31^{+0.05}_{-0.05}$ & $-13.85^{+0.07}_{-0.07}$ & $1.19^{+0.08}_{-0.08}$ & $117.0^{+6.91}_{-7.03}$ & $-2.65^{+0.05}_{-0.04}$ & $0.1^{+0.02}_{-0.02}$ & 804 & HESS J1825-137 (0.163) \\
4HWC J1825-1337 & $276.46^{+0.01}_{-0.01}$ & $-13.63^{+0.01}_{-0.01}$ & $0.27^{+0.01}_{-0.01}$ & $19.3^{+3.35}_{-3.19}$ & $-1.76^{+0.14}_{-0.04}$ & $0.15^{+0.01}_{-0.03}$ & 3261 & HESS J1825-137 (0.148) \\
4HWC J1825-1256 & $276.48^{+0.01}_{-0.01}$ & $-12.94^{+0.02}_{-0.02}$ & $0.16^{+0.01}_{-0.01}$ & $9.47^{+1.2}_{-1.1}$ & $-2.39^{+0.04}_{-0.04}$ & - & 902 & HESS J1826-130 (0.13) \\
4HWC J1826-1445 & $276.53^{+0.02}_{-0.02}$ & $-14.76^{+0.03}_{-0.03}$ & $0.11^{+0.03}_{-0.04}$ & $6.79^{+1.61}_{-1.39}$ & $-2.69^{+0.08}_{-0.08}$ & - & 156 & LS 5039 (0.067) \\
4HWC J1830-1006 & $277.72^{+0.03}_{-0.03}$ & $-10.11^{+0.03}_{-0.03}$ & $0.71^{+0.03}_{-0.03}$ & $31.9^{+2.97}_{-3.02}$ & $-2.16^{+0.11}_{-0.1}$ & $0.13^{+0.03}_{-0.02}$ & 359 & HESS J1831-098 (0.249) \\
4HWC J1834-0828 & $278.6^{+0.03}_{-0.03}$ & $-8.48^{+0.04}_{-0.04}$ & $0.36^{+0.03}_{-0.03}$ & $18.4^{+1.71}_{-1.71}$ & $-2.77^{+0.07}_{-0.06}$ & $0.05^{+0.03}_{-0.02}$ & 359 & HESS J1834-087 (0.294) \\
4HWC J1837-0650 & $279.35^{+0.01}_{-0.01}$ & $-6.84^{+0.01}_{-0.01}$ & $0.37^{+0.01}_{-0.01}$ & $43.5^{+1.84}_{-1.84}$ & $-2.37^{+0.05}_{-0.05}$ & $0.15^{+0.02}_{-0.02}$ & 4305 & HESS J1837-069 (0.13) \\
4HWC J1840-0536 & $280.1^{+0.02}_{-0.02}$ & $-5.61^{+0.02}_{-0.02}$ & $0.5^{+0.01}_{-0.01}$ & $46.0^{+1.74}_{-1.64}$ & $-2.36^{+0.04}_{-0.04}$ & $0.11^{+0.02}_{-0.01}$ & 4600 & HESS J1841-055 (0.144) \\
4HWC J1843-0330 & $280.96^{+0.02}_{-0.02}$ & $-3.51^{+0.03}_{-0.02}$ & $0.44^{+0.01}_{-0.01}$ & $14.2^{+3.22}_{-2.85}$ & $-1.49^{+0.19}_{-0.17}$ & $0.27^{+0.04}_{-0.04}$ & 2757 & HESS J1843-033 (0.047) \\
4HWC J1846-0235 & $281.67^{+0.07}_{-0.08}$ & $-2.6^{+0.07}_{-0.09}$ & $0.3^{+0.05}_{-0.09}$ & $7.84^{+1.8}_{-1.91}$ & $-2.78^{+0.04}_{-0.04}$ & - & 237 & HESS J1846-029 (0.384) \\
4HWC J1847+0051 & $281.77^{+0.29}_{-0.31}$ & $-0.87^{+0.43}_{-0.43}$ & $2.84^{+0.35}_{-0.32}$ & $46.8^{+6.12}_{-5.43}$ & $-2.93^{+0.04}_{-0.04}$ & - & 191 & HESS J1848-018 (0.989) \\
4HWC J1848-0140 & $282.11^{+0.02}_{-0.02}$ & $-1.67^{+0.04}_{-0.03}$ & $0.43^{+0.02}_{-0.02}$ & $13.6^{+1.54}_{-1.73}$ & $-2.03^{+0.14}_{-0.11}$ & $0.19^{+0.04}_{-0.03}$ & 269 & HESS J1848-018 (0.126) \\
4HWC J1848+0000$^{\dagger}$ & $282.24^{+0.01}_{-0.01}$ & $0.01^{+0.01}_{-0.01}$ & $0.08^{+0.02}_{-0.02}$ & $1.55^{+0.11}_{-0.13}$ & $-2.18^{+0.06}_{-0.06}$ & - & 399 & IGR J18490-0000 (0.033) \\
4HWC J1851+0002 & $282.85^{+0.03}_{-0.03}$ & $-0.05^{+0.02}_{-0.02}$ & $0.46^{+0.02}_{-0.02}$ & $8.08^{+1.74}_{-1.56}$ & $-1.38^{+0.21}_{-0.2}$ & $0.3^{+0.05}_{-0.05}$ & 686 & HESS J1852-000 (0.254) \\
4HWC J1854+0120 & $283.58^{+0.09}_{-0.1}$ & $1.34^{+0.11}_{-0.12}$ & $0.88^{+0.08}_{-0.07}$ & $22.5^{+2.55}_{-2.2}$ & $-2.73^{+0.02}_{-0.02}$ & - & 273 & 2HWC J1852+013* (0.571) \\
4HWC J1857+0247 & $284.34^{+0.01}_{-0.01}$ & $2.8^{+0.02}_{-0.02}$ & $0.27^{+0.02}_{-0.02}$ & $13.0^{+1.02}_{-0.95}$ & $-1.88^{+0.14}_{-0.13}$ & $0.38^{+0.06}_{-0.05}$ & 593 & HESS J1857+026 (0.139) \\
4HWC J1857+0200 & $284.48^{+0.02}_{-0.02}$ & $2.0^{+0.02}_{-0.02}$ & $0.18^{+0.02}_{-0.02}$ & $4.63^{+0.64}_{-0.59}$ & $-2.45^{+0.04}_{-0.04}$ & - & 273 & HESS J1858+020 (0.138) \\
4HWC J1858+0752 & $284.51^{+0.04}_{-0.02}$ & $7.88^{+0.05}_{-0.02}$ & - & $0.49^{+0.27}_{-0.21}$ & $-2.4^{+0.19}_{-0.17}$ & - & 44 & - \\
4HWC J1858+0344 & $284.74^{+0.05}_{-0.05}$ & $3.74^{+0.05}_{-0.05}$ & $0.63^{+0.04}_{-0.04}$ & $13.9^{+2.75}_{-2.57}$ & $-2.68^{+0.05}_{-0.05}$ & - & 228 & 1LHAASO J1858+0330 (0.269) \\
4HWC J1906+0613 & $286.55^{+0.06}_{-0.06}$ & $6.22^{+0.06}_{-0.06}$ & $1.22^{+0.06}_{-0.06}$ & $43.0^{+2.17}_{-2.16}$ & $-2.47^{+0.05}_{-0.04}$ & $0.08^{+0.02}_{-0.02}$ & 25 & MGRO J1908+06 (0.426) \\
4HWC J1908+0845 & $287.03^{+0.06}_{-0.06}$ & $8.76^{+0.09}_{-0.09}$ & $0.76^{+0.08}_{-0.07}$ & $11.5^{+1.39}_{-1.28}$ & $-2.72^{+0.03}_{-0.03}$ & - & 378 & 2HWC J1907+084* (0.351) \\
4HWC J1908+0616 & $287.06^{+0.01}_{-0.01}$ & $6.27^{+0.01}_{-0.01}$ & $0.39^{+0.01}_{-0.01}$ & $9.96^{+1.32}_{-1.27}$ & $-1.64^{+0.11}_{-0.1}$ & $0.18^{+0.02}_{-0.02}$ & 227 & LHAASO J1908+0621 (0.076) \\
4HWC J1910+0503 & $287.62^{+0.0}_{-0.0}$ & $5.06^{+0.001}_{-0.001}$ & - & $0.67^{+0.2}_{-0.18}$ & $-2.28^{+0.1}_{-0.1}$ & - & 101 & SS 433 (0.341) \\
4HWC J1912+1013 & $288.23^{+0.03}_{-0.02}$ & $10.22^{+0.03}_{-0.03}$ & $0.38^{+0.02}_{-0.02}$ & $10.6^{+0.9}_{-0.89}$ & $-2.33^{+0.1}_{-0.08}$ & $0.18^{+0.04}_{-0.03}$ & 777 & HESS J1912+101 (0.075) \\
4HWC J1913+0457 & $288.37^{+0.02}_{-0.02}$ & $4.96^{+0.02}_{-0.03}$ & - & $0.5^{+0.19}_{-0.16}$ & $-2.27^{+0.13}_{-0.12}$ & - & 77 & SS 433 (0.411) \\
4HWC J1914+1151 & $288.68^{+0.02}_{-0.02}$ & $11.85^{+0.02}_{-0.02}$ & $0.1^{+0.03}_{-0.03}$ & $1.48^{+0.13}_{-0.12}$ & $-2.45^{+0.04}_{-0.04}$ & - & 167 & 2HWC J1914+117* (0.131) \\
4HWC J1915+1113 & $288.86^{+0.04}_{-0.04}$ & $11.23^{+0.05}_{-0.05}$ & $0.4^{+0.03}_{-0.03}$ & $3.71^{+0.24}_{-0.22}$ & $-2.56^{+0.03}_{-0.03}$ & - & 164 & - \\
4HWC J1918+1343 & $289.73^{+0.14}_{-0.14}$ & $13.72^{+0.18}_{-0.18}$ & $1.44^{+0.08}_{-0.08}$ & $13.2^{+0.91}_{-0.85}$ & $-2.73^{+0.03}_{-0.03}$ & - & 175 & 2HWC J1921+131 (0.823) \\
4HWC J1922+1405 & $290.74^{+0.02}_{-0.02}$ & $14.09^{+0.02}_{-0.02}$ & $0.08^{+0.02}_{-0.02}$ & $2.54^{+0.29}_{-0.28}$ & $-2.62^{+0.05}_{-0.05}$ & - & 304 & W51 (0.105) \\
4HWC J1923+1631 & $290.93^{+0.08}_{-0.08}$ & $16.53^{+0.08}_{-0.08}$ & $0.9^{+0.04}_{-0.04}$ & $8.98^{+0.48}_{-0.46}$ & $-2.62^{+0.02}_{-0.02}$ & - & 270 & 1LHAASO JJ1924+1609 (0.412) \\
4HWC J1928+1747 & $292.14^{+0.02}_{-0.02}$ & $17.8^{+0.01}_{-0.02}$ & $0.18^{+0.02}_{-0.02}$ & $1.17^{+0.58}_{-0.49}$ & $-1.22^{+0.03}_{-0.03}$ & $0.29^{+0.01}_{-0.01}$ & 498 & 2HWC J1928+177 (0.023) \\
4HWC J1928+1843 & $292.16^{+0.02}_{-0.02}$ & $18.73^{+0.03}_{-0.03}$ & $0.47^{+0.02}_{-0.02}$ & $6.43^{+0.23}_{-0.23}$ & $-2.5^{+0.02}_{-0.02}$ & - & 455 & SNR G054.1+00.3 (0.494) \\
4HWC J1930+1852 & $292.59^{+0.1}_{-0.1}$ & $18.88^{+0.003}_{-0.003}$ & $0.03^{+0.002}_{-0.01}$ & $0.81^{+0.07}_{-0.06}$ & $-2.64^{+0.06}_{-0.06}$ & - & 33 & SNR G054.1+00.3 (0.045) \\
4HWC J1931+1655 & $292.81^{+0.03}_{-0.03}$ & $16.92^{+0.03}_{-0.03}$ & $0.18^{+0.03}_{-0.03}$ & $1.26^{+0.14}_{-0.12}$ & $-2.53^{+0.04}_{-0.04}$ & - & 77 & 1LHAASO J1931+1653 (0.026) \\
4HWC J1932+1917 & $293.1^{+0.02}_{-0.03}$ & $19.3^{+0.03}_{-0.02}$ & - & $0.87^{+0.23}_{-0.23}$ & $-2.66^{+0.13}_{-0.11}$ & - & 59 & - \\
4HWC J1937+2143 & $294.36^{+0.17}_{-0.16}$ & $21.73^{+0.18}_{-0.18}$ & $1.15^{+0.14}_{-0.12}$ & $7.5^{+1.2}_{-1.08}$ & $-2.66^{+0.05}_{-0.05}$ & - & 125 & 1LHAASO J1937+2128 (0.253) \\
4HWC J1945+2434 & $296.37^{+0.17}_{-0.18}$ & $24.58^{+0.16}_{-0.16}$ & $1.66^{+0.12}_{-0.11}$ & $24.7^{+2.72}_{-2.52}$ & $-2.61^{+0.07}_{-0.06}$ & $0.11^{+0.04}_{-0.03}$ & 340 & - \\
4HWC J1952+2610 & $298.01^{+0.17}_{-0.17}$ & $26.18^{+0.18}_{-0.2}$ & $0.99^{+0.14}_{-0.12}$ & $6.61^{+1.22}_{-1.07}$ & $-2.69^{+0.06}_{-0.06}$ & - & 106 & 3HWC J1951+266 (0.446) \\
4HWC J1952+2925 & $298.11^{+0.04}_{-0.03}$ & $29.42^{+0.03}_{-0.03}$ & $0.18^{+0.04}_{-0.03}$ & $2.04^{+0.33}_{-0.3}$ & $-2.53^{+0.06}_{-0.06}$ & - & 190 & 2HWC J1953+294 (0.163) \\
4HWC J1953+2837$^{\dagger}$ & $298.5^{+0.04}_{-0.03}$ & $28.62^{+0.03}_{-0.03}$ & $0.2^{+0.04}_{-0.03}$ & $1.13^{+0.14}_{-0.13}$ & $-2.33^{0.07}_{-0.07}$ & - & 164.1 & 0FGL J1954.4+2838 (0.12) \\
4HWC J1954+3253 & $298.61^{+0.05}_{-0.05}$ & $32.89^{+0.03}_{-0.04}$ & $0.19^{+0.04}_{-0.04}$ & $2.01^{+0.36}_{-0.36}$ & $-2.65^{+0.08}_{-0.07}$ & - & 105 & 1LHAASO J1954+3253 (0.021) \\
4HWC J1958+2851 & $299.6^{+0.05}_{-0.05}$ & $28.85^{+0.04}_{-0.04}$ & $0.33^{+0.04}_{-0.04}$ & $2.16^{+0.44}_{-0.4}$ & $-2.34^{+0.06}_{-0.06}$ & - & 269 & 0FGL J1958.1+2848 (0.083) \\
4HWC J2005+3056 & $301.47^{+0.05}_{-0.05}$ & $30.93^{+0.04}_{-0.04}$ & $0.25^{+0.04}_{-0.03}$ & $1.52^{+0.38}_{-0.33}$ & $-2.43^{+0.08}_{-0.08}$ & - & 124 & 3HWC J2005+311 (0.232) \\
4HWC J2006+3355 & $301.57^{+0.1}_{-0.1}$ & $33.92^{+0.08}_{-0.08}$ & $1.28^{+0.08}_{-0.07}$ & $25.4^{+2.1}_{-1.98}$ & $-2.33^{+0.08}_{-0.07}$ & $0.15^{+0.03}_{-0.03}$ & 619 & 2HWC J2006+341 (0.263) \\
4HWC J2015+3708 & $303.99^{+0.02}_{-0.03}$ & $37.14^{+0.03}_{-0.03}$ & - & $0.89^{+0.28}_{-0.28}$ & $-2.62^{+0.14}_{-0.71}$ & - & 48 & VER J2016+371 (0.063) \\
4HWC J2018+3641$^{\dagger}$ & $304.67^{+0.02}_{-0.02}$ & $36.69^{+0.01}_{-0.01}$ & $0.23^{+0.01}_{-0.01}$ & $7.95^{+0.26}_{-0.26}$ & $-2.32^{+0.026}_{-0.02}$ & - & 2591 & MGRO J2019+37 (0.143) \\
4HWC J2021+3650 & $305.28^{+0.02}_{-0.02}$ & $36.85^{+0.01}_{-0.01}$ & $0.15^{+0.01}_{-0.01}$ & $1.73^{+0.66}_{-0.1}$ & $-1.21^{+0.03}_{-0.27}$ & $0.25^{+0.08}_{-0.01}$ & 851 & - \\
4HWC J2021+4036 & $305.29^{+0.14}_{-0.14}$ & $40.61^{+0.18}_{-0.18}$ & $0.2^{+0.04}_{-0.03}$ & $4.63^{+0.97}_{-0.91}$ & $-2.94^{+0.08}_{-0.09}$ & - & 147 & 0FGL J2021.5+4026 (0.2) \\
4HWC J2022+3715 & $305.69^{+0.26}_{-0.24}$ & $37.26^{+0.23}_{-0.2}$ & $1.61^{+0.22}_{-0.18}$ & $19.3^{+3.2}_{-2.6}$ & $-2.67^{+0.05}_{-0.04}$ & - & 200 & VER J2019+368 (0.948) \\
4HWC J2026+3327 & $306.61^{+0.08}_{-0.08}$ & $33.47^{+0.05}_{-0.06}$ & $0.26^{+0.08}_{-0.07}$ & $1.47^{+0.43}_{-0.38}$ & $-2.58^{+0.1}_{-0.09}$ & - & 54 & - \\
4HWC J2029+3641 & $307.39^{+0.06}_{-0.06}$ & $36.68^{+0.05}_{-0.04}$ & - & $0.11^{+0.04}_{-0.05}$ & $-2.03^{+0.11}_{-0.06}$ & - & 28 & - \\
4HWC J2030+4056 & $307.64^{+0.13}_{-0.13}$ & $40.95^{+0.1}_{-0.1}$ & $1.81^{+0.08}_{-0.08}$ & $55.3^{+3.64}_{-3.51}$ & $-2.48^{+0.06}_{-0.05}$ & $0.11^{+0.02}_{-0.02}$ & 731 & LHAASO J2032+4102 (0.427) \\
4HWC J2031+4128 & $307.95^{+0.02}_{-0.02}$ & $41.47^{+0.01}_{-0.01}$ & $0.23^{+0.01}_{-0.01}$ & $4.31^{+1.02}_{-0.96}$ & $-1.4^{+0.26}_{-0.22}$ & $0.28^{+0.06}_{-0.05}$ & 1283 & PSR J2032+4127 (0.095) \\
4HWC J2043+4353 & $310.82^{+0.34}_{-0.62}$ & $43.9^{+0.23}_{-0.41}$ & $0.87^{+0.31}_{-0.22}$ & $5.81^{+3.54}_{-1.86}$ & $-2.61^{+0.11}_{-0.65}$ & - & 52 & 3HWC J2043+443 (0.404) \\
4HWC J2108+5156 & $317.15^{+0.05}_{-0.04}$ & $51.94^{+0.03}_{-0.02}$ & - & $1.16^{+0.53}_{-0.43}$ & $-2.44^{+0.72}_{-0.13}$ & - & 73 & LHAASO J2108+5157 (0.013) \\
4HWC J2228+6058 & $337.06^{+0.07}_{-0.07}$ & $60.97^{+0.04}_{-0.04}$ & $0.31^{+0.03}_{-0.03}$ & $14.4^{+2.98}_{-2.76}$ & $-2.58^{+0.06}_{-0.06}$ & - & 284 & SNR G106.3+02.7 (0.118) \\
4HWC J2234+5904$^{\dagger}$ & $338.62^{+0.19}_{-0.19}$ & $59.08^{+0.13}_{-0.12}$ & $0.65^{+0.06}_{-0.08}$ & $10.1^{+1.41}_{-2.78}$ & $-2.59^{+0.09}_{-0.09}$ & - & 67.9 & -

    \enddata
\end{deluxetable*}


\vspace{-0.75cm}

Figure~\ref{fig:Galplot1} shows a point source at the Galactic Center. The white section represents the limit of data reconstructed by HAWC for this catalog. Figure~\ref{fig:Galplot2} shows the most source-dense region of the catalog. The maximum color bar threshold of $14\sigma$ causes the extremely bright regions from $l$ = 20\degree \space to 40\degree \space to merge into a long strip of emission along the plane. The ALPS methodology is able to disentangle the complicated and bright emission in this region and detects a significant number of sources not found in older HAWC analyses such as the 3HWC catalog, which was run on Pass~4 HAWC data. There are several sources in this portion of the sky, including several larger extended sources centered around the Plane. These sources could be a result of under-predicting the actual diffuse flux while using the HERMES template. 

In Figure~\ref{fig:Galplot3}, two regions of emission are modeled with point or minimally extended sources surrounded by a larger extended source. One such region near $l$ = 80\degree \space is modeled by two smaller sources corresponding to the pulsar wind nebula (PWN) TeV J2032+4130 and supernova remnant Gamma-Cygni, as well as a very large extended source modeling the heavily studied Cygnus Cocoon. The nearby region near $l=76\degree \space$ similarly has several point-like and smaller extended sources with a large extended source overlapping. Much work remains to identify if the extended source is related to the SNRs or middle-aged pulsars in the region or if it represents some bulk region of over-dense TeV emission due to unresolved sources and/or uncertainty in the diffuse emission modeling.

Figures~\ref{fig:Galplot4} and \ref{fig:Galplot5} are less complex, showing a handful of points and small extended sources. Figure~\ref{fig:Galplot6} shows several emission associated with sources like the Crab nebula and the Geminga and Monogem pulsars, alongside other sources. One interesting result of the inclusion of the Geminga pulsar in three separate ROIs is that two sources of significantly different sizes are fit to the emission. The doubled sources could be attributed to the source modeling of a symmetric Gaussian distribution being insufficient to fully explain the true spatial distribution of the TeV Halo. The spatially coincident sources were not determined to be similar enough by the automated disambiguation code and were both included in the final catalog, but likely represent a single true gamma-ray source more accurately modeled by a diffusion-based spatial template similar to the work in \citet{GemingaDiffusion}. Figures~\ref{fig:Offplane1} through \ref{fig:Offplane7} highlight all of the sources found more than 10\degree \space off the Plane. 

\subsection{Associating Catalog Sources with Known Sources}
\label{sec:AssocMethod}
To associate the sources quoted in the 4HWC list with other catalogs, a method similar to the one used in \citet{1LHAASO} was adopted. A variable search radius for each source is constructed as:
\begin{equation}
    r_{search} = \sqrt{(0.3\degree)^2 + (\sigma_{ext})^2 + (\delta_{stat})^2} \enspace ,
\end{equation}
 where the first fixed term represents a minimum search radius of 0.3\degree \space to account for any possible pointing differences between instruments, as well as possible energy-dependent morphology. The second term is the extension of a source in 4HWC (zero for point-like sources), allowing larger sources to search further from their center for a possible match. The final term is the statistical uncertainty from the final step of the ALPS fitting process, which allows poorly resolved sources to search a greater distance for a match, even if the source extension is smaller. A maximum value of 1\degree \space is placed on the searching radius to match the maximum searching radius of previous catalogs, such as 3HWC \citep{3HWC}.

This search process is run on two sets of catalogs. Any source with multiple sources within the search radius reports only the closest result. The first is a TeV-focused set including TeVCat \citep{TeVCat}, 1LHAASO \citep{1LHAASO}, the HGPS \citep{HGPS}, and 3HWC \citep{3HWC}. TeVCat is the most comprehensive TeV source catalog and contains most, but not all of the sources in 1LHAASO, HGPS, and 3HWC. The canonical name quoted in TeVCat is used to identify the associated source whenever possible, along with the angular distance between the sources when associating TeV sources. The final column of Table \ref{table:sourceList} contains the results of this search process on these four catalogs. The second set of catalogs is focused on physical sources to help trace the origin of sources and includes the High-Mass X-ray Binary Catalog (HXMB Cat. \citep{HMXB}), the Low-Mass X-ray Binary Catalog (LMXB Cat. \citep{LMXB}), the Australian Telescope National Facility (ATNF) Pulsar Catalog \citep{ATNF}, the Third \textit{Fermi} Pulsar Catalog \citep{FermiPulsar}, and the catalog of High-Energy Observations of Supernova Remnants (SNRCat \citep{SNRcat}). The final column of Table \ref{table:sourceListPhysical} contains the results of this search process on these five catalogs. 


\clearpage

\subsection{Comparison with Previous Wide-Field-of-View TeV Catalogs}

Both the previous iterations of the HWC catalog and the recent 1LHAASO catalog share significant overlap and notable differences with 4HWC. The new methodology and improved data reconstruction of 4HWC represent a major improvement in the ability to disentangle complicated source morphologies compared to 3HWC. The source fitting methodology of 1LHASSO and 4HWC are similar but differ in significant ways, especially when considering the initial source seed determination process. While the 1LHAASO method assumes an extended source initially, the ALPS method starts with a simpler point source assumption when determining seed locations. Additionally, the termination point for the seed-adding phase in 4HWC is based on the TS of the added source model, while in 1LHAASO it is based on the maximum pixel value in the residual maps between steps. Most significantly 1LHAASO represents a combination of two instruments in different energy ranges, while 4HWC uses a single instrument over the entire energy range. 

Despite these differences, the results of these two catalogs have significant agreement in the overlapping fields of view of the two observatories. Out of the 85 4HWC sources, 22 have no association in the 1LHAASO catalog using the variable search criteria defined earlier in Section~\ref{sec:AssocMethod}. Of the remaining sources with a spatially nearby source, 48 have matching extension parameters within the $2\sigma$ uncertainty of the 4HWC source, including the statistical and systematic effects. Only 15 sources have significantly different sizes if they are spatially associated. It is difficult to draw direct conclusions from the similarity or lack thereof between sources, as so many factors impact the modeling of both datasets. However, in general, the agreement between the 4HWC and 1LHAASO catalogs is higher than 4HWC and 3HWC. This is likely due to the 3HWC bias towards point sources over extended sources.

\subsubsection{Comparing 3HWC and 4HWC}
Using a fixed 1\degree \space search radius to associate the 3HWC source list with 4HWC, similarly to the 2HWC comparison perfomed in the 3HWC catalog paper, shows that of the 65 3HWC sources, 12 do not have a counterpart in 4HWC. The details of these missing sources that are no longer found in HAWC data using the pipeline are included in Table

Reversing the comparison and now using the variable search radius defined in Section~\ref{sec:AssocMethod} to compare 4HWC sources to 3HWC, we find 32 4HWC sources without a 3HWC counterpart. Comparing just the overall numerical difference between the 3HWC and 4HWC catalogs underestimates the improvement between the two iterations. The previous 3HWC catalog used a method to find local maxima in several HAWC all-sky maps made using different extension assumptions detailed in \citet{3HWC}. The order of fitting these pre-made maps heavily favored a point-source assumption over extended models. 

\startlongtable
\begin{deluxetable*}{c||c|c|c||c|c||c}
    \tabletypesize{\tiny}
    \tablecaption{Sources from 3HWC no longer found by the automated pipeline in HAWC Pass~5 data.}
    \label{table:source3HWCMissing}
    \tablehead{
        \multicolumn{1}{c||}{} & \multicolumn{3}{c||}{Location and Extension} & \multicolumn{2}{c||}{Spectral Parameters} & \multicolumn{1}{c}{} \\
        \multicolumn{1}{c||}{Name} & \colhead{R.A.} & \colhead{Dec.} & \multicolumn{1}{c||}{Extension} & \colhead{Flux} & \multicolumn{1}{c||}{Index} & \multicolumn{1}{c}{TS} \\
        \multicolumn{1}{c||}{} & \colhead{[\degree]} & \colhead{[\degree]} & \multicolumn{1}{c||}{[\degree]} & \colhead{$10^{-15}\frac{1}{\mathrm{TeV} \mathrm{cm}^{2} \mathrm{s}}$} & \multicolumn{1}{c||}{} & \multicolumn{1}{c}{}
    }
    \startdata
      3HWC J1940+237 & $295.04$  & $23.765$ & 0.0 & $4.0$  & $-3.17$ & 27 \\
3HWC J1950+242 & $297.694$ & $ 24.26$ & 0.0 & $4.3$  & $-2.49$ & 25 \\
3HWC J1743+149 & $265.824$ & $ 14.94$ & 0.0 & $4.0$  & $-2.37$ & 26 \\
3HWC J1857+051 & $284.33$  & $ 5.19$  & 0.0 & $5.9$  & $-3.03$ & 35 \\
3HWC J2010+345 & $302.69$  & $34.545$ & 0.0 & $5.4$  & $-2.91$ & 28 \\
3HWC J0633+191 & $98.437$  & $19.115$ & 0.0 & $4.8$  & $-2.64$ & 27 \\
3HWC J1915+164 & $288.76$  & $16.412$ & 0.0 & $4.6$  & $-2.60$ & 27 \\
3HWC J2022+431 & $305.52$  & $43.15$  & 0.0 & $6.0$  & $-2.34$ & 29 \\
3HWC J2023+324 & $305.81$  & $32.442$ & 1.0 & $13.8$ & $-2.70$ & 31 \\
3HWC J0630+186 & $97.69$   & $18.681$ & 0.0 & $5.1$  & $-2.21$ & 39 \\
3HWC J1937+193 & $294.3$   & $19.31$  & 0.0 & $4.2$  & $-2.90$ & 25 \\
3HWC J1918+159 & $289.68$  & $15.907$ & 0.0 & $5.1$  & $-2.49$ & 32 
    \enddata
\end{deluxetable*}

\clearpage

\begin{figure*}[bth!]
    \centering
    \includegraphics[width=0.60\textwidth]{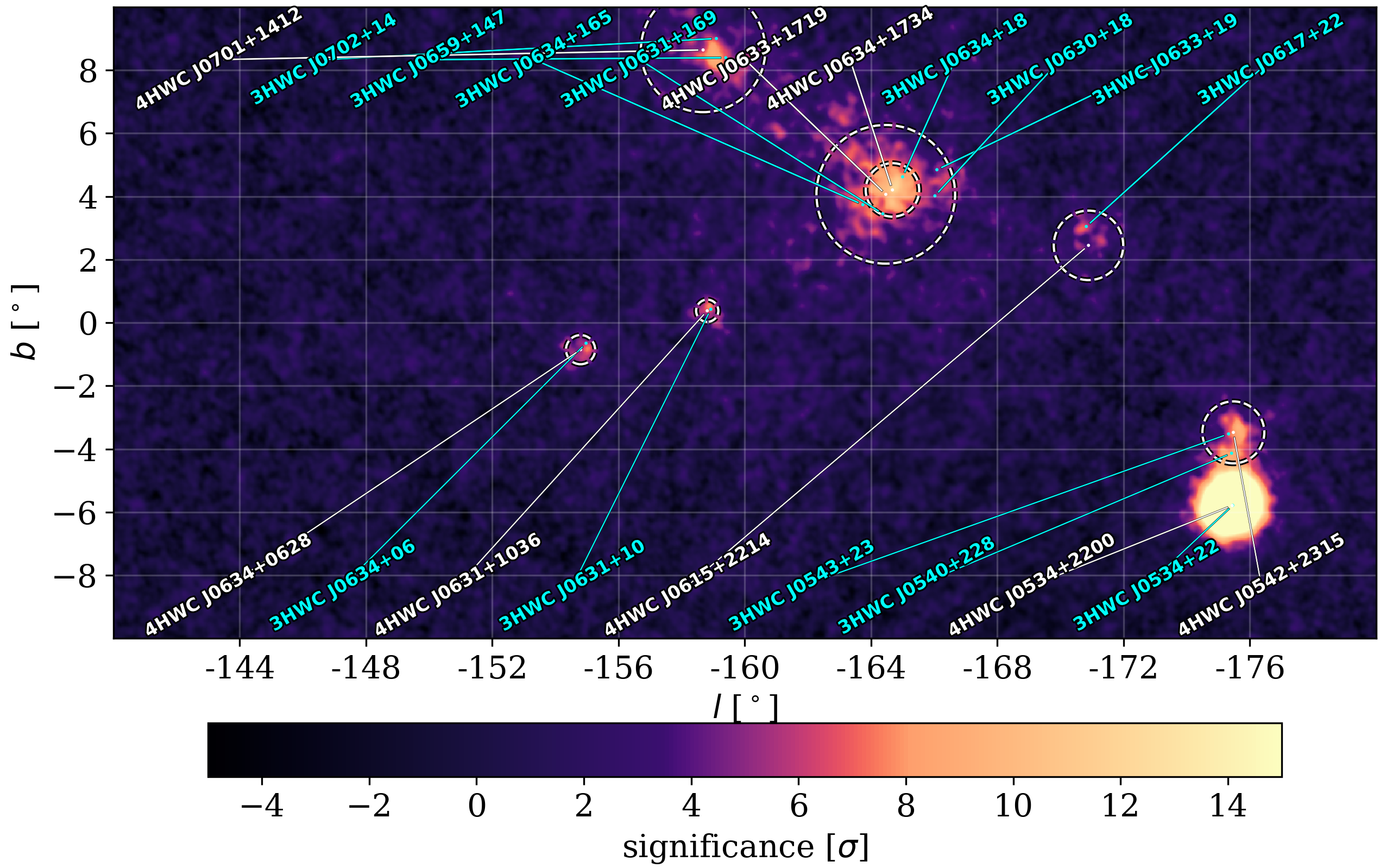}
    \caption{Comparison of 3HWC (cyan) and 4HWC (white) sources in the Crab and Geminga and Monogem regions.}
    \label{fig:3v4HWC200}
\end{figure*}

Looking more closely at the sources not present in 4HWC from 3HWC, nine of these sources are near the Galactic Plane and typically represent one of two major differences between the data and modeling. Four of these nine appear to be in clusters of point sources near regions of emission modeled by one larger extended source in 4HWC that is more than 1\degree \space away from the point source in question. Even some of the sources that do report a 4HWC counterpart similarly show this behavior, resulting in cases like that near the Crab in the bottom right and near Geminga and Monogem at the top of Figure~\ref{fig:3v4HWC200}. The source that shows as a single extended source above the Crab in 4HWC was modeled with two separate point sources in 3HWC. Even more obvious is the Geminga and Monogem region, which in 3HWC was fit with seven point sources, while the same region is fit with three extended sources in 4HWC. As previously discussed, even the improvement of 4HWC modeling fits two different sized extended sources to Geminga, which may be due to the symmetric Gaussian spatial model being insufficient to explain the long tail of the diffusion of the Geminga TeV Halo. 

The other five of the nine near-Galactic-Plane sources are point sources located in regions that show little or no significant emission in the data maps. These 3HWC sources missing in 4HWC may be a result of the improved angular resolution and background suppression of HAWC Pass 5 data compared to Pass 4. Due to the peak-finding algorithm used in 3HWC, poor angular resolution and significant background can cause a local maximum to appear in the data, which will appear as a source in 3HWC. However, once the background is suppressed and the emission is more localized, the local maxima become insignificant. The remaining three missing sources include two near the just-discussed Geminga Monogem region and one source 20\degree \space off the Plane, which now shows no significant emission in the data used for this catalog. Of the sources in 4HWC but not 3HWC, many of these sources appear along the Galactic Plane, where the improved declination range and modeling power of the 4HWC catalog were most effective. The ALPS method was able to disambiguate many more sources from the dense emission closer to the Galactic Center.   

\begin{figure}[h!]
    \centering
    \includegraphics[width=0.48\textwidth]{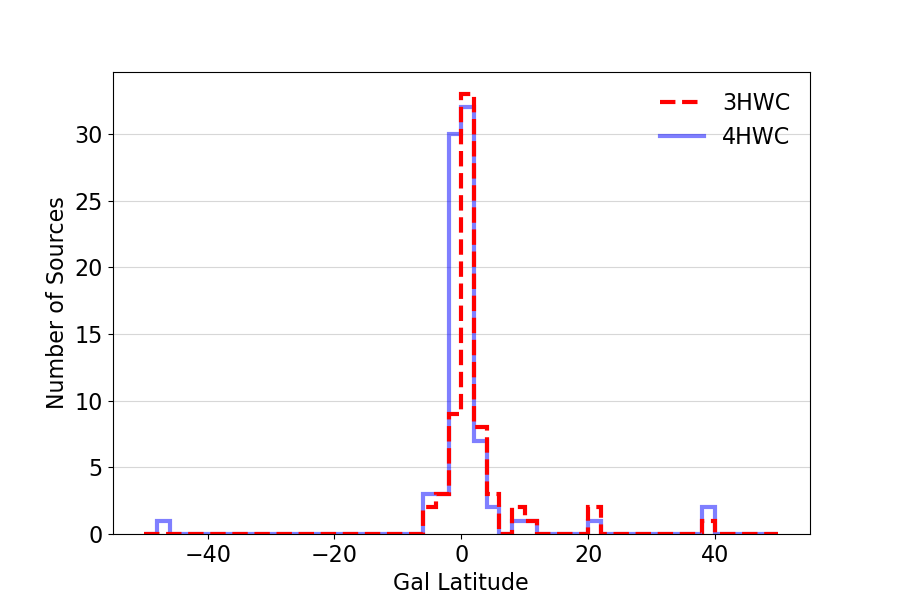}
    \caption{Latitude distribution of 3HWC and 4HWC compared.}
    \label{fig:FullBDist3v4}
\end{figure}

\begin{figure}[h!]
    \centering
    \includegraphics[width=0.48\textwidth]{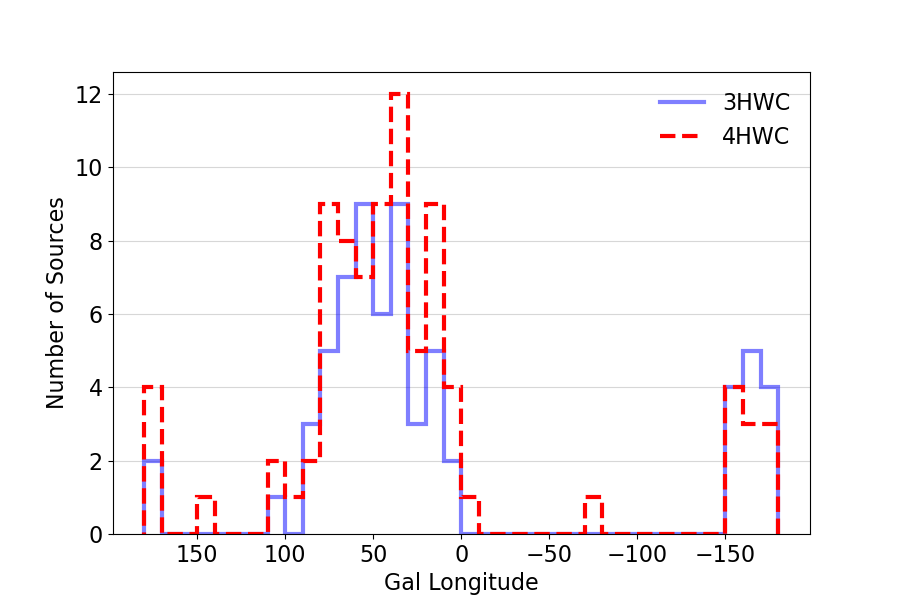}
    \caption{Longitude distribution of 3HWC and 4HWC compared.}
    \label{fig:FullLDist3v4}
\end{figure}

In total, the 3HWC catalog only contained six extended sources, while we report 68 sources as extended in 4HWC. In addition to the clear difference in the number of extended sources between the two iterations of the HWC catalog, there are subtle differences in the source distributions (Figures~\ref{fig:FullBDist3v4} and \ref{fig:FullLDist3v4}). This is most noticeable in the longitude distribution shown in Figure~\ref{fig:FullLDist3v4}. The increased declination range of the new Pass 5 maps allows sources closer to the Galactic Center to be fit.  


In the 3HWC paper, we discovered a pointing issue that significantly impacted sources in the Southern sky. The furthest-South sources in 3HWC had systematic offsets of nearly 0.4\degree \space when compared to IACT components of associated sources. When limiting the 3HWC source list to only sources in the Southern sky, an average $-0.21$\degree \space offset exists, but performing the same analysis for the 4HWC sources shows an average systematic offset of 0.019\degree, which is within the systematic uncertainty calculated in Section~\ref{sec:systematic}.
\newpage 

\subsubsection{Comparing 1LHAASO and 4HWC}

\begin{figure}
    \centering
    \includegraphics[trim={0 0cm 0 0},width=0.45\textwidth]{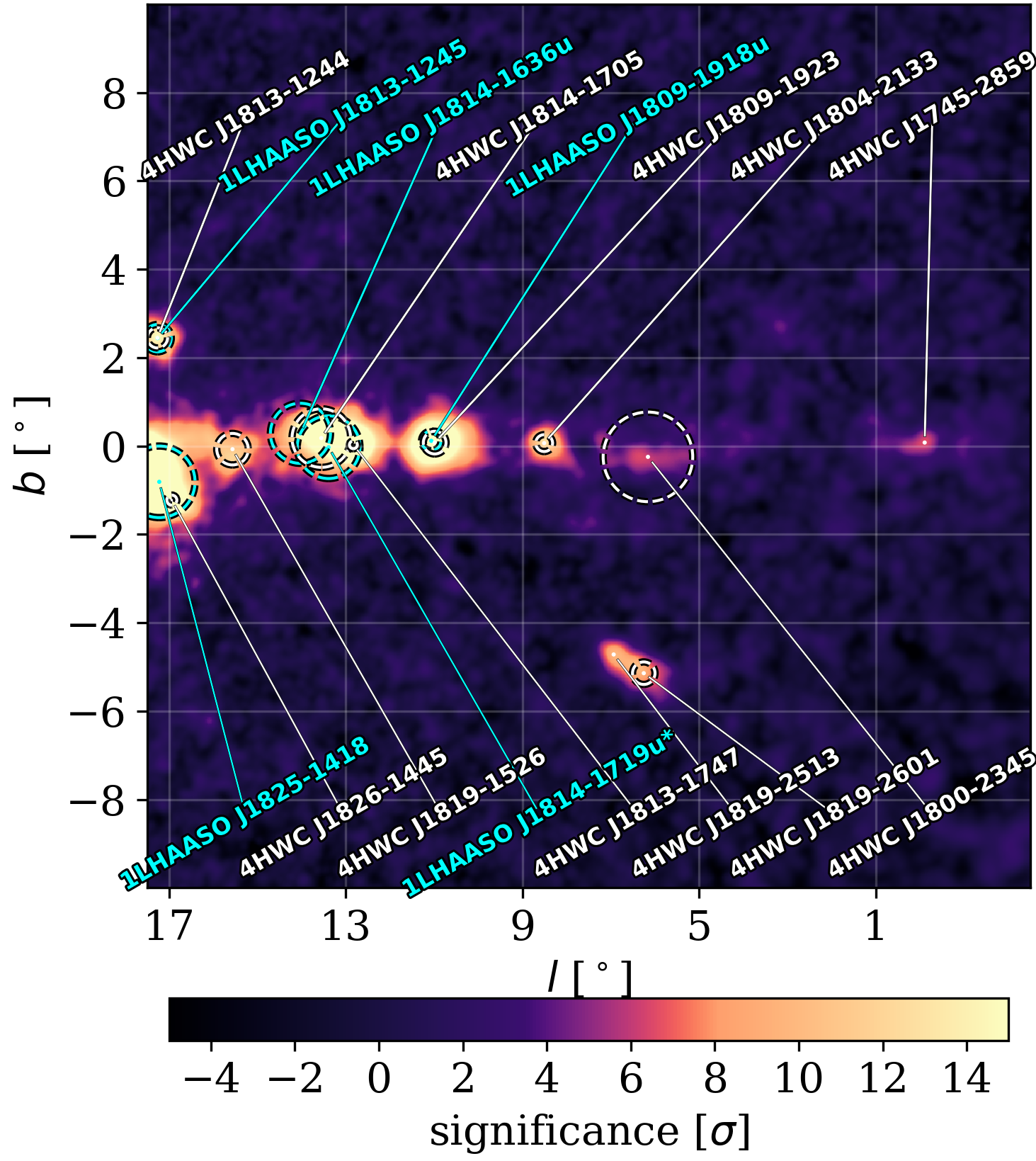}
    \caption{HAWC significance map near the Galactic Center with 1LHAASO (cyan) and 4HWC (white) source positions and sizes overlaid.}
    \label{fig:GalCenHAWCLHAASO}
\end{figure}

\begin{figure*}
    \centering
    \includegraphics[trim={0 0cm 0 0cm},width=0.45\textwidth]{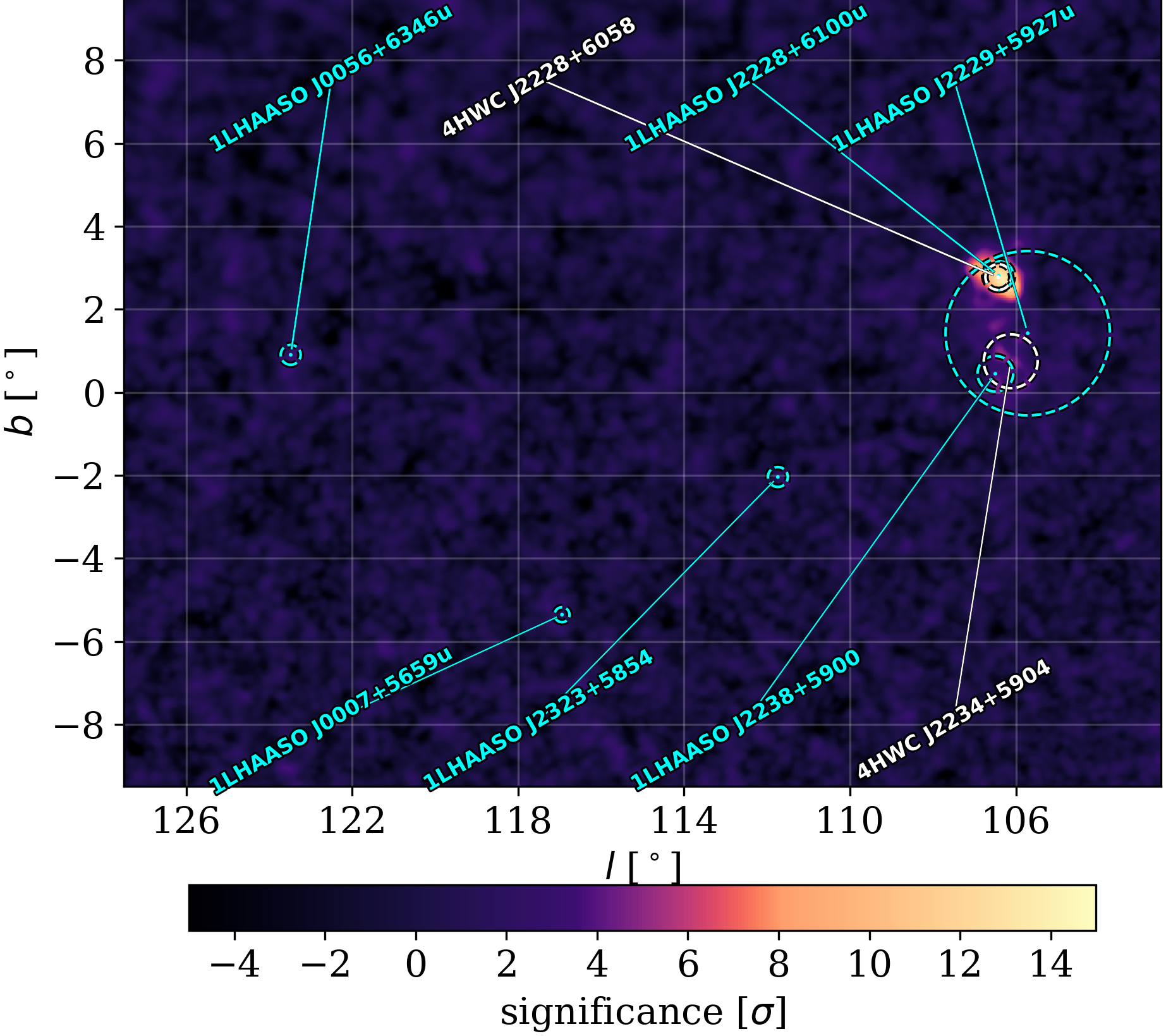}
    \caption{HAWC significance map near the Boomerang Nebula with 1LHAASO (cyan) and 4HWC (white) source positions and sizes overlaid.}
    \label{fig:BoomerangHAWCLHAASO}
\end{figure*}

 1LHAASO detects 90 sources between its two instruments WCDA and KM2A, and 4HWC detects 85. Of these, 1LHAASO detects 65 sources to be extended at the $3\sigma$ level while 4HWC detects 68 at the $4\sigma$ level. Twenty-five of 1LHAASO sources do not associate with a 4HWC source using the variable search radius, while 22 of 4HWC sources do not associate with a 1LHAASO source. Many of these differing sources are in the region of the sky where the instruments do not significantly overlap or one of the instruments is at a large zenith. The region close to the Galactic Center, as shown in Figure~\ref{fig:GalCenHAWCLHAASO}, is more favorable for the latitude of HAWC, while a region farther North, like that in Figure~\ref{fig:BoomerangHAWCLHAASO}, is more favorable for LHAASO.

\begin{figure*}
    \centering
    \includegraphics[trim={0cm 0cm 0cm 0cm},width=0.45\textwidth]{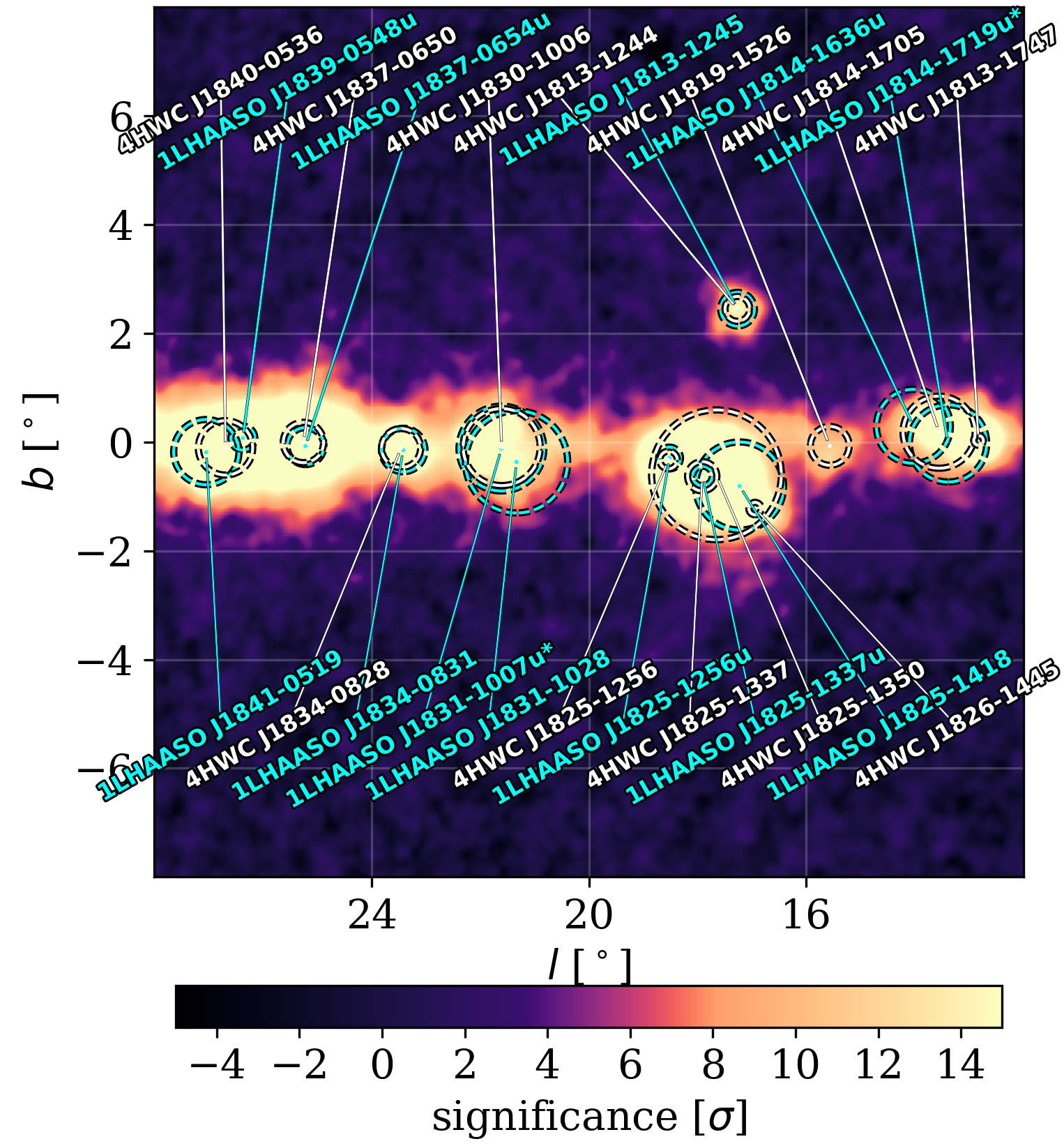}
    \caption{HAWC significance map centered on longitude 20\degree \space with 1LHAASO (cyan) and 4HWC (white) source positions and sizes overlaid.}
    \label{fig:galHAWCLHAASO}
\end{figure*}

When comparing a region with zenith angles below 40\degree \space for both experiments, the overall agreement between the two catalogs is clear. In Figure~\ref{fig:galHAWCLHAASO}, nearly all 4HWC and 1LHAASO sources are in strong agreement. Several 4HWC sources are paired with two 1LHAASO sources that typically correspond to the WCDA and KM2A components that were not merged in the 1LHAASO catalog. This agreement between the two catalogs in this overlapping region of the sky is a powerful systematic cross-check of the validity of the sources in both catalogs.   

\clearpage
\startlongtable
\begin{deluxetable*}{c||c|c|c||c|c||c}
    \tabletypesize{\tiny}
    \tablecaption{Sources from 4HWC not found in the 1LHAAO catalog}
    \label{table:sourceLHAASOMissing}
    \tablehead{
        \multicolumn{1}{c||}{} & \multicolumn{3}{c||}{Location and Extension} & \multicolumn{2}{c||}{Spectral Parameters} & \multicolumn{1}{c}{} \\
        \multicolumn{1}{c||}{Name} & \colhead{R.A.} & \colhead{Dec.} & \multicolumn{1}{c||}{Extension} & \colhead{Flux} & \multicolumn{1}{c||}{Index} & \multicolumn{1}{c}{TS} \\
        \multicolumn{1}{c||}{} & \colhead{[\degree]} & \colhead{[\degree]} & \multicolumn{1}{c||}{[\degree]} & \colhead{$10^{-15}\frac{1}{\mathrm{TeV} \mathrm{cm}^{2} \mathrm{s}}$} & \multicolumn{1}{c||}{} & \colhead{}
    }
    \startdata
      4HWC J0347-1405 & $56.82^{+0.35}_{-0.35}$ & $-14.09^{+0.08}_{-0.07}$  & - & $3.28^{+0.92}_{-0.73}$ & $-3.5^{+0.15}_{-0.06}$ & 30 \\
4HWC J0538+2804 & $84.58^{+0.03}_{-0.09}$ & $28.08^{+0.05}_{-0.04}$  & - & $0.73^{+0.45}_{-0.31}$ & $-2.87^{+0.11}_{-1.09}$ & 27 \\
4HWC J0856+2910 & $134.12^{+0.04}_{-0.07}$ & $29.18^{+0.04}_{-0.04}$  & - & $0.91^{+0.19}_{-0.2}$ & $-2.95^{+0.14}_{-1.04}$ & 31 \\
4HWC J1230+1223 & $187.67^{+0.04}_{-0.03}$ & $12.39^{+0.04}_{-0.03}$  & - & $0.59^{+0.18}_{-0.17}$ & $-2.63^{+0.13}_{-0.59}$ & 31 \\
4HWC J1745-2859 & $266.26^{+0.07}_{-0.06}$ & $-28.98^{+0.04}_{-0.04}$  & - & $28.0^{+17.1}_{-8.87}$ & $-2.95^{+0.14}_{-0.87}$ & 48 \\
4HWC J1800-2345 & $270.14^{+0.18}_{-0.19}$ & $-23.75^{+0.18}_{-0.18}$ & $1.01^{+0.12}_{-0.11}$ & $40.6^{+9.94}_{-8.99}$ & $-2.69^{+0.08}_{-0.07}$ & 106 \\
4HWC J1804-2133 & $271.1^{+0.04}_{-0.04}$ & $-21.56^{+0.04}_{-0.04}$ & $0.19^{+0.04}_{-0.04}$ & $19.4^{+3.4}_{-3.42}$ & $-2.88^{+0.08}_{-0.07}$ & 164 \\
4HWC J1813-1244 & $273.36^{+0.03}_{-0.03}$ & $-12.74^{+0.03}_{-0.03}$ & $0.23^{+0.03}_{-0.02}$ & $7.01^{+1.19}_{-1.14}$ & $-2.59^{+0.06}_{-0.06}$ & 267 \\
4HWC J1819-1526 & $274.81^{+0.06}_{-0.06}$ & $-15.44^{+0.07}_{-0.06}$ & $0.36^{+0.07}_{-0.06}$ & $11.9^{+3.25}_{-2.78}$ & $-2.73^{+0.07}_{-0.07}$ & 149 \\
4HWC J1819-2513 & $274.85^{+0.03}_{-0.04}$ & $-25.23^{+0.05}_{-0.03}$  & - & $0.89^{+0.54}_{-0.37}$ & $-1.99^{+0.22}_{-0.42}$ & 84 \\
4HWC J1819-2601 & $274.92^{+0.04}_{-0.04}$ & $-26.03^{+0.06}_{-0.06}$ & $0.26^{+0.04}_{-0.04}$ & $4.5^{+2.83}_{-1.89}$ & $-2.23^{+0.15}_{-0.14}$ & 123 \\
4HWC J1826-1445 & $276.53^{+0.02}_{-0.02}$ & $-14.76^{+0.03}_{-0.03}$ & $0.11^{+0.03}_{-0.04}$ & $6.79^{+1.61}_{-1.39}$ & $-2.69^{+0.08}_{-0.08}$ & 156 \\
4HWC J1846-0235 & $281.67^{+0.07}_{-0.08}$ & $-2.6^{+0.07}_{-0.09}$ & $0.3^{+0.05}_{-0.09}$ & $7.84^{+1.8}_{-1.91}$ & $-2.78^{+0.04}_{-0.04}$ & 237 \\
4HWC J1858+0752 & $284.51^{+0.04}_{-0.02}$ & $7.88^{+0.05}_{-0.02}$  & - & $0.49^{+0.27}_{-0.21}$ & $-2.4^{+0.19}_{-0.17}$ & 44 \\
4HWC J1915+1113 & $288.86^{+0.04}_{-0.04}$ & $11.23^{+0.05}_{-0.05}$ & $0.4^{+0.03}_{-0.03}$ & $3.71^{+0.24}_{-0.22}$ & $-2.56^{+0.03}_{-0.03}$ & 164 \\
4HWC J1918+1343 & $289.73^{+0.14}_{-0.14}$ & $13.72^{+0.18}_{-0.18}$ & $1.44^{+0.08}_{-0.08}$ & $13.2^{+0.91}_{-0.85}$ & $-2.73^{+0.03}_{-0.03}$ & 175 \\
4HWC J1932+1917 & $293.1^{+0.02}_{-0.03}$ & $19.3^{+0.03}_{-0.02}$  & - & $0.87^{+0.23}_{-0.23}$ & $-2.66^{+0.13}_{-0.11}$ & 59 \\
4HWC J2015+3708 & $303.99^{+0.02}_{-0.03}$ & $37.14^{+0.03}_{-0.03}$  & - & $0.89^{+0.28}_{-0.28}$ & $-2.62^{+0.14}_{-0.71}$ & 48 \\
4HWC J2026+3327 & $306.61^{+0.08}_{-0.08}$ & $33.47^{+0.05}_{-0.06}$ & $0.26^{+0.08}_{-0.07}$ & $1.47^{+0.43}_{-0.38}$ & $-2.58^{+0.1}_{-0.09}$ & 54 \\
4HWC J2029+3641 & $307.39^{+0.06}_{-0.06}$ & $36.68^{+0.05}_{-0.04}$  & - & $0.11^{+0.04}_{-0.05}$ & $-2.03^{+0.11}_{-0.06}$ & 28 \\ 
4HWC J2043+4353 & $310.82^{+0.34}_{-0.62}$ & $43.9^{+0.23}_{-0.41}$ & $0.87^{+0.31}_{-0.22}$ & $5.81^{+3.54}_{-1.86}$ & $-2.61^{+0.11}_{-0.65}$ & 52 \\
4HWC J2234+5904$^{\dagger}$ & $338.62^{+0.19}_{-0.19}$ & $59.08^{+0.13}_{-0.12}$ & $0.65^{+0.06}_{-0.08}$ & $10.1^{+1.41}_{-2.78}$ & $-2.59^{+0.09}_{-0.09}$ & 67.9
    \enddata
\end{deluxetable*}
\vspace{0.5cm}
\subsection{Association with Non-TeV Catalogs}

In addition to comparing and associating with previous TeV Catalogs, comparisons with multi-wavelength and physical source catalogs provides significant insight into the source characteristics of the 4HWC catalog. The sources in 4HWC are compared and associated with five additional catalogs: LMXB Cat. \citep{LMXB}, HMXB Cat. \citep{HMXB}, the Third \textit{Fermi} Pulsar Catalog \citep{FermiPulsar}, the ATNF Pulsar Catalog \citep{ATNF}, and SNRCat \citep{SNRcat}. 
\vspace{0.5cm}
\subsubsection{PWN and TeV Halo Candidate Association}

The two largest catalogs of the five are the Fermi third pulsar catalog and the ATNF pulsar catalog. In the 3HWC catalog, emission from PWNs and TeV Halos represented the largest source class of associated sources; this is also true for the 4HWC catalog. Associating with only the two pulsar catalogs results in 67 associated sources of the 85 total. However, not all pulsars in either catalog, but the ATNF especially, are likely to have either pulsar wind nebula or TeV Halos with detectable gamma-ray emission in HAWC data. As such, a chance probability cut based on the methodology used in \citet{Mattox_1997} and \citet{1LHAASO} is used to estimate the likelihood that a pulsar is spatially coincident with a 4HWC source by pure chance. This probability is calculated as:
\begin{equation}
    P_{chance} = 1 - e^{-r^2/r_0^2}\,,
\end{equation}
where $r_0$ is defined as $\left(\pi\rho(\dot{E})\right)^{-1/2}$ and $\rho(\dot{E})$ is the number density of pulsars with spindown power greater than the pulsar being considered. For this work, a radius of 5\degree \space around the pulsar in question was searched to calculate the number density. Using this chance probability, the closest pulsar may not be the one with the lowest chance probability. In that case, the pulsar with the smallest chance probability value below 0.01 that is within the search radius of a source is chosen instead. Sources with no pulsar with a chance lower than 0.01 are considered unassociated. Performing the association between just the two pulsar catalogs combined and 4HWC yields 50 of the 67 sources associated after the cut. The results of these associations are included in Table~\ref{table:sourceListPulsar}.
The distribution of associations between the two catalogs changes significantly after the cut as well. Before, 44 of the 67 associations were with the ATNF catalog, but after, only 26 still associate with it. In the common case of the same pulsar being present in both catalogs, the association is considered to be with the \textit{Fermi} catalog due to its more similar energy range to HAWC. The \textit{Fermi} catalog had 23 associations before and 24 after. The significant reduction in ATNF associations is expected and helps validate the chance criterion's performance as the \textit{Fermi} catalog should, and does, contain a much more energetic subset of the pulsars in the ATNF catalog. 

In addition to this chance probability criterion, a subset of both the ATNF and \textit{Fermi} pulsar catalogs is constructed to correspond to likely TeV Halo candidates. As in the 3HWC catalog, an age and spindown-power cut was used to broadly define the category based on the growing number of confirmed TeV Halos observed by HAWC, H.E.S.S., and LHAASO. Middle-aged pulsars aged 100--400 kyr with a spindown power of at least 1\% of the Geminga pulsar were selected to be marked as TeV Halo candidates. Due to the previous chance criterion cut the spindown power requirement is largely unnecessary, but is retained to better replicate the previous TeV Halo candidate listing from the 3HWC. When associating with both catalogs, eight of the 26 ATNF pulsars and 12 of the 24 \textit{Fermi} pulsars are marked as TeV Halo candidates. The conservative choice of 100~kyr as the lower limit of TeV Halo candidacy is chosen to match the 3HWC, but pulsars as young as a few tens of kyr are capable of producing a TeV Halo \citep{TeVHaloAge}. In total, about 40\% of the pulsar associations are linked with a possible TeV Halo after applying the chance probability cut. Before this cut, only 23\% were marked as linked with a TeV Halo. This result is unsurprising as the chance probability cut removes underpowered pulsars that are unlikely to produce a detectable TeV Halo regardless of age.
\startlongtable
\begin{deluxetable*}{c||c|c|c||c|c|c|c||c||c}
    \tabletypesize{\tiny}
    \tablecaption{Sources within 4HWC most closely associated with pulsars in the ATNF and Third \textit{Fermi} Pulsar Catalogs. The relevant pulsar spindown power ($\dot{E}$), age, and distance, when available, are provided. If a pulsar from the catalog passes the criteria to be a Halo Candidate, it is labeled as such in the Type column. Otherwise, it is labeled as a Pulsar regardless of the actual acceleration mechanism (e.g., a PWN).}
    \label{table:sourceListPulsar}
    \tablehead{
        \multicolumn{1}{c||}{} & \multicolumn{3}{c||}{Location and Extension} & \multicolumn{4}{c||}{Pulsar Characteristics} & \multicolumn{1}{c||}{} & \colhead{}  \\
        \multicolumn{1}{c||}{Name} & \colhead{R.A.} & \colhead{Dec.} & \multicolumn{1}{c||}{Extension} & \colhead{Type} & \colhead{$\dot{E}$} & \colhead{Age} & \multicolumn{1}{c||}{Dist} & \multicolumn{1}{c||}{TS} & \colhead{Source Assoc. (distance)} \\
        \multicolumn{1}{c||}{} & \colhead{[\degree]} & \colhead{[\degree]} & \multicolumn{1}{c||}{[\degree]} & \colhead{} & \colhead{erg/s} & \colhead{kyr} & \multicolumn{1}{c||}{kpc} & \multicolumn{1}{c||}{} & \colhead{[\degree]}
    }
    \startdata
      4HWC J0341+5258 & $55.49^{+0.06}_{-0.06}$ & $52.98^{+0.04}_{-0.04}$ & $0.18^{+0.03}_{-0.03}$ & Pulsar & 7.32e+31 & 2280 & 1.705 & 122 & B0339+53 (0.394) \\
4HWC J0534+2200 & $83.63^{+0.001}_{-0.001}$ & $22.01^{+0.001}_{-0.002}$ & - & Pulsar & 4.35e+38 & 1.27 & - & 82677 & J0534+2200 (0.005) \\
4HWC J0538+2804 & $84.58^{+0.03}_{-0.09}$ & $28.08^{+0.05}_{-0.04}$ & - & Pulsar & 4.94e+34 & 618.0 & 0.946 & 27 & J0538+2817 (0.207) \\
4HWC J0542+2315 & $85.74^{+0.09}_{-0.1}$ & $23.26^{+0.07}_{-0.08}$ & $0.98^{+0.07}_{-0.07}$ & Halo Candidate & 4.09e+34 & 253.0 & 1.565 & 411 & B0540+23 (0.226) \\
4HWC J0615+2214 & $93.79^{+0.13}_{-0.14}$ & $22.24^{+0.13}_{-0.14}$ & $1.1^{+0.12}_{-0.11}$ & Pulsar & 6.24e+34 & 89.3 & 1.744 & 181 & B0611+22 (0.335) \\
4HWC J0622+3759 & $95.56^{+0.08}_{-0.08}$ & $37.99^{+0.08}_{-0.07}$ & $0.5^{+0.09}_{-0.05}$ & Halo Candidate & 2.71e+34 & 208.0 & - & 170 & J0622+3749 (0.173) \\
4HWC J0631+1036 & $97.79^{+0.06}_{-0.06}$ & $10.61^{+0.05}_{-0.06}$ & $0.35^{+0.05}_{-0.05}$ & Pulsar & 1.70e+35 & 44.39 & - & 91 & J0631+1036 (0.077) \\
4HWC J0633+1719 & $98.48^{+0.08}_{-0.08}$ & $17.32^{+0.07}_{-0.07}$ & $2.2^{+0.07}_{-0.06}$ & Halo Candidate & 3.25e+34 & 342.36 & - & 1721 & J0633+1746 (0.447) \\
4HWC J0634+0628 & $98.56^{+0.06}_{-0.06}$ & $6.48^{+0.06}_{-0.06}$ & $0.46^{+0.1}_{-0.05}$ & Pulsar & 1.19e+35 & 59.21 & - & 150 & J0633+0632 (0.136) \\
4HWC J0701+1412 & $105.4^{+0.1}_{-0.1}$ & $14.22^{+0.08}_{-0.08}$ & $1.97^{+0.08}_{-0.08}$ & Halo Candidate & 3.80e+34 & 110.96 & - & 1048 & J0659+1414 (0.451) \\
4HWC J1740+0950 & $265.04^{+0.04}_{-0.02}$ & $9.85^{+0.02}_{-0.02}$ & - & Halo Candidate & 2.30e+35 & 114.45 & - & 89 & J1740+1000 (0.168) \\
4HWC J1804-2133 & $271.1^{+0.04}_{-0.04}$ & $-21.56^{+0.04}_{-0.04}$ & $0.19^{+0.04}_{-0.04}$ & Pulsar & 2.22e+36 & 15.8 & 3.425 & 164 & B1800-21 (0.148) \\
4HWC J1809-1923 & $272.38^{+0.02}_{-0.02}$ & $-19.39^{+0.02}_{-0.02}$ & $0.26^{+0.02}_{-0.02}$ & Pulsar & 1.78e+36 & 51.4 & 3.269 & 1371 & J1809-1917 (0.106) \\
4HWC J1813-1244 & $273.36^{+0.03}_{-0.03}$ & $-12.74^{+0.03}_{-0.03}$ & $0.23^{+0.03}_{-0.02}$ & Pulsar & 6.24e+36 & 43.38 & - & 267 & J1813-1246 (0.023) \\
4HWC J1825-1256 & $276.48^{+0.01}_{-0.01}$ & $-12.94^{+0.02}_{-0.02}$ & $0.16^{+0.01}_{-0.01}$ & Pulsar & 3.57e+36 & 14.41 & - & 902 & J1826-1256 (0.054) \\
4HWC J1834-0828 & $278.6^{+0.03}_{-0.03}$ & $-8.48^{+0.04}_{-0.04}$ & $0.36^{+0.03}_{-0.03}$ & Halo Candidate & 5.84e+35 & 147.0 & 4.377 & 359 & B1830-08 (0.188) \\
4HWC J1857+0200 & $284.48^{+0.02}_{-0.02}$ & $2.0^{+0.02}_{-0.02}$ & $0.18^{+0.02}_{-0.02}$ & Halo Candidate & 2.21e+34 & 164.0 & 5.93 & 273 & B1855+02 (0.212) \\
4HWC J1908+0616 & $287.06^{+0.01}_{-0.01}$ & $6.27^{+0.01}_{-0.01}$ & $0.39^{+0.01}_{-0.01}$ & Pulsar & 2.83e+36 & 19.5 & 2.576 & 227 & J1907+0602 (0.248) \\
4HWC J1912+1013 & $288.23^{+0.03}_{-0.02}$ & $10.22^{+0.03}_{-0.03}$ & $0.38^{+0.02}_{-0.02}$ & Halo Candidate & 2.88e+36 & 168.25 & - & 777 & J1913+1011 (0.112) \\
4HWC J1914+1151 & $288.68^{+0.02}_{-0.02}$ & $11.85^{+0.02}_{-0.02}$ & $0.1^{+0.03}_{-0.03}$ & Halo Candidate & 5.39e+35 & 116.0 & 14.003 & 167 & J1915+1150 (0.144) \\
4HWC J1923+1631 & $290.93^{+0.08}_{-0.08}$ & $16.53^{+0.08}_{-0.08}$ & $0.9^{+0.04}_{-0.04}$ & Pulsar & 2.56e+34 & 977.0 & 5.059 & 270 & J1924+1639 (0.153) \\
4HWC J1928+1747 & $292.14^{+0.02}_{-0.02}$ & $17.8^{+0.01}_{-0.02}$ & $0.18^{+0.02}_{-0.02}$ & Pulsar & 1.60e+36 & 82.6 & - & 498 & J1928+1746 (0.047) \\
4HWC J1928+1843 & $292.16^{+0.02}_{-0.02}$ & $18.73^{+0.03}_{-0.03}$ & $0.47^{+0.02}_{-0.02}$ & Pulsar & 1.16e+37 & 2.89 & 6.169 & 455 & J1930+1852 (0.487) \\
4HWC J1930+1852 & $292.59^{+0.1}_{-0.1}$ & $18.88^{+0.003}_{-0.003}$ & $0.03^{+0.002}_{-0.01}$ & Pulsar & 1.16e+37 & 2.89 & 6.169 & 33 & J1930+1852 (0.038) \\
4HWC J1932+1917 & $293.1^{+0.02}_{-0.03}$ & $19.3^{+0.03}_{-0.02}$ & - & Pulsar & 4.074e+35 & 35.41 & - & 59 & J1932+1916 (0.032) \\
4HWC J1953+2837 & $298.5^{+0.04}_{-0.03}$ & $28.62^{+0.03}_{-0.03}$ & $0.2^{+0.04}_{-0.03}$ & Pulsar & 1.05e+36 & 69.42 & - & 186 & J1954+2836 (0.084) \\
4HWC J1958+2851 & $299.6^{+0.05}_{-0.05}$ & $28.85^{+0.04}_{-0.04}$ & $0.33^{+0.04}_{-0.04}$ & Pulsar & 3.41e+35 & 21.74 & - & 269 & J1958+2846 (0.111) \\
4HWC J2005+3056 & $301.47^{+0.05}_{-0.05}$ & $30.93^{+0.04}_{-0.04}$ & $0.25^{+0.04}_{-0.03}$ & Halo Candidate & 2.24e+35 & 104.28 & - & 124 & J2006+3102 (0.123) \\
4HWC J2018+3641 & $304.67^{+0.02}_{-0.02}$ & $36.69^{+0.01}_{-0.01}$ & $0.23^{+0.01}_{-0.01}$ & Pulsar & 1.16e+34 & 1942.7 & - & 2262 & J2017+3625 (0.333) \\
4HWC J2021+3650 & $305.28^{+0.02}_{-0.02}$ & $36.85^{+0.01}_{-0.01}$ & $0.15^{+0.01}_{-0.01}$ & Pulsar & 3.35e+36 & 17.33 & - & 851 & J2021+3651 (0.007) \\
4HWC J2021+4036 & $305.29^{+0.14}_{-0.14}$ & $40.61^{+0.18}_{-0.18}$ & $0.2^{+0.04}_{-0.03}$ & Pulsar & 1.16e+35 & 76.9 & - & 147 & J2021+4026 (0.182) \\
4HWC J2026+3327 & $306.61^{+0.08}_{-0.08}$ & $33.47^{+0.05}_{-0.06}$ & $0.26^{+0.08}_{-0.07}$ & Halo Candidate & 3.47E+34 & 157.63 & - & 54 & J2028+3332 (0.478) \\
4HWC J2029+3641 & $307.39^{+0.06}_{-0.06}$ & $36.68^{+0.05}_{-0.04}$ & - & Halo Candidate & 3.206e+34 & 387.04 & - & 28 & J2030+3641 (0.115) \\
4HWC J2031+4128 & $307.95^{+0.02}_{-0.02}$ & $41.47^{+0.01}_{-0.01}$ & $0.23^{+0.01}_{-0.01}$ & Halo Candidate & 1.56e+35 & 195.25 & - & 1283 & J2032+4127 (0.109) \\

    \enddata
\end{deluxetable*}
\vspace{-1.5cm}
\subsubsection{Full Physical Source Association}

We combine the pulsar catalogs, refined using the chance probability criterion, with the other three previously mentioned catalogs. Using this combined catalog, we provide a distribution of likely associated source class labels. The results of these associations are included in Table~\ref{table:sourceListPhysical}

Three sources, 4HWC J1819-2601, 4HWC J1910+0503, and 4HWC J1913+0457, were manually labeled to account for their known or likely association to the jets of the low-mass X-ray binary V4641 Sgr and high-mass X-ray binary SS~433. The LMXB and HMXB catalogs report the center of the microquasars, while these sources are associated with the jets that may extend far from the central engine of these two systems \citep{SS433,V4641}. 
After performing the association with all of these catalogs, a total of 33 4HWC sources best associate with pulsars, of which 14 are marked as TeV Halo candidates. An additional 22 associate most closely to an SNR. A further nine associate with a high-mass X-ray binary and six with a low-mass X-ray binary. The remaining 15 are unassociated, however, extragalactic sources like Markarian 421 and 501, along with M87, are not included in any of the five catalogs searched, so they appear as unassociated. At least five of the unassociated sources are $>$$20\degree$ from the Galactic Plane and thus likely have extragalactic origins. 

\startlongtable
\begin{deluxetable*}{c||c|c|c|c|c||c||c|c}
    \tabletypesize{\tiny}
    \tablecaption{4HWC Source List with location parameters of each source in R.A./Dec. as well as Galactic longitude and latitude. The TS and category of association are provided alongside the label and angular distance between the physical source association. The association names come from the relevant source catalog, so a mix of proper names and coordinate-based names is used depending on the origin. The Type column designates the catalog of origin for the source association if any is made.}
    \label{table:sourceListPhysical}
    \tablehead{
        \multicolumn{1}{c||}{} & \multicolumn{5}{c||}{Location and Extension} & \multicolumn{1}{c||}{} & \colhead{} \\
        \multicolumn{1}{c||}{Name} & \colhead{R.A.} & \colhead{Dec.} & \colhead{Longitude} & \colhead{Latitude} & \multicolumn{1}{c||}{Extension} & \multicolumn{1}{c||}{TS} & \colhead{Type} & \colhead{Source Assoc. (distance)} \\
        \multicolumn{1}{c||}{} & \colhead{[\degree]} & \colhead{[\degree]} & \colhead{[\degree]} & \colhead{[\degree]} & \multicolumn{1}{c||}{[\degree]} & \multicolumn{1}{c||}{} & \colhead{} & \colhead{[\degree]}
    }
    \startdata
      4HWC J0341+5258 & $55.49^{+0.06}_{-0.06}$ & $52.98^{+0.04}_{-0.04}$ & $147.01$ & $-1.73$ & $0.18^{+0.03}_{-0.03}$ & 122 & - & - \\
4HWC J0347-1405 & $56.82^{+0.35}_{-0.35}$ & $-14.09^{+0.08}_{-0.07}$ & $204.34$ & $-47.08$ & - & 30 & - & - \\
4HWC J0533+3534 & $83.44^{+0.02}_{-0.02}$ & $35.57^{+0.02}_{-0.03}$ & $173.05$ & $1.41$ & - & 61 & SNR & G172.8+01.5 (0.269) \\
4HWC J0534+2200 & $83.63^{+0.001}_{-0.001}$ & $22.01^{+0.001}_{-0.002}$ & $184.56$ & $-5.79$ & - & 82677 & Pulsar & J0534+2200 (0.005) \\
4HWC J0538+2804 & $84.58^{+0.03}_{-0.09}$ & $28.08^{+0.05}_{-0.04}$ & $179.88$ & $-1.81$ & - & 27 & Pulsar & J0538+2817 (0.207) \\
4HWC J0542+2315 & $85.74^{+0.09}_{-0.1}$ & $23.26^{+0.07}_{-0.08}$ & $184.53$ & $-3.48$ & $0.98^{+0.07}_{-0.07}$ & 411 & Pulsar & B0540+23 (0.226) \\
4HWC J0615+2214 & $93.79^{+0.13}_{-0.14}$ & $22.24^{+0.13}_{-0.14}$ & $189.12$ & $2.45$ & $1.1^{+0.12}_{-0.11}$ & 181 & Pulsar & B0611+22 (0.335) \\
4HWC J0622+3759 & $95.56^{+0.08}_{-0.08}$ & $37.99^{+0.08}_{-0.07}$ & $175.73$ & $11.04$ & $0.5^{+0.09}_{-0.05}$ & 170 & Pulsar & J0622+3749 (0.173) \\
4HWC J0631+1036 & $97.79^{+0.06}_{-0.06}$ & $10.61^{+0.05}_{-0.06}$ & $201.19$ & $0.38$ & $0.35^{+0.05}_{-0.05}$ & 91 & Pulsar & J0631+1036 (0.077) \\
4HWC J0633+1719 & $98.48^{+0.08}_{-0.08}$ & $17.32^{+0.07}_{-0.07}$ & $195.54$ & $4.06$ & $2.2^{+0.07}_{-0.06}$ & 1721 & Pulsar & J0633+1746 (0.447) \\
4HWC J0634+1734 & $98.51^{+0.06}_{-0.06}$ & $17.57^{+0.06}_{-0.06}$ & $195.33$ & $4.2$ & $0.85^{+0.06}_{-0.05}$ & 486 & Pulsar & J0633+1746 (0.199) \\
4HWC J0634+0628 & $98.56^{+0.06}_{-0.06}$ & $6.48^{+0.06}_{-0.06}$ & $205.21$ & $-0.85$ & $0.46^{+0.1}_{-0.05}$ & 150 & Pulsar & J0633+0632 (0.136) \\
4HWC J0701+1412 & $105.4^{+0.1}_{-0.1}$ & $14.22^{+0.08}_{-0.08}$ & $201.32$ & $8.64$ & $1.97^{+0.08}_{-0.08}$ & 1048 & Pulsar & J0659+1414 (0.451) \\
4HWC J0856+2910 & $134.12^{+0.04}_{-0.07}$ & $29.18^{+0.04}_{-0.04}$ & $195.99$ & $38.72$ & - & 31 & - & - \\
4HWC J1104+3810 & $166.15^{+0.004}_{-0.004}$ & $38.18^{+0.004}_{-0.004}$ & $179.88$ & $65.07$ & $0.09^{+0.005}_{-0.005}$ & 10782 & - & - \\
4HWC J1230+1223 & $187.67^{+0.04}_{-0.03}$ & $12.39^{+0.04}_{-0.03}$ & $283.65$ & $74.48$ & - & 31 & - & - \\
4HWC J1654+3944 & $253.51^{+0.01}_{-0.01}$ & $39.74^{+0.01}_{-0.01}$ & $63.58$ & $38.82$ & - & 782 & - & - \\
4HWC J1740+0950 & $265.04^{+0.04}_{-0.02}$ & $9.85^{+0.02}_{-0.02}$ & $33.83$ & $20.26$ & - & 89 & Pulsar & J1740+1000 (0.168) \\
4HWC J1745-2859 & $266.26^{+0.07}_{-0.06}$ & $-28.98^{+0.04}_{-0.04}$ & $359.9$ & $0.09$ & - & 48 & LMXB & GRS 1741.9-2853 (0.07) \\
4HWC J1800-2345 & $270.14^{+0.18}_{-0.19}$ & $-23.75^{+0.18}_{-0.18}$ & $6.17$ & $-0.25$ & $1.01^{+0.12}_{-0.11}$ & 106 & SNR & G006.4-00.1 (0.321) \\
4HWC J1804-2133 & $271.1^{+0.04}_{-0.04}$ & $-21.56^{+0.04}_{-0.04}$ & $8.51$ & $0.06$ & $0.19^{+0.04}_{-0.04}$ & 164 & Pulsar & B1800-21 (0.148) \\
4HWC J1809-1923 & $272.38^{+0.02}_{-0.02}$ & $-19.39^{+0.02}_{-0.02}$ & $10.99$ & $0.07$ & $0.26^{+0.02}_{-0.02}$ & 1371 & Pulsar & J1809-1917 (0.106) \\
4HWC J1813-1244 & $273.36^{+0.03}_{-0.03}$ & $-12.74^{+0.03}_{-0.03}$ & $17.27$ & $2.45$ & $0.23^{+0.03}_{-0.02}$ & 267 & Pulsar & J1813-1246 (0.023) \\
4HWC J1813-1747 & $273.36^{+0.01}_{-0.01}$ & $-17.8^{+0.01}_{-0.01}$ & $12.83$ & $0.03$ & $0.09^{+0.02}_{-0.02}$ & 614 & SNR & G012.8-00.0 (0.049) \\
4HWC J1814-1705 & $273.58^{+0.04}_{-0.04}$ & $-17.09^{+0.05}_{-0.05}$ & $13.55$ & $0.18$ & $0.66^{+0.04}_{-0.04}$ & 717 & LMXB & GX 13+1 (0.084) \\
4HWC J1819-1526 & $274.81^{+0.06}_{-0.06}$ & $-15.44^{+0.07}_{-0.06}$ & $15.57$ & $-0.07$ & $0.36^{+0.07}_{-0.06}$ & 149 & SNR & G015.4+00.1 (0.301) \\
4HWC J1819-2513 & $274.85^{+0.03}_{-0.04}$ & $-25.23^{+0.05}_{-0.03}$ & $6.94$ & $-4.71$ & - & 84 & LMXB & V4641 Sgr (0.178) \\
4HWC J1819-2601 & $274.92^{+0.04}_{-0.04}$ & $-26.03^{+0.06}_{-0.06}$ & $6.25$ & $-5.14$ & $0.26^{+0.04}_{-0.04}$ & 123 & - & - \\
4HWC J1825-1350 & $276.31^{+0.05}_{-0.05}$ & $-13.85^{+0.07}_{-0.07}$ & $17.65$ & $-0.61$ & $1.19^{+0.08}_{-0.08}$ & 804 & SNR & G018.0-00.7 (0.363) \\
4HWC J1825-1337 & $276.46^{+0.01}_{-0.01}$ & $-13.63^{+0.01}_{-0.01}$ & $17.91$ & $-0.63$ & $0.27^{+0.01}_{-0.01}$ & 3261 & SNR & G018.0-00.7 (0.108) \\
4HWC J1825-1256 & $276.48^{+0.01}_{-0.01}$ & $-12.94^{+0.02}_{-0.02}$ & $18.53$ & $-0.33$ & $0.16^{+0.01}_{-0.01}$ & 902 & Pulsar & J1826-1256 (0.054) \\
4HWC J1826-1445 & $276.53^{+0.02}_{-0.02}$ & $-14.76^{+0.03}_{-0.03}$ & $16.95$ & $-1.22$ & $0.11^{+0.03}_{-0.04}$ & 156 & HMXB & LS 5039 (0.089) \\
4HWC J1830-1006 & $277.72^{+0.03}_{-0.03}$ & $-10.11^{+0.03}_{-0.03}$ & $21.6$ & $-0.09$ & $0.71^{+0.03}_{-0.03}$ & 359 & SNR & G021.5-00.1 (0.041) \\
4HWC J1834-0828 & $278.6^{+0.03}_{-0.03}$ & $-8.48^{+0.04}_{-0.04}$ & $23.45$ & $-0.11$ & $0.36^{+0.03}_{-0.03}$ & 359 & Pulsar & B1830-08 (0.188) \\
4HWC J1837-0650 & $279.35^{+0.01}_{-0.01}$ & $-6.84^{+0.01}_{-0.01}$ & $25.25$ & $-0.01$ & $0.37^{+0.01}_{-0.01}$ & 4305 & HMXB & AX J1838.0-0655 (0.184) \\
4HWC J1840-0536 & $280.1^{+0.02}_{-0.02}$ & $-5.61^{+0.02}_{-0.02}$ & $26.68$ & $-0.11$ & $0.5^{+0.01}_{-0.01}$ & 4600 & SNR & G026.6-00.1 (0.093) \\
4HWC J1843-0330 & $280.96^{+0.02}_{-0.02}$ & $-3.51^{+0.03}_{-0.02}$ & $28.94$ & $0.09$ & $0.44^{+0.01}_{-0.01}$ & 2757 & LMXB & SWIFT J1843.5-0343 (0.217) \\
4HWC J1846-0235 & $281.67^{+0.07}_{-0.08}$ & $-2.6^{+0.07}_{-0.09}$ & $30.08$ & $-0.13$ & $0.3^{+0.05}_{-0.09}$ & 237 & HMXB & IGR J18462-0223 (0.252) \\
4HWC J1847+0051 & $281.77^{+0.29}_{-0.31}$ & $-0.87^{+0.43}_{-0.43}$ & $31.66$ & $0.57$ & $2.84^{+0.35}_{-0.32}$ & 191 & HMXB & Swift J1845.7-0037 (0.361) \\
4HWC J1848-0140 & $282.11^{+0.02}_{-0.02}$ & $-1.67^{+0.04}_{-0.03}$ & $31.1$ & $-0.09$ & $0.43^{+0.02}_{-0.02}$ & 269 & LMXB & MAXI J1848-015 (0.202) \\
4HWC J1848+0000 & $282.24^{+0.01}_{-0.01}$ & $0.01^{+0.01}_{-0.01}$ & $32.66$ & $0.56$ & $0.08^{+0.02}_{-0.02}$ & 272 & SNR & G032.6+00.5 (0.032) \\
4HWC J1851+0002 & $282.85^{+0.03}_{-0.03}$ & $-0.05^{+0.02}_{-0.02}$ & $32.88$ & $-0.01$ & $0.46^{+0.02}_{-0.02}$ & 686 & SNR & G032.8-00.1 (0.083) \\
4HWC J1854+0120 & $283.58^{+0.09}_{-0.1}$ & $1.34^{+0.11}_{-0.12}$ & $34.45$ & $-0.03$ & $0.88^{+0.08}_{-0.07}$ & 273 & SNR & G034.7-00.4 (0.421) \\
4HWC J1857+0247 & $284.34^{+0.01}_{-0.01}$ & $2.8^{+0.02}_{-0.02}$ & $36.1$ & $-0.04$ & $0.27^{+0.02}_{-0.02}$ & 593 & SNR & G036.0+00.1 (0.134) \\
4HWC J1857+0200 & $284.48^{+0.02}_{-0.02}$ & $2.0^{+0.02}_{-0.02}$ & $35.45$ & $-0.53$ & $0.18^{+0.02}_{-0.02}$ & 273 & Pulsar & B1855+02 (0.212) \\
4HWC J1858+0752 & $284.51^{+0.04}_{-0.02}$ & $7.88^{+0.05}_{-0.02}$ & $40.7$ & $2.13$ & - & 44 & - & - \\
4HWC J1858+0344 & $284.74^{+0.05}_{-0.05}$ & $3.74^{+0.05}_{-0.05}$ & $37.12$ & $0.03$ & $0.63^{+0.04}_{-0.04}$ & 228 & HMXB & XTE J1858+034 (0.307) \\
4HWC J1906+0613 & $286.55^{+0.06}_{-0.06}$ & $6.22^{+0.06}_{-0.06}$ & $40.15$ & $-0.43$ & $1.22^{+0.06}_{-0.06}$ & 25 & SNR & G040.5-00.5 (0.378) \\
4HWC J1908+0845 & $287.03^{+0.06}_{-0.06}$ & $8.76^{+0.09}_{-0.09}$ & $42.62$ & $0.31$ & $0.76^{+0.08}_{-0.07}$ & 378 & SNR & G042.8+00.6 (0.376) \\
4HWC J1908+0616 & $287.06^{+0.01}_{-0.01}$ & $6.27^{+0.01}_{-0.01}$ & $40.43$ & $-0.86$ & $0.39^{+0.01}_{-0.01}$ & 227 & Pulsar & J1907+0602 (0.248) \\
4HWC J1910+0503 & $287.62^{+0.0}_{-0.0}$ & $5.06^{+0.001}_{-0.001}$ & $39.61$ & $-1.91$ & - & 101 & HMXB & SS 433 (0.341) \\
4HWC J1912+1013 & $288.23^{+0.03}_{-0.02}$ & $10.22^{+0.03}_{-0.03}$ & $44.46$ & $-0.06$ & $0.38^{+0.02}_{-0.02}$ & 777 & Pulsar & J1913+1011 (0.112) \\
4HWC J1913+0457 & $288.37^{+0.02}_{-0.02}$ & $4.96^{+0.02}_{-0.03}$ & $39.86$ & $-2.62$ & - & 77 & HMXB & SS 433 (0.411) \\
4HWC J1914+1151 & $288.68^{+0.02}_{-0.02}$ & $11.85^{+0.02}_{-0.02}$ & $46.11$ & $0.3$ & $0.1^{+0.03}_{-0.03}$ & 167 & Pulsar & J1915+1150 (0.144) \\
4HWC J1915+1113 & $288.86^{+0.04}_{-0.04}$ & $11.23^{+0.05}_{-0.05}$ & $45.65$ & $-0.14$ & $0.4^{+0.03}_{-0.03}$ & 164 & SNR & G045.7-00.4 (0.26) \\
4HWC J1918+1343 & $289.73^{+0.14}_{-0.14}$ & $13.72^{+0.18}_{-0.18}$ & $48.24$ & $0.27$ & $1.44^{+0.08}_{-0.08}$ & 175 & - & - \\
4HWC J1922+1405 & $290.74^{+0.02}_{-0.02}$ & $14.09^{+0.02}_{-0.02}$ & $49.03$ & $-0.42$ & $0.08^{+0.02}_{-0.02}$ & 304 & SNR & G049.2-00.7 (0.223) \\
4HWC J1923+1631 & $290.93^{+0.08}_{-0.08}$ & $16.53^{+0.08}_{-0.08}$ & $51.27$ & $0.57$ & $0.9^{+0.04}_{-0.04}$ & 270 & Pulsar & J1924+1639 (0.153) \\
4HWC J1928+1747 & $292.14^{+0.02}_{-0.02}$ & $17.8^{+0.01}_{-0.02}$ & $52.94$ & $0.16$ & $0.18^{+0.02}_{-0.02}$ & 498 & Pulsar & J1928+1746 (0.047) \\
4HWC J1928+1843 & $292.16^{+0.02}_{-0.02}$ & $18.73^{+0.03}_{-0.03}$ & $53.76$ & $0.58$ & $0.47^{+0.02}_{-0.02}$ & 455 & Pulsar & J1930+1852 (0.487) \\
4HWC J1930+1852 & $292.59^{+0.1}_{-0.1}$ & $18.88^{+0.003}_{-0.003}$ & $54.09$ & $0.3$ & $0.03^{+0.002}_{-0.01}$ & 33 & Pulsar & J1930+1852 (0.038) \\
4HWC J1931+1655 & $292.81^{+0.03}_{-0.03}$ & $16.92^{+0.03}_{-0.03}$ & $52.47$ & $-0.83$ & $0.18^{+0.03}_{-0.03}$ & 77 & - & - \\
4HWC J1932+1917 & $293.1^{+0.02}_{-0.03}$ & $19.3^{+0.03}_{-0.02}$ & $54.69$ & $0.08$ & - & 59 & Pulsar & J1932+1916 (0.032) \\
4HWC J1937+2143 & $294.36^{+0.17}_{-0.16}$ & $21.73^{+0.18}_{-0.18}$ & $57.39$ & $0.23$ & $1.15^{+0.14}_{-0.12}$ & 125 & SNR & G057.2+00.8 (0.649) \\
4HWC J1945+2434 & $296.37^{+0.17}_{-0.18}$ & $24.58^{+0.16}_{-0.16}$ & $60.78$ & $0.03$ & $1.66^{+0.12}_{-0.11}$ & 340 & - & - \\
4HWC J1952+2610 & $298.01^{+0.17}_{-0.17}$ & $26.18^{+0.18}_{-0.2}$ & $62.91$ & $-0.43$ & $0.99^{+0.14}_{-0.12}$ & 106 & HMXB & IGR J19498+2534 (0.811) \\
4HWC J1952+2925 & $298.11^{+0.04}_{-0.03}$ & $29.42^{+0.03}_{-0.03}$ & $65.73$ & $1.15$ & $0.18^{+0.04}_{-0.03}$ & 190 & SNR & G065.7+01.2 (0.068) \\
4HWC J1953+2837 & $298.5^{+0.04}_{-0.03}$ & $28.62^{+0.03}_{-0.03}$ & $65.22$ & $0.45$ & $0.2^{+0.04}_{-0.03}$ & 186 & Pulsar & J1954+2836 (0.084) \\
4HWC J1954+3253 & $298.61^{+0.05}_{-0.05}$ & $32.89^{+0.03}_{-0.04}$ & $68.93$ & $2.56$ & $0.19^{+0.04}_{-0.04}$ & 105 & SNR & G069.0+02.7 (0.279) \\
4HWC J1958+2851 & $299.6^{+0.05}_{-0.05}$ & $28.85^{+0.04}_{-0.04}$ & $65.92$ & $-0.26$ & $0.33^{+0.04}_{-0.04}$ & 269 & Pulsar & J1958+2846 (0.111) \\
4HWC J2005+3056 & $301.47^{+0.05}_{-0.05}$ & $30.93^{+0.04}_{-0.04}$ & $68.54$ & $-0.53$ & $0.25^{+0.04}_{-0.03}$ & 124 & Pulsar & J2006+3102 (0.123) \\
4HWC J2006+3355 & $301.57^{+0.1}_{-0.1}$ & $33.92^{+0.08}_{-0.08}$ & $71.11$ & $1.01$ & $1.28^{+0.08}_{-0.07}$ & 619 & - & - \\
4HWC J2015+3708 & $303.99^{+0.02}_{-0.03}$ & $37.14^{+0.03}_{-0.03}$ & $74.88$ & $1.12$ & - & 48 & SNR & G074.9+01.2 (0.066) \\
4HWC J2018+3641 & $304.67^{+0.02}_{-0.02}$ & $36.69^{+0.01}_{-0.01}$ & $74.82$ & $0.42$ & $0.23^{+0.01}_{-0.01}$ & 2262 & Pulsar & J2017+3625 (0.333) \\
4HWC J2021+3650 & $305.28^{+0.02}_{-0.02}$ & $36.85^{+0.01}_{-0.01}$ & $75.22$ & $0.11$ & $0.15^{+0.01}_{-0.01}$ & 851 & Pulsar & J2021+3651 (0.007) \\
4HWC J2021+4036 & $305.29^{+0.14}_{-0.14}$ & $40.61^{+0.18}_{-0.18}$ & $78.32$ & $2.24$ & $0.2^{+0.04}_{-0.03}$ & 147 & Pulsar & J2021+4026 (0.182) \\
4HWC J2022+3715 & $305.69^{+0.26}_{-0.24}$ & $37.26^{+0.23}_{-0.2}$ & $75.75$ & $0.07$ & $1.61^{+0.22}_{-0.18}$ & 200 & SNR & G075.2+00.1 (0.572) \\
4HWC J2026+3327 & $306.61^{+0.08}_{-0.08}$ & $33.47^{+0.05}_{-0.06}$ & $73.08$ & $-2.72$ & $0.26^{+0.08}_{-0.07}$ & 54 & Pulsar & J2028+3332 (0.478) \\
4HWC J2029+3641 & $307.39^{+0.06}_{-0.06}$ & $36.68^{+0.05}_{-0.04}$ & $76.06$ & $-1.37$ & - & 28 & Pulsar & J2030+3641 (0.115) \\
4HWC J2030+4056 & $307.64^{+0.13}_{-0.13}$ & $40.95^{+0.1}_{-0.1}$ & $79.63$ & $0.98$ & $1.81^{+0.08}_{-0.08}$ & 731 & HMXB & Cyg X-3 (0.473) \\
4HWC J2031+4128 & $307.95^{+0.02}_{-0.02}$ & $41.47^{+0.01}_{-0.01}$ & $80.19$ & $1.1$ & $0.23^{+0.01}_{-0.01}$ & 1283 & Pulsar & J2032+4127 (0.109) \\
4HWC J2043+4353 & $310.82^{+0.34}_{-0.62}$ & $43.9^{+0.23}_{-0.41}$ & $83.4$ & $0.89$ & $0.87^{+0.31}_{-0.22}$ & 52 & - & - \\
4HWC J2108+5156 & $317.15^{+0.05}_{-0.04}$ & $51.94^{+0.03}_{-0.02}$ & $92.27$ & $2.86$ & - & 73 & - & - \\
4HWC J2228+6058 & $337.06^{+0.07}_{-0.07}$ & $60.97^{+0.04}_{-0.04}$ & $106.42$ & $2.77$ & $0.31^{+0.03}_{-0.03}$ & 284 & SNR & G106.3+02.7 (0.235) \\
4HWC J2234+5904 & $338.62^{+0.19}_{-0.19}$ & $59.08^{+0.13}_{-0.12}$ & $106.13$ & $0.75$ & $0.65^{+0.06}_{-0.08}$ & 90 & - & - \\

    \enddata
\end{deluxetable*}
\vspace{-1.5cm}
\subsection{Analyzing Unassociated 4HWC Sources}

Of the 85 sources detected above 25~TS in the catalog, 11 have no identified TeV counterpart in TeVCat: 4HWC J0347-1405, 4HWC J0538+2804, 4HWC J0856+2910, 4HWC J1858+0752, 4HWC J1915+1113, 4HWC J1932+1917, 4HWC J1945+2434, 4HWC J2021+3650, 4HWC J2026+3327, 4HWC J2029+3641, and 4HWC J2234+5904. Of these 11 sources 4HWC J0347-1405, 4HWC J0856+2910, 4HWC J1858+0752, 4HWC J1945+2434, 4HWC J2234+5904 have no Physical counterpart.

Of these sources, two (4HWC J0347-1405 and 4HWC J0856+2910) are very far from the Galactic Plane and are likely be extragalactic sources. They both lack any association nearby with any of the catalogs searched, as well as 4FGL \citep{4FGL}. Their soft spectrum and distance from the Plane leads to the belief that they are extragalactic in origin. Follow-up studies will be needed to confirm their existence and any possible physical origin. At 27 and 31~TS, respectively, these sources are close to the detection threshold of 25 TS set for the catalog. 

4HWC J0538+2804 is very near the Galactic anti-center at (179.88,$-1.816$). Its index of around $-2.5$ and proximity to the Galactic Plane points to a Galactic origin for this source. PSR J0538+2817 in the ATNF catalog is only 0.207 \degree \space away and is a likely candidate for the source of the emission as it passes the 0.01 chance probability cut applied to the pulsar catalog.  

4HWC J1858+0752 has no TeV counterpart within 1\degree, but there is a 4FGL source (4FGL J1857.5+0756) within 0.14\degree, which places it firmly within association territory. The spectral properties of the power law fits of the two sources is compatible within the $2\sigma$ uncertainty bands. However, the 4FGL source has evidence of a log-parabola curved spectrum at $2.75\sigma$ which predicts incompatibly low flux at TeV energies for the observed emission in HAWC. The source origin is unlabeled in the 4FGL catalog. There is also a pulsar approximately 0.35\degree \space away from the location of 4HWC J1858+0752, but it fails the 0.01 probability chance criterion and there is no evidence other than the moderate proximity to link the TeV component to the pulsar directly.

4HWC J1915+1113 also only has a single gamma-ray association in 4FGL. The 4FGL source J1916.3+1108 is about 0.25\degree\, away, with an SNR at the same location. The 4HWC source morphology fits a nearly 0.4\degree \space extent while the 4FGL source quotes an average extent of 0.15\degree. The spectra match up well within $1\sigma$, so it's plausible that this source is the TeV counterpart to the 4FGL source, but needs a proper follow-up analysis. 

4HWC J1932+1917 has a strong spatial match with PSR J1932+1916 at a distance of less than 0.04\degree \space and 4FGL J1932.3+1916 less than 0.05\degree. The power-law-spectrum fits of the two sources matches strongly in both normalization and spectral index. However, 4FGL J1932.3+1916 is much better explained by a log-parabola spectrum in \textit{Fermi}-LAT data with a significance of ~$16\sigma$. This log-parabola spectrum cuts off strongly before 100 GeV and is incompatible with the flux seen in HAWC. 4HWC J1932+1917 is a point source in the catalog fit and 4FGL J1932.3+1916 reports a size of 0.01\degree \space to 0.02\degree. The spatial and morphological link between the new TeV source and PSR J1932+1916 is very strong, but more detailed multi-wavelength analysis is necessary to confirm or rule out any link between the pulsar and TeV emission in its vicinity. Although it is not cited in TeVCat, the HAWC publication \citep{J1932} has already modeled this source and performed associations with the nearby pulsar and 4FGL source in more detail. The paper indicates that the HAWC source J1932+1917 likely corresponds to a PWN around the pulsar PSR J1932+1916.

4HWC J1945+2434 has a 4FGL source (J1946.1+2436c) only 0.171\degree \space away with an index of $-2.42$, as well as a 1LHAASO source at the same location. However, this 1LHAASO source was not included in TeVCat. The 1LHAASO source fit was marked as heavily impacted by the diffuse modeling, and this may be the reason for its non-inclusion into TeVCat.   

4HWC J2021+3650 has a 4FGL source J2021.1+365 extremely close by--- 0.007\degree \space away. The \textit{Fermi} source has a moderately hard spectrum at $-2.34$, making it possible for the emission to extend up into the TeV band. There is a close 1LHAASO source at approximately 0.35\degree\, away, but the LHAASO source is an extended source offset towards other emission, while J2021+3650 is a point source located in a small pocket of emission in the HAWC data maps. 

4HWC J2026+3327 has a 4FGL source approximately 0.3\degree \space away and a pulsar component nearly 0.48\degree \space away, leading to a weak possible association with either source. The fit source has large uncertainties for the observed flux and position, along with being an extended source, so an association 0.3\degree away is not unreasonable, but little data exists for any other emission in the region. 

4HWC J2029+3641 has two nearby candidates: 4FGL J2030.0+3641 around 0.12\degree \space away and a pulsar PSR J2030+3641 similarly around 0.12\degree \space away. PSR J2030+3641 meets the criteria for a TeV Halo Candidate in this work, marking this source as a possible TeV Halo. This provides quite strong evidence to link the \textit{Fermi}-LAT source and the pulsar to this unassociated TeV emission. The observed spectrum from the HAWC source is harder than the \textit{Fermi} source, but the uncertainties for the source are quite large, so they still overlap. Extrapolating the HAWC power-law spectrum beyond its sensitive range into the GeV band predicts emission on the scale of the sensitivity of the \textit{Fermi}-LAT. A dedicated analysis will be needed to identify if this emission is a TeV Halo or PWN related to PSR J2030+3641. 

4HWC J2234+5904 is another fully unassociated source in the catalog. This source lies several degrees west of the Boomerang Nebula along the line of zero Galactic latitude. There are no nearby sources in any of the catalogs searched in this work. The source is a moderately sized extended source with nearly 68 TS significance in the catalog.\\
\vspace{-0.75cm}

\section{Conclusion}

The 4HWC survey represents the most sensitive catalog from the HAWC Collaboration in terms of both source modeling and the live time of the dataset. With a total of 85 sources, this is the most comprehensive catalog using HAWC gamma-ray data. The number of sources quoted as extended in the 4HWC is a major departure from the previous 3HWC, with 68 of the 85 sources extended at the $4\sigma$ level. Additionally, this catalog improves on the spectral modeling of the previous HAWC catalogs by testing the spectral curvature at the $4\sigma$ level. 4HWC has strong agreement with other gamma-ray catalogs included in TeVCat and presents 11 new gamma-ray sources not reported in TeVCat. Additionally, this catalog shows many sources near confirmed and candidate TeV Halos as well as X-ray binary systems. Sources reported without physical or gamma-ray counterparts represent exciting candidates for follow-up analysis by current and future astrophysical observatories. The overlapping FoV of the LHAASO and HAWC observatories provides a strong systematic cross-check of the two catalogs using wide-FoV observatories. Future experiments like CTAO \citep{CTA} and SWGO \citep{SWGO} represent exciting future opportunities to observe even greater regions of the gamma-ray sky as well as a larger range of energies with better spatial and energy resolution. From the two proposed iterations of SWGO, SWGO-A is expected to produce nearly as many sources in one year as 4HWC, while the full SWGO array expects 4--5 times as many sources in one year \citep{SWGO2025}.

\bibliography{sample631}{}

\begin{thebibliography}{}
\expandafter\ifx\csname natexlab\endcsname\relax\def\natexlab#1{#1}\fi
\providecommand{\url}[1]{\href{#1}{#1}}
\providecommand{\dodoi}[1]{doi:~\href{http://doi.org/#1}{\nolinkurl{#1}}}
\providecommand{\doeprint}[1]{\href{http://ascl.net/#1}{\nolinkurl{http://ascl.net/#1}}}
\providecommand{\doarXiv}[1]{\href{https://arxiv.org/abs/#1}{\nolinkurl{https://arxiv.org/abs/#1}}}

\bibitem[{Abe {et~al.}(2024)Abe, Abhir, Abhishek, Acero, Acharyya, Adam, Aguasca-Cabot, Agudo, Aguirre-Santaella, Alfaro, Alvarez-Crespo, Alves~Batista, Amans, Amato, Ambrosi, Ambrosino, Angüner, Aramo, Arcaro, Arrabito, Asano, Ascasíbar, Aschersleben, Augusto~Stuani, Backes, Balazs, Balbo, Ballet, Baquero~Larriva, Barbosa~Martins, Barres~de Almeida, Barrio, Batković, Batzofin, Baxter, Becerra~González, Beck, Beiske, Belmont, Benbow, Bernardini, Bernete, Bernlöhr, Berti, Bertucci, Beshley, Bhattacharjee, Bhattacharyya, Bi, Biederbeck, Biland, Bissaldi, Biteau, Blanch, Blazek, Bocchino, Boisson, Bolmont, Bonneau~Arbeletche, Bonnoli, Bonollo, Bordas, Bosnjak, Bottacini, Braiding, Bronzini, Brose, Brown, Brun, Brunelli, Bucciantini, Bulgarelli, Burelli, Burmistrov, Burton, Burtovoi, Bylund, Calisse, Campoy-Ordaz, Cantlay, Caproni, Capuzzo-Dolcetta, Caraveo, Caroff, Carosi, Carosi, Carquin, Carrasco, Cascone, Cassol, Castrejon, Castro-Tirado, Cerasole, Cerruti, Chadwick, Chambery, Chaty, Chen, Chernyakova,
  Chiavassa, Chytka, Cifuentes, Coimbra~Araujo, Conforti, Conte, Contreras, Cortina, Costa, Costantini, Cotter, Crestan, Cristofari, Cuevas, Curtis-Ginsberg, D'Aì, D'Amico, D'Ammando, Dadina, Dalchenko, David, Dazzi, de~Bony~de Lavergne, De~Caprio, De~Frondat~Laadim, de~Gouveia Dal~Pino, De~Lotto, De~Lucia, de~Martino, de~Menezes, de~Naurois, de~Ona~Wilhelmi, de~Souza, del Peral, Delgado~Giler, Delgado, Dell'aiera, Della~Valle, della Volpe, Depaoli, Di~Girolamo, Di~Piano, Di~Pierro, Di~Tria, Di~Venere, Díaz, Diebold, Dinesh, Djannati-Ataï, Djuvsland, Domínguez, Dominik, Donini, Dörner, Doro, dos Anjos, Dournaux, Duangchan, Dubos, Dubus, Duffy, Dumora, Dwarkadas, Ebr, Eckner, Egberts, Einecke, Elsässer, Emery, Errando, Escanuela, Escarate, Escobar~Godoy, Escudero, Esposito, Evoli, Falceta-Goncalves, Fattorini, Fegan, Feijen, Feng, Ferrand, Ferrarotto, Fiandrini, Fiasson, Filipovic, Fioretti, Fiori, Flores, Foffano, Font~Guiteras, Fontaine, Fröse, Fukazawa, Fukui, Funk, Furniss, Gaggero, Galanti, Galaz,
  Gallant, Gallozzi, Gammaldi, Garczarczyk, Gasbarra, Gasparrini, Gaug, Ghalumyan, Giarrusso, Giesbrecht, Giglietto, Giordano, Giuffrida, Giuliani, Glicenstein, Glombitza, Godinovic, Goldoni, González, Goulart~Coelho, Granot, Grasso, Grau, Gréaux, Green, Green, Greenshaw, Grenier, Grolleron, Grondin, Gueta, Gunji, Hackfeld, Hadasch, Hanlon, Hara, Harvey, Hassan, Hayashi, Heckmann, Heller, Hermann, Hernández~Cadena, Hervet, Hinton, Hiroshima, Hnatyk, Hnatyk, Hofmann, Holder, Holler, Horan, Horvath, Hovatta, Hrabovsky, Iarlori, Inada, Incardona, Inoue, Iocco, Iori, Jamrozy, Janecek, Jankowsky, Jarnot, Jean, Jiménez~Martínez, Jin, Juramy-Gilles, Jurysek, Kagaya, Kalekin, Kantzas, Karas, Katagiri, Kataoka, Kaufmann, Kazanas, Kerszberg, Khélifi, Kieda, Kissmann, Kleiner, Kluge, Kluźniak, Knödlseder, Kobayashi, Kohri, Komin, Kornecki, Kosack, Kostunin, Kowal, Kubo, Kushida, La~Barbera, La~Palombara, Láinez, Lamastra, Lapington, Laporte, Lazarević, Lazendic-Galloway, Lemoine-Goumard, Lenain, Leone, Leto,
  Leuschner, Lindfors, Linhoff, Liodakis, Lombardi, Longo, López-Coto, López-Moya, López-Oramas, Loporchio, Lozano~Bahilo, Lucarelli, Luque-Escamilla, Lyard, Macias, Mackey, Maier, Malyshev, Mandat, Manicò, Marcowith, Marinos, Mariotti, Markoff, Marquez, Marsella, Martí, Martin, Martínez, Martínez, Martinez, Marty, Mas-Aguilar, Mastropietro, Maurin, Mazin, McKeague, Mello, Menchiari, Mereghetti, Mestre, Meunier, Meyer, Miceli, Miceli, Michailidis, Michałowski, Miener, Miranda, Mitchell, Mizuno, Moderski, Mohrmann, Molero, Molfese, Molina, Montaruli, Moralejo, Morcuende, Morik, Morlino, Morselli, Moulin, Moya~Zamanillo, Mukherjee, Munari, Murach, Muraczewski, Muraishi, Nagataki, Nakamori, Nemmen, Nickel, Niemiec, Nieto, Nievas~Rosillo, Nikołajuk, Nikolić, Noda, Nosek, Novosyadlyj, Novotny, Nozaki, Ohishi, Ohtani, Okumura, Olive, Olmi, Ong, Orienti, Orito, Orlandini, Orlando, Orlando, Ostrowski, Oya, Pagano, Pagliaro, Palatiello, Panebianco, Paneque, Pantaleo, Paoletti, Paredes, Parmiggiani, Patel,
  Patricelli, Pavlović, Pech, Pecimotika, Peresano, Pérez-Romero, Pérez-Torres, Peron, Persic, Petrucci, Petruk, Piano, Pierre, Pietropaolo, Pihet, Pintore, Pittori, Plard, Podobnik, Pohl, Pons, Ponti, Prandini, Principe, Priyadarshi, Produit, Prokhorov, Pueschel, Pühlhofer, Pumo, Punch, Queiroz, Quirrenbach, Rando, Ravel, Razzaque, Regeard, Reichherzer, Reimer, Reimer, Remy, Renaud, Reposeur, Rhode, Ribeiro, Ribó, Richtler, Rico, Rieger, Rigoselli, Rizi, Roache, Rodriguez~Fernandez, Rodríguez-Vázquez, Romano, Romeo, Rosado, Rosales~de Leon, Rowell, Rudak, Ruiter, Rulten, Russo, Sadeh, Saha, Saito, Salzmann, Sánchez-Conde, Sangiorgi, Sano, Santander, Santangelo, Santos-Lima, Sapienza, Šarić, Sarkar, Saturni, Scherer, Schiavone, Schipani, Schleicher, Schovanek, Schubert, Schussler, Schwanke, Schwefer, Seglar~Arroyo, Seitenzahl, Sergijenko, Servillat, Sguera, Sharma, Siejkowski, Siqueira, Sizun, Sliusar, Slowikowska, Sol, Spencer, Spiga, Stamerra, Stanič, Starling, Stawarz, Steinmassl, Steppa,
  Stolarczyk, Suda, Suomijärvi, Tajima, Takeishi, Tanaka, Tavecchio, Tavernier, Terada, Terrier, Teshima, Tian, Tibaldo, Tibolla, Torradeflot, Torres, Tothill, Toussenel, Touzard, Travnicek, Tripodo, Trois, Tsiahina, Tutone, Umana, Vaclavek, Vacula, Vallania, van Eldik, Vassiliev, Vazquez~Acosta, Vecchi, Ventura, Vercellone, Verna, Viana, Viaux, Vigliano, Vignatti, Vigorito, Villanueva, Vink, Vitale, Vodeb, Voisin, Vorobiov, Voutsinas, Vovk, Vuillaume, Waegebaert, Wagner, Walter, Wechakama, White, Wierzcholska, Williams, Wohlleben, Yamazaki, Yang, Yoshida, Yoshikoshi, Zacharias, Zaharijas, Zampieri, Zanin, Zavrtanik, Zavrtanik, Zdziarski, Zech, Zhdanov, Ziętara, Živec, Zuriaga-Puig, De~la Torre~Luque, Guillemot, Smith, \& Consortium}]{OverlapCTA}
Abe, S., Abhir, J., Abhishek, A., {et~al.} 2024, Journal of Cosmology and Astroparticle Physics, 2024, 081, \dodoi{10.1088/1475-7516/2024/10/081}

\bibitem[{Abeysekara {et~al.}(2023)Abeysekara, Albert, Alfaro, Alvarez, Álvarez, Araya, Arteaga-Velázquez, Arunbabu, Rojas, Solares, Babu, Barber, Becerril, Belmont-Moreno, BenZvi, Blanco, Braun, Brisbois, Caballero-Mora, Martínez, Capistrán, Carramiñana, Casanova, Castillo, Chaparro-Amaro, Cotti, Cotzomi, {de León}, {de la Fuente}, {de León}, {De Young}, Hernandez, Dingus, DuVernois, Durocher, Díaz-Vélez, Ellsworth, Engel, Espinoza, Fan, Fang, Fick, Fleischhack, Flores, Fraija, García-González, Garcia-Torales, Garfias, Giacinti, Goksu, González, González-Muñoz, Goodman, Harding, Hernandez, Hernandez, Hinton, Hona, Huang, Hueyotl-Zahuantitla, Hui, Humensky, Hüntemeyer, Iriarte, Imran, Jardin-Blicq, Joshi, Kaufmann, Kieda, Kunde, Lara, Lauer, Lee, Lennarz, Vargas, Linnemann, Longinotti, Luis-Raya, Lundeen, Malone, Marandon, Marinelli, Martinez, Martínez-Castellanos, Martínez-Castro, Martínez-Huerta, Matthews, Miranda-Romagnoli, Montaruli, Morales-Soto, Moreno, Mostafá, Nayerhoda, Nellen,
  Newbold, Nisa, Noriega-Papaqui, Oceguera-Becerra, Olivera-Nieto, Omodei, Peisker, Araujo, Pérez-Pérez, Ponce, Pretz, Rho, Rosa-González, Ruiz-Velasco, Salazar, Salazar-Gallegos, Greus, Sandoval, Schneider, Schoorlemmer, Serna-Franco, Sinnis, Smith, Son, Woodle, Springer, Taboada, Tepe, Tibolla, Tollefson, Torres, Torres-Escobedo, Turner, Ureña-Mena, Ukwatta, Varela, Vargas-Magaña, Villaseñor, Wang, Watson, Werner, Westerhoff, Willox, Wisher, Wood, Yodh, Zaborov, Zepeda, \& Zhou}]{NimPaper}
Abeysekara, A., Albert, A., Alfaro, R., {et~al.} 2023, Nuclear Instruments and Methods in Physics Research Section A: Accelerators, Spectrometers, Detectors and Associated Equipment, 1052, 168253, \dodoi{https://doi.org/10.1016/j.nima.2023.168253}

\bibitem[{Abeysekara {et~al.}(2017)Abeysekara, Albert, Alfaro, Alvarez, Álvarez, Arceo, Arteaga-Velázquez, Solares, Barber, Baughman, Bautista-Elivar, Gonzalez, Becerril, Belmont-Moreno, BenZvi, Berley, Bernal, Braun, Brisbois, Caballero-Mora, Capistrán, Carramiñana, Casanova, Castillo, Cotti, Cotzomi, León, Fuente, León, Hernandez, Dingus, DuVernois, Díaz-Vélez, Ellsworth, Engel, Fiorino, Fraija, García-González, Garfias, Gerhardt, Muñoz, González, Goodman, Hampel-Arias, Harding, Hernandez, Hernandez-Almada, Hinton, Hui, Hüntemeyer, Iriarte, Jardin-Blicq, Joshi, Kaufmann, Kieda, Lara, Lauer, Lee, Lennarz, Vargas, Linnemann, Longinotti, Raya, Luna-García, López-Coto, Malone, Marinelli, Martinez, Martinez-Castellanos, Martínez-Castro, Martínez-Huerta, Matthews, Miranda-Romagnoli, Moreno, Mostafá, Nellen, Newbold, Nisa, Noriega-Papaqui, Pelayo, Pretz, Pérez-Pérez, Ren, Rho, Rivière, Rosa-González, Rosenberg, Ruiz-Velasco, Salazar, Greus, Sandoval, Schneider, Schoorlemmer, Sinnis, Smith,
  Springer, Surajbali, Taboada, Tibolla, Tollefson, Torres, Ukwatta, Vianello, Villaseñor, Weisgarber, Westerhoff, Wisher, Wood, Yapici, Younk, Zepeda, \& Zhou}]{2HWC}
Abeysekara, A.~U., Albert, A., Alfaro, R., {et~al.} 2017, The Astrophysical Journal, 843, 40, \dodoi{10.3847/1538-4357/aa7556}

\bibitem[{Abeysekara {et~al.}(2018)Abeysekara, Albert, Alfaro, Alvarez, {\'A}lvarez, Arceo, Arteaga-Vel{\'a}zquez, Avila~Rojas, Ayala~Solares, Belmont-Moreno, BenZvi, Brisbois, Caballero-Mora, Capistr{\'a}n, Carrami{\~{n}}ana, Casanova, Castillo, Cotti, Cotzomi, Couti{\~{n}}o~de Le{\'o}n, De~Le{\'o}n, De~la Fuente, D{\'i}az-V{\'e}lez, Dichiara, Dingus, DuVernois, Ellsworth, Engel, Espinoza, Fang, Fleischhack, Fraija, Galv{\'a}n-G{\'a}mez, Garc{\'i}a-Gonz{\'a}lez, Garfias, Gonz{\'a}lez-Mu{\~{n}}oz, Gonz{\'a}lez, Goodman, Hampel-Arias, Harding, Hernandez, Hinton, Hona, Hueyotl-Zahuantitla, Hui, H{\"u}ntemeyer, Iriarte, Jardin-Blicq, Joshi, Kaufmann, Kar, Kunde, Lauer, Lee, Le{\'o}n~Vargas, Li, Linnemann, Longinotti, Luis-Raya, L{\'o}pez-Coto, Malone, Marinelli, Martinez, Martinez-Castellanos, Mart{\'i}nez-Castro, Matthews, Miranda-Romagnoli, Moreno, Mostaf{\'a}, Nayerhoda, Nellen, Newbold, Nisa, Noriega-Papaqui, Pretz, P{\'e}rez-P{\'e}rez, Ren, Rho, Rivi{\`e}re, Rosa-Gonz{\'a}lez, Rosenberg, Ruiz-Velasco,
  Salesa~Greus, Sandoval, Schneider, Schoorlemmer, Seglar~Arroyo, Sinnis, Smith, Springer, Surajbali, Taboada, Tibolla, Tollefson, Torres, Vianello, Villase{\~{n}}or, Weisgarber, Werner, Westerhoff, Wood, Yapici, Yodh, Zepeda, Zhang, \& Zhou}]{SS433}
---. 2018, Nature, 562, 82, \dodoi{10.1038/s41586-018-0565-5}

\bibitem[{Abeysekara {et~al.}(2019)Abeysekara, Albert, Alfaro, Alvarez, Álvarez, Camacho, Arceo, Arteaga-Velázquez, Arunbabu, Rojas, Solares, Baghmanyan, Belmont-Moreno, BenZvi, Brisbois, Caballero-Mora, Capistrán, Carramiñana, Casanova, Cotti, Cotzomi, de~León, Fuente, León, Dichiara, Dingus, DuVernois, Díaz-Vélez, Ellsworth, Engel, Espinoza, Fick, Fleischhack, Fraija, Galván-Gámez, García-González, Garfias, González, Goodman, Harding, Hernandez, Hinton, Hona, Hueyotl-Zahuantitla, Hui, Hüntemeyer, Iriarte, Jardin-Blicq, Joshi, Kaufmann, Kieda, Lara, Lee, Vargas, Linnemann, Longinotti, Luis-Raya, Lundeen, Malone, Marinelli, Martinez, Martinez-Castellanos, Martínez-Castro, Martínez-Huerta, Matthews, Miranda-Romagnoli, Morales-Soto, Moreno, Mostafá, Nayerhoda, Nellen, Newbold, Nisa, Noriega-Papaqui, Peisker, Pérez-Pérez, Pretz, Ren, Rho, Rivière, Rosa-González, Rosenberg, Ruiz-Velasco, Salazar, Greus, Sandoval, Schneider, Schoorlemmer, Arroyo, Sinnis, Smith, Springer, Surajbali, Tabachnick,
  Tanner, Tibolla, Tollefson, Torres, Weisgarber, Westerhoff, Wood, Yapici, Zepeda, Zhou, \& Collaboration}]{HAWCCrab2019}
---. 2019, The Astrophysical Journal, 881, 134, \dodoi{10.3847/1538-4357/ab2f7d}

\bibitem[{Ackermann {et~al.}(2017)Ackermann, Ajello, Baldini, Ballet, Barbiellini, Bastieri, Bellazzini, Bissaldi, Bloom, Bonino, Bottacini, Brandt, Bregeon, Bruel, Buehler, Cameron, Caragiulo, Caraveo, Castro, Cavazzuti, Cecchi, Charles, Chekhtman, Cheung, Chiaro, Ciprini, Cohen, Costantin, Costanza, Cutini, D’Ammando, de~Palma, Desiante, Digel, Lalla, Mauro, Venere, Favuzzi, Fegan, Ferrara, Franckowiak, Fukazawa, Funk, Fusco, Gargano, Gasparrini, Giglietto, Giordano, Giroletti, Green, Grenier, Grondin, Guillemot, Guiriec, Harding, Hays, Hewitt, Horan, Hou, Jóhannesson, Kamae, Kuss, La~Mura, Larsson, Lemoine-Goumard, Li, Longo, Loparco, Lubrano, Magill, Maldera, Malyshev, Manfreda, Mazziotta, Michelson, Mitthumsiri, Mizuno, Monzani, Morselli, Moskalenko, Negro, Nuss, Ohsugi, Omodei, Orienti, Orlando, Ormes, Paliya, Paneque, Perkins, Persic, Pesce-Rollins, Petrosian, Piron, Porter, Principe, Rainò, Rando, Razzano, Razzaque, Reimer, Reimer, Reposeur, Sgrò, Simone, Siskind, Spada, Spandre, Spinelli, Suson,
  Tak, Thayer, Thompson, Torres, Tosti, Troja, Vianello, Wood, \& Wood}]{FermiExt}
Ackermann, M., Ajello, M., Baldini, L., {et~al.} 2017, The Astrophysical Journal, 843, 139, \dodoi{10.3847/1538-4357/aa775a}

\bibitem[{Ahnen {et~al.}(2017)Ahnen, Ansoldi, Antonelli, Arcaro, Babić, Banerjee, Bangale, {Barres de Almeida}, Barrio, {Becerra González}, Bednarek, Bernardini, Berti, Bhattacharyya, Biasuzzi, Biland, Blanch, Bonnefoy, Bonnoli, Carosi, Carosi, Chatterjee, Colin, Colombo, Contreras, Cortina, Covino, Cumani, {Da Vela}, Dazzi, {De Angelis}, {De Lotto}, {de Oña Wilhelmi}, {Di Pierro}, Doert, Domínguez, {Dominis Prester}, Dorner, Doro, Einecke, {Eisenacher Glawion}, Elsaesser, Engelkemeier, {Fallah Ramazani}, Fernández-Barral, Fidalgo, Fonseca, Font, Fruck, Galindo, {García López}, Garczarczyk, Gaug, Giammaria, Godinović, Gora, Griffiths, Guberman, Hadasch, Hahn, Hassan, Hayashida, Herrera, Hose, Hrupec, Hughes, Ishio, Konno, Kubo, Kushida, Kuveždić, Lelas, Lindfors, Lombardi, Longo, López, Maggio, Majumdar, Makariev, Maneva, Manganaro, Mannheim, Maraschi, Mariotti, Martínez, Mazin, Menzel, Minev, Mirzoyan, Moralejo, Moreno, Moretti, Neustroev, Niedzwiecki, {Nievas Rosillo}, Nilsson, Ninci, Nishijima,
  Noda, Nogués, Paiano, Palacio, Paneque, Paoletti, Paredes, Paredes-Fortuny, Pedaletti, Peresano, Perri, Persic, {Prada Moroni}, Prandini, Puljak, Garcia, Reichardt, Rhode, Ribó, Rico, Rugliancich, Saito, Satalecka, Schroeder, Schweizer, Sillanpää, Sitarek, Šnidarić, Sobczynska, Stamerra, Strzys, Surić, Takalo, Tavecchio, Temnikov, Terzić, Tescaro, Teshima, Torres, Torres-Albà, Treves, Vanzo, {Vazquez Acosta}, Vovk, Ward, Will, \& Zarić}]{Magic}
Ahnen, M., Ansoldi, S., Antonelli, L., {et~al.} 2017, Astroparticle Physics, 94, 29, \dodoi{https://doi.org/10.1016/j.astropartphys.2017.08.001}

\bibitem[{Albert {et~al.}(2019)Albert, Alfaro, Ashkar, Alvarez, Álvarez, Arteaga-Velázquez, Solares, Arceo, Bellido, BenZvi, Bretz, Brisbois, Brown, Brun, Caballero-Mora, Carosi, Carramiñana, Casanova, Chadwick, Cotter, León, Cristofari, Dasso, de~la Fuente, Dingus, Desiati, de~O.~Salles, de~Souza, Dorner, Díaz-Vélez, García-González, DuVernois, Sciascio, Engel, Fleischhack, Fraija, Funk, Glicenstein, Gonzalez, González, Goodman, Harding, Haungs, Hinton, Hona, Hoyos, Huentemeyer, Iriarte, Jardin-Blicq, Joshi, Kaufmann, Kawata, Kunwar, Lefaucheur, Lenain, Link, López-Coto, Marandon, Mariotti, Martínez-Castro, Martínez-Huerta, Mostafá, Nayerhoda, Nellen, de~Oña~Wilhelmi, Parsons, Patricelli, Pichel, Piel, Prandini, Pueschel, Procureur, Reisenegger, Rivière, Rodriguez, Rovero, Rowell, Ruiz-Velasco, Sandoval, Santander, Sako, Sako, Satalecka, Schoorlemmer, Schüssler, Seglar-Arroyo, Smith, Spencer, Surajbali, Tabachnick, Taylor, Tibolla, Torres, Vallage, Viana, Watson, Weisgarber, Werner, White,
  Wischnewski, Yang, Zepeda, \& Zhou}]{SWGO}
Albert, A., Alfaro, R., Ashkar, H., {et~al.} 2019, Science Case for a Wide Field-of-View Very-High-Energy Gamma-Ray Observatory in the Southern Hemisphere.
\newblock \doarXiv{1902.08429}

\bibitem[{Albert {et~al.}(2020)Albert, Alfaro, Alvarez, Camacho, Arteaga-Velázquez, Arunbabu, Rojas, Solares, Baghmanyan, Belmont-Moreno, BenZvi, Brisbois, Caballero-Mora, Capistrán, Carramiñana, Casanova, Cotti, de~León, Fuente, Hernandez, Diaz-Cruz, Dingus, DuVernois, Durocher, Díaz-Vélez, Ellsworth, Engel, Espinoza, Fan, Fang, Alonso, Fleischhack, Fraija, Galván-Gámez, Garcia, García-González, Garfias, Giacinti, González, Goodman, Harding, Hernandez, Hinton, Hona, Huang, Hueyotl-Zahuantitla, Hüntemeyer, Iriarte, Jardin-Blicq, Joshi, Kieda, Lara, Lee, Vargas, Linnemann, Longinotti, Luis-Raya, Lundeen, López-Coto, Malone, Marandon, Martinez, Martinez-Castellanos, Martínez-Castro, Matthews, Miranda-Romagnoli, Morales-Soto, Moreno, Mostafá, Nayerhoda, Nellen, Newbold, Nisa, Noriega-Papaqui, Olivera-Nieto, Omodei, Peisker, Araujo, Pérez-Pérez, Ren, Rho, Rivière, Rosa-González, Ruiz-Velasco, Salazar, Greus, Sandoval, Schneider, Schoorlemmer, Serna, Sinnis, Smith, Springer, Surajbali,
  Tollefson, Torres, Torres-Escobedo, Ukwatta, Ureña-Mena, Weisgarber, Werner, Willox, Zepeda, Zhou, León, Álvarez, \& Collaboration)}]{3HWC}
Albert, A., Alfaro, R., Alvarez, C., {et~al.} 2020, The Astrophysical Journal, 905, 76, \dodoi{10.3847/1538-4357/abc2d8}

\bibitem[{Albert {et~al.}(2023)Albert, Alfaro, Alvarez, Arteaga-Velázquez, Rojas, Solares, Babu, Belmont-Moreno, Brisbois, Caballero-Mora, Capistrán, Carramiñana, Casanova, Chaparro-Amaro, Cotti, Cotzomi, de~León, De~la Fuente, de~León, Hernandez, Díaz-Vélez, Dingus, DuVernois, Durocher, Engel, Espinoza, Fan, Alonso, Fraija, García-González, Garfias, Goksu, González, Goodman, Harding, Hernandez, Hinton, Hona, Huang, Hueyotl-Zahuantitla, Hüntemeyer, Iriarte, Jardin-Blicq, Joshi, Kaufmann, Kieda, Lee, Vargas, Linnemann, Longinotti, Luis-Raya, López-Coto, Malone, Marandon, Martinez, Martínez-Castro, Matthews, Miranda-Romagnoli, Morales-Soto, Moreno, Mostafá, Nayerhoda, Nellen, Newbold, Nisa, Noriega-Papaqui, Olivera-Nieto, Omodei, Peisker, Araujo, Pérez-Pérez, Rho, Rosa-González, Ruiz-Velasco, Salazar, Salazar-Gallegos, Greus, Sandoval, Schneider, Serna-Franco, Smith, Son, Springer, Tibolla, Tollefson, Torres, Torres-Escobedo, Turner, Ureña-Mena, Villaseñor, Wang, Werner, Willox, Zhou, \&
  collaboration)}]{J1932}
---. 2023, The Astrophysical Journal, 942, 96, \dodoi{10.3847/1538-4357/ac8de3}

\bibitem[{Albert {et~al.}(2024{\natexlab{a}})Albert, Alfaro, Alvarez, Andrés, Arteaga-Velázquez, Avila~Rojas, Ayala~Solares, Babu, Belmont-Moreno, Bernal, Caballero-Mora, Capistrán, Carramiñana, Carreón, Casanova, Cotti, Cotzomi, Coutiño~de León, De~la Fuente, de~León, Depaoli, Di~Lalla, Díaz~Hernández, Dingus, DuVernois, Engel, Ergin, Espinoza, Fan, Fang, Fraija, Fraija, García-González, Garfias, Goksu, González, Goodman, Groetsch, Harding, Hernández-Cadena, Herzog, Hinton, Huang, Hueyotl-Zahuantitla, Hüntemeyer, Iriarte, Kaufmann, Lara, Lee, León~Vargas, Linnemann, Longinotti, Luis-Raya, Malone, Martínez-Castro, Matthews, Miranda-Romagnoli, Montes, Moreno, Mostafá, Nellen, Nisa, Noriega-Papaqui, Olivera-Nieto, Omodei, Osorio-Archila, Pérez~Araujo, Pérez-Pérez, Rho, Rosa-González, Ruiz-Velasco, Salazar, Salazar-Gallegos, Sandoval, Schneider, Schwefer, Serna-Franco, Smith, Son, Springer, Tibolla, Tollefson, Torres, Torres-Escobedo, Turner, Ureña-Mena, Varela, Wang, Watson, Whitaker,
  Willox, Wu, Yu, Yun-Cárcamo, Zhou, \& Collaboration}]{Pass5Perf}
---. 2024{\natexlab{a}}, The Astrophysical Journal, 972, 144, \dodoi{10.3847/1538-4357/ad5f2d}

\bibitem[{Albert {et~al.}(2024{\natexlab{b}})Albert, Alfaro, Alvarez, Arteaga-Velázquez, Avila~Rojas, Ayala~Solares, Babu, Belmont-Moreno, Bernal, Caballero-Mora, Capistrán, Carramiñana, Casanova, Cotti, Cotzomi, Coutiño~de León, de~la Fuente, Depaoli, Di~Lalla, Diaz~Hernandez, Dingus, DuVernois, Durocher, Díaz-Vélez, Engel, Espinoza, Fan, Fang, Fraija, García-González, Garfias, Goksu, González, Goodman, Groetsch, Harding, Hernández-Cadena, Herzog, Hüntemeyer, Huang, Hueyotl-Zahuantitla, Iriarte, Joshi, Kaufmann, Kieda, Lara, Lee, Lee, León~Vargas, Linnemann, Longinotti, Luis-Raya, Malone, Martinez, Martínez-Castro, Matthews, Miranda-Romagnoli, Montes, Morales-Soto, Moreno, Mostafá, Nayerhoda, Nellen, Noriega-Papaqui, Olivera-Nieto, Omodei, Pérez~Araujo, Pérez-Pérez, Rho, Rosa-González, Salazar, Salazar-Gallegos, Sandoval, Schneider, Schwefer, Serna-Franco, Son, Springer, Tibolla, Tollefson, Torres, Torres-Escobedo, Turner, Urea-Mena, Varela, Villaseñor, Wang, Watson, Willox, Wu,
  Yun-Cárcamo, Zhou, de~León, (for~theHAWC Collaboration), \& Di~Mauro}]{GemingaDiffusion}
---. 2024{\natexlab{b}}, The Astrophysical Journal, 974, 246, \dodoi{10.3847/1538-4357/ad738e}

\bibitem[{Alfaro {et~al.}(2024)Alfaro, Alvarez, Arteaga-Vel{\'a}zquez, Avila~Rojas, Ayala~Solares, Babu, Belmont-Moreno, Caballero-Mora, Capistr{\'a}n, Carrami{\~{n}}ana, Casanova, Cotti, Cotzomi, Couti{\~{n}}o~de Le{\'o}n, De~la Fuente, Depaoli, Di~Lalla, Diaz~Hernandez, Dingus, DuVernois, Durocher, D{\'i}az-V{\'e}lez, Engel, Espinoza, Fan, Fang, Fraija, Fraija, Garc{\'i}a-Gonz{\'a}lez, Garfias, Gonzalez~Mu{\~{n}}oz, Gonz{\'a}lez, Goodman, Groetsch, Harding, Herzog, Hinton, Huang, Hueyotl-Zahuantitla, H{\"u}ntemeyer, Iriarte, Joshi, Kaufmann, Kieda, de~Le{\'o}n, Lee, Le{\'o}n~Vargas, Linnemann, Longinotti, Luis-Raya, Malone, Martinez, Mart{\'i}nez-Castro, Matthews, Miranda-Romagnoli, Morales-Soto, Moreno, Mostaf{\'a}, Nayerhoda, Nellen, Newbold, Nisa, Noriega-Papaqui, Olivera-Nieto, Omodei, Osorio, P{\'e}rez~Araujo, P{\'e}rez-P{\'e}rez, Rho, Rosa-Gonz{\'a}lez, Ruiz-Velasco, Salazar, Salazar-Gallegos, Sandoval, Schneider, Serna-Franco, Smith, Son, Springer, Tibolla, Tollefson, Torres, Torres-Escobedo, Turner,
  Ure{\~{n}}a-Mena, Varela, Villase{\~{n}}or, Wang, Watson, Willox, Yun-C{\'a}rcamo, \& Zhou}]{V4641}
Alfaro, R., Alvarez, C., Arteaga-Vel{\'a}zquez, J.~C., {et~al.} 2024, Nature, 634, 557, \dodoi{10.1038/s41586-024-07995-9}

\bibitem[{Amenomori {et~al.}(2019)Amenomori, Bao, Bi, Chen, Chen, Chen, Chen, Chen, Cirennima, Cui, Danzengluobu, Ding, Fang, Fang, Feng, Feng, Feng, Gao, Gou, Guo, He, He, Hibino, Hotta, Hu, Hu, Huang, Jia, Jiang, Jin, Kajino, Kasahara, Katayose, Kato, Kato, Kawata, Kozai, Labaciren, Le, Li, Li, Li, Lin, Liu, Liu, Liu, Liu, Lou, Lu, Meng, Mitsui, Munakata, Nakamura, Nanjo, Nishizawa, Ohnishi, Ohta, Ozawa, Qian, Qu, Saito, Sakata, Sako, Sengoku, Shao, Shibata, Shiomi, Sugimoto, Takita, Tan, Tateyama, Torii, Tsuchiya, Udo, Wang, Wu, Xue, Yagisawa, Yamamoto, Yang, Yuan, Zhai, Zhang, Zhang, Zhang, Zhang, Zhang, Zhang, Zhang, Zhaxisangzhu, \& Zhou}]{Tibet}
Amenomori, M., Bao, Y.~W., Bi, X.~J., {et~al.} 2019, Phys. Rev. Lett., 123, 051101, \dodoi{10.1103/PhysRevLett.123.051101}

\bibitem[{Atkins {et~al.}(2003)Atkins, Benbow, Berley, Blaufuss, Bussons, Coyne, Delay, DeYoung, Dingus, Dorfan, Ellsworth, Falcone, Fleysher, Fleysher, Gisler, Gonzalez, Goodman, Haines, Hays, Hoffman, Kelley, Laird, McCullough, McEnery, Miller, Mincer, Morales, Nemethy, Noyes, Ryan, Samuelson, Schneider, Shen, Shoup, Sinnis, Smith, Sullivan, Tumer, Wang, Wascko, Williams, Westerhoff, Wilson, Xu, \& Yodh}]{MilagroDirectIntegration}
Atkins, R., Benbow, W., Berley, D., {et~al.} 2003, The Astrophysical Journal, 595, 803, \dodoi{10.1086/377498}

\bibitem[{{Avakyan, A.} {et~al.}(2023){Avakyan, A.}, {Neumann, M.}, {Zainab, A.}, {Doroshenko, V.}, {Wilms, J.}, \& {Santangelo, A.}}]{LMXB}
{Avakyan, A.}, {Neumann, M.}, {Zainab, A.}, {et~al.} 2023, A\&A, 675, A199, \dodoi{10.1051/0004-6361/202346522}

\bibitem[{{Ballet} {et~al.}(2023){Ballet}, {Bruel}, {Burnett}, {Lott}, \& {The Fermi-LAT collaboration}}]{4FGL}
{Ballet}, J., {Bruel}, P., {Burnett}, T.~H., {Lott}, B., \& {The Fermi-LAT collaboration}. 2023, arXiv e-prints, arXiv:2307.12546, \dodoi{10.48550/arXiv.2307.12546}

\bibitem[{Balzer {et~al.}(2014)Balzer, Füßling, Gajdus, Göring, Lopatin, {de Naurois}, Schlenker, Schwanke, \& Stegmann}]{HESS}
Balzer, A., Füßling, M., Gajdus, M., {et~al.} 2014, Astroparticle Physics, 54, 67, \dodoi{https://doi.org/10.1016/j.astropartphys.2013.11.007}

\bibitem[{Bartoli {et~al.}(2013)Bartoli, Bernardini, \& Collaboration)}]{ARGO}
Bartoli, B., Bernardini, P., \& Collaboration), T. A.-Y. 2013, The Astrophysical Journal, 779, 27, \dodoi{10.1088/0004-637X/779/1/27}

\bibitem[{Cao {et~al.}(2024)Cao, Aharonian, An, Axikegu, Bai, Bao, Bastieri, Bi, Bi, Cai, Cao, Cao, Cao, Chang, Chang, Chen, Chen, Chen, Chen, Chen, Chen, Chen, Chen, Chen, Chen, Chen, Chen, Cheng, Cheng, Cui, Cui, Cui, Cui, Dai, Dai, Dai, Danzengluobu, della Volpe, Dong, Duan, Fan, Fan, Fang, Fang, Feng, Feng, Feng, Feng, Feng, Gabici, Gao, Gao, Gao, Gao, Gao, Gao, Ge, Geng, Giacinti, Gong, Gou, Gu, Guo, Guo, Guo, Guo, Han, He, He, He, He, He, Heller, Hor, Hou, Hou, Hou, Hu, Hu, Hu, Huang, Huang, Huang, Huang, Huang, Huang, Huang, Ji, Jia, Jia, Jiang, Jiang, Jiang, Jin, Kang, Ke, Kuleshov, Kurinov, Li, Li, Li, Li, Li, Li, Li, Li, Li, Li, Li, Li, Li, Li, Li, Li, Li, Li, Li, Liang, Liang, Lin, Liu, Liu, Liu, Liu, Liu, Liu, Liu, Liu, Liu, Liu, Liu, Liu, Liu, Liu, Lu, Luo, Lv, Ma, Ma, Ma, Mao, Min, Mitthumsiri, Mu, Nan, Neronov, Ou, Pang, Pattarakijwanich, Pei, Qi, Qi, Qiao, Qin, Ruffolo, Sáiz, Semikoz, Shao, Shao, Shchegolev, Sheng, Shu, Song, Stenkin, Stepanov, Su, Sun, Sun, Sun, Tam, Tang, Tang, Tian, Wang,
  Wang, Wang, Wang, Wang, Wang, Wang, Wang, Wang, Wang, Wang, Wang, Wang, Wang, Wang, Wang, Wang, Wang, Wang, Wang, Wang, Wei, Wei, Wei, Wen, Wu, Wu, Wu, Wu, Wu, Xi, Xia, Xia, Xiang, Xiao, Xiao, Xin, Xin, Xing, Xiong, Xu, Xu, Xu, Xu, Xue, Yan, Yan, Yan, Yang, Yang, Yang, Yang, Yang, Yang, Yang, Yang, Yang, Yao, Yao, Ye, Yin, Yin, You, You, Yu, Yuan, Yue, Zeng, Zeng, Zeng, Zha, Zhang, Zhang, Zhang, Zhang, Zhang, Zhang, Zhang, Zhang, Zhang, Zhang, Zhang, Zhang, Zhang, Zhang, Zhang, Zhang, Zhang, Zhang, Zhao, Zhao, Zhao, Zhao, Zhao, Zheng, Zhou, Zhou, Zhou, Zhou, Zhou, Zhou, Zhou, Zhu, Zhu, Zhu, Zhu, Zuo, \& Collaboration)}]{1LHAASO}
Cao, Z., Aharonian, F., An, Q., {et~al.} 2024, The Astrophysical Journal Supplement Series, 271, 25, \dodoi{10.3847/1538-4365/acfd29}

\bibitem[{Consortium(2019)}]{CTA}
Consortium, T.~C. 2019, Science with the Cherenkov Telescope Array (WORLD SCIENTIFIC), \dodoi{10.1142/10986}

\bibitem[{Dame \& Thaddeus(2003)}]{dame2003largeextensioncfagalactic}
Dame, T.~M., \& Thaddeus, P. 2003, A Large Extension of the CfA Galactic CO Survey.
\newblock \doarXiv{astro-ph/0310102}

\bibitem[{{Dundovic, A.} {et~al.}(2021){Dundovic, A.}, {Evoli, C.}, {Gaggero, D.}, \& {Grasso, D.}}]{HERMES}
{Dundovic, A.}, {Evoli, C.}, {Gaggero, D.}, \& {Grasso, D.} 2021, A\&A, 653, A18, \dodoi{10.1051/0004-6361/202140801}

\bibitem[{Evoli {et~al.}(2017)Evoli, Gaggero, Vittino, Bernardo, Mauro, Ligorini, Ullio, \& Grasso}]{DRAGON1}
Evoli, C., Gaggero, D., Vittino, A., {et~al.} 2017, Journal of Cosmology and Astroparticle Physics, 2017, 015, \dodoi{10.1088/1475-7516/2017/02/015}

\bibitem[{Evoli {et~al.}(2018)Evoli, Gaggero, Vittino, Di~Mauro, Grasso, \& Mazziotta}]{DRAGON2}
---. 2018, Journal of Cosmology and Astroparticle Physics, 2018, 006, \dodoi{10.1088/1475-7516/2018/07/006}

\bibitem[{Ferrand \& Safi-Harb(2012)}]{SNRcat}
Ferrand, G., \& Safi-Harb, S. 2012, Advances in Space Research, 49, 1313, \dodoi{https://doi.org/10.1016/j.asr.2012.02.004}

\bibitem[{{Giacinti, G.} {et~al.}(2020){Giacinti, G.}, {Mitchell, A. M. W.}, {López-Coto, R.}, {Joshi, V.}, {Parsons, R. D.}, \& {Hinton, J. A.}}]{TeVHaloAge}
{Giacinti, G.}, {Mitchell, A. M. W.}, {López-Coto, R.}, {et~al.} 2020, A\&A, 636, A113, \dodoi{10.1051/0004-6361/201936505}

\bibitem[{{G{\'o}rski} {et~al.}(2005){G{\'o}rski}, {Hivon}, {Banday}, {Wandelt}, {Hansen}, {Reinecke}, \& {Bartelmann}}]{HealPix}
{G{\'o}rski}, K.~M., {Hivon}, E., {Banday}, A.~J., {et~al.} 2005, \apj, 622, 759, \dodoi{10.1086/427976}

\bibitem[{{H.E.S.S. Collaboration} {et~al.}(2018){H.E.S.S. Collaboration}, {Abdalla, H.}, {Abramowski, A.}, {Aharonian, F.}, {Ait Benkhali, F.}, {Angüner, E. O.}, {Arakawa, M.}, {Arrieta, M.}, {Aubert, P.}, {Backes, M.}, {Balzer, A.}, {Barnard, M.}, {Becherini, Y.}, {Becker Tjus, J.}, {Berge, D.}, {Bernhard, S.}, {Bernlöhr, K.}, {Blackwell, R.}, {Böttcher, M.}, {Boisson, C.}, {Bolmont, J.}, {Bonnefoy, S.}, {Bordas, P.}, {Bregeon, J.}, {Brun, F.}, {Brun, P.}, {Bryan, M.}, {Büchele, M.}, {Bulik, T.}, {Capasso, M.}, {Carrigan, S.}, {Caroff, S.}, {Carosi, A.}, {Casanova, S.}, {Cerruti, M.}, {Chakraborty, N.}, {Chaves, R. C. G.}, {Chen, A.}, {Chevalier, J.}, {Colafrancesco, S.}, {Condon, B.}, {Conrad, J.}, {Davids, I. D.}, {Decock, J.}, {Deil, C.}, {Devin, J.}, {deWilt, P.}, {Dirson, L.}, {Djannati-Ataï, A.}, {Domainko, W.}, {Donath, A.}, {Drury, L. O’C.}, {Dutson, K.}, {Dyks, J.}, {Edwards, T.}, {Egberts, K.}, {Eger, P.}, {Emery, G.}, {Ernenwein, J.-P.}, {Eschbach, S.}, {Farnier, C.}, {Fegan, S.},
  {Fernandes, M. V.}, {Fiasson, A.}, {Fontaine, G.}, {Förster, A.}, {Funk, S.}, {Füßling, M.}, {Gabici, S.}, {Gallant, Y. A.}, {Garrigoux, T.}, {Gast, H.}, {Gaté, F.}, {Giavitto, G.}, {Giebels, B.}, {Glawion, D.}, {Glicenstein, J. F.}, {Gottschall, D.}, {Grondin, M.-H.}, {Hahn, J.}, {Haupt, M.}, {Hawkes, J.}, {Heinzelmann, G.}, {Henri, G.}, {Hermann, G.}, {Hinton, J. A.}, {Hofmann, W.}, {Hoischen, C.}, {Holch, T. L.}, {Holler, M.}, {Horns, D.}, {Ivascenko, A.}, {Iwasaki, H.}, {Jacholkowska, A.}, {Jamrozy, M.}, {Jankowsky, D.}, {Jankowsky, F.}, {Jingo, M.}, {Jouvin, L.}, {Jung-Richardt, I.}, {Kastendieck, M. A.}, {Katarzyński, K.}, {Katsuragawa, M.}, {Katz, U.}, {Kerszberg, D.}, {Khangulyan, D.}, {Khélifi, B.}, {King, J.}, {Klepser, S.}, {Klochkov, D.}, {Kluźniak, W.}, {Komin, Nu.}, {Kosack, K.}, {Krakau, S.}, {Kraus, M.}, {Krüger, P. P.}, {Laffon, H.}, {Lamanna, G.}, {Lau, J.}, {Lees, J.-P.}, {Lefaucheur, J.}, {Lemière, A.}, {Lemoine-Goumard, M.}, {Lenain, J.-P.}, {Leser, E.}, {Lohse, T.}, {Lorentz,
  M.}, {Liu, R.}, {López-Coto, R.}, {Lypova, I.}, {Marandon, V.}, {Malyshev, D.}, {Marcowith, A.}, {Mariaud, C.}, {Marx, R.}, {Maurin, G.}, {Maxted, N.}, {Mayer, M.}, {Meintjes, P.J.}, {Meyer, M.}, {Mitchell, A. M. W.}, {Moderski, R.}, {Mohamed, M.}, {Mohrmann, L.}, {Morå, K.}, {Moulin, E.}, {Murach, T.}, {Nakashima, S.}, {de Naurois, M.}, {Ndiyavala, H.}, {Niederwanger, F.}, {Niemiec, J.}, {Oakes, L.}, {O’Brien, P.}, {Odaka, H.}, {Ohm, S.}, {Ostrowski, M.}, {Oya, I.}, {Padovani, M.}, {Panter, M.}, {Parsons, R. D.}, {Paz Arribas, M.}, {Pekeur, N. W.}, {Pelletier, G.}, {Perennes, C.}, {Petrucci, P.-O.}, {Peyaud, B.}, {Piel, Q.}, {Pita, S.}, {Poireau, V.}, {Poon, H.}, {Prokhorov, D.}, {Prokoph, H.}, {Pühlhofer, G.}, {Punch, M.}, {Quirrenbach, A.}, {Raab, S.}, {Rauth, R.}, {Reimer, A.}, {Reimer, O.}, {Renaud, M.}, {de los Reyes, R.}, {Rieger, F.}, {Rinchiuso, L.}, {Romoli, C.}, {Rowell, G.}, {Rudak, B.}, {Rulten, C. B.}, {Safi-Harb, S.}, {Sahakian, V.}, {Saito, S.}, {Sanchez, D. A.}, {Santangelo, A.},
  {Sasaki, M.}, {Schandri, M.}, {Schlickeiser, R.}, {Schüssler, F.}, {Schulz, A.}, {Schwanke, U.}, {Schwemmer, S.}, {Seglar-Arroyo, M.}, {Settimo, M.}, {Seyffert, A. S.}, {Shafi, N.}, {Shilon, I.}, {Shiningayamwe, K.}, {Simoni, R.}, {Sol, H.}, {Spanier, F.}, {Spir-Jacob, M.}, {Stawarz, Ł.}, {Steenkamp, R.}, {Stegmann, C.}, {Steppa, C.}, {Sushch, I.}, {Takahashi, T.}, {Tavernet, J.-P.}, {Tavernier, T.}, {Taylor, A. M.}, {Terrier, R.}, {Tibaldo, L.}, {Tiziani, D.}, {Tluczykont, M.}, {Trichard, C.}, {Tsirou, M.}, {Tsuji, N.}, {Tuffs, R.}, {Uchiyama, Y.}, {van der Walt, D. J.}, {van Eldik, C.}, {van Rensburg, C.}, {van Soelen, B.}, {Vasileiadis, G.}, {Veh, J.}, {Venter, C.}, {Viana, A.}, {Vincent, P.}, {Vink, J.}, {Voisin, F.}, {Völk, H. J.}, {Vuillaume, T.}, {Wadiasingh, Z.}, {Wagner, S. J.}, {Wagner, P.}, {Wagner, R. M.}, {White, R.}, {Wierzcholska, A.}, {Willmann, P.}, {Wörnlein, A.}, {Wouters, D.}, {Yang, R.}, {Zaborov, D.}, {Zacharias, M.}, {Zanin, R.}, {Zdziarski, A. A.}, {Zech, A.}, {Zefi, F.},
  {Ziegler, A.}, {Zorn, J.}, \& {Żywucka, N.}}]{HGPS}
{H.E.S.S. Collaboration}, {Abdalla, H.}, {Abramowski, A.}, {et~al.} 2018, A\&A, 612, A1, \dodoi{10.1051/0004-6361/201732098}

\bibitem[{{HI4PI Collaboration:} {et~al.}(2016){HI4PI Collaboration:}, {Ben Bekhti, N.}, {Flöer, L.}, {Keller, R.}, {Kerp, J.}, {Lenz, D.}, {Winkel, B.}, {Bailin, J.}, {Calabretta, M. R.}, {Dedes, L.}, {Ford, H. A.}, {Gibson, B. K.}, {Haud, U.}, {Janowiecki, S.}, {Kalberla, P. M. W.}, {Lockman, F. J.}, {McClure-Griffiths, N. M.}, {Murphy, T.}, {Nakanishi, H.}, {Pisano, D. J.}, \& {Staveley-Smith, L.}}]{HI4PI}
{HI4PI Collaboration:}, {Ben Bekhti, N.}, {Flöer, L.}, {et~al.} 2016, A\&A, 594, A116, \dodoi{10.1051/0004-6361/201629178}

\bibitem[{Holder {et~al.}(2008)Holder, Acciari, Aliu, Arlen, Beilicke, Benbow, Bradbury, Buckley, Bugaev, Butt, Byrum, Cannon, Celik, Cesarini, Ciupik, Chow, Cogan, Colin, Cui, Daniel, Ergin, Falcone, Fegan, Finley, Finnegan, Fortin, Fortson, Furniss, Gillanders, Grube, Guenette, Gyuk, Hanna, Hays, Horan, Hui, Humensky, Imran, Kaaret, Karlsson, Kertzman, Kieda, Kildea, Konopelko, Krawczynski, Krennrich, Lang, LeBohec, Maier, McCann, McCutcheon, Moriarty, Mukherjee, Nagai, Niemiec, Ong, Pandel, Perkins, Pohl, Quinn, Ragan, Reyes, Reynolds, Rose, Schroedter, Sembroski, Smith, Steele, Swordy, Toner, Valcarcel, Vassiliev, Wagner, Wakely, Ward, Weekes, Weinstein, White, Williams, Wissel, Wood, \& Zitzer}]{Veritas}
Holder, J., Acciari, V.~A., Aliu, E., {et~al.} 2008, AIP Conference Proceedings, 1085, 657, \dodoi{10.1063/1.3076760}

\bibitem[{Manchester {et~al.}(2005)Manchester, Hobbs, Teoh, \& Hobbs}]{ATNF}
Manchester, R.~N., Hobbs, G.~B., Teoh, A., \& Hobbs, M. 2005, The Astronomical Journal, 129, 1993, \dodoi{10.1086/428488}

\bibitem[{Mattox {et~al.}(1997)Mattox, Schachter, Molnar, Hartman, \& Patnaik}]{Mattox_1997}
Mattox, J.~R., Schachter, J., Molnar, L., Hartman, R.~C., \& Patnaik, A.~R. 1997, The Astrophysical Journal, 481, 95, \dodoi{10.1086/304039}

\bibitem[{{Neumann, M.} {et~al.}(2023){Neumann, M.}, {Avakyan, A.}, {Doroshenko, V.}, \& {Santangelo, A.}}]{HMXB}
{Neumann, M.}, {Avakyan, A.}, {Doroshenko, V.}, \& {Santangelo, A.} 2023, A\&A, 677, A134, \dodoi{10.1051/0004-6361/202245728}

\bibitem[{Remy {et~al.}(2020)Remy, Gallant, \& Renaud}]{OverlapHESS}
Remy, Q., Gallant, Y., \& Renaud, M. 2020, Astroparticle Physics, 122, 102462, \dodoi{https://doi.org/10.1016/j.astropartphys.2020.102462}

\bibitem[{Smith {et~al.}(2023)Smith, Abdollahi, Ajello, Bailes, Baldini, Ballet, Baring, Bassa, Gonzalez, Bellazzini, Berretta, Bhattacharyya, Bissaldi, Bonino, Bottacini, Bregeon, Bruel, Burgay, Burnett, Cameron, Camilo, Caputo, Caraveo, Cavazzuti, Chiaro, Ciprini, Clark, Cognard, Corongiu, Orestano, Crnogorcevic, Cuoco, Cutini, D’Ammando, de~Angelis, DeCesar, De~Gaetano, de~Menezes, Deneva, de~Palma, Di~Lalla, Dirirsa, Di~Venere, Domínguez, Dumora, Fegan, Ferrara, Fiori, Fleischhack, Flynn, Franckowiak, Freire, Fukazawa, Fusco, Galanti, Gammaldi, Gargano, Gasparrini, Giacchino, Giglietto, Giordano, Giroletti, Green, Grenier, Guillemot, Guiriec, Gustafsson, Harding, Hays, Hewitt, Horan, Hou, Jankowski, Johnson, Johnson, Johnston, Kataoka, Keith, Kerr, Kramer, Kuss, Latronico, Lee, Li, Li, Limyansky, Longo, Loparco, Lorusso, Lovellette, Lower, Lubrano, Lyne, Maan, Maldera, Manchester, Manfreda, Marelli, Martí-Devesa, Mazziotta, McEnery, Mereu, Michelson, Mickaliger, Mitthumsiri, Mizuno, Moiseev, Monzani,
  Morselli, Negro, Nemmen, Nieder, Nuss, Omodei, Orienti, Orlando, Ormes, Palatiello, Paneque, Panzarini, Parthasarathy, Persic, Pesce-Rollins, Pillera, Poon, Porter, Possenti, Principe, Rainò, Rando, Ransom, Ray, Razzano, Razzaque, Reimer, Reimer, Renault-Tinacci, Romani, Sánchez-Conde, Parkinson, Scotton, Serini, Sgrò, Shannon, Sharma, Shen, Siskind, Spandre, Spinelli, Stappers, Stephens, Suson, Tabassum, Tajima, Tak, Theureau, Thompson, Tibolla, Torres, Valverde, Venter, Wadiasingh, Wang, Wang, Wang, Weltevrede, Wood, Yan, Zaharijas, Zhang, \& Zhu}]{FermiPulsar}
Smith, D.~A., Abdollahi, S., Ajello, M., {et~al.} 2023, The Astrophysical Journal, 958, 191, \dodoi{10.3847/1538-4357/acee67}

\bibitem[{SWGO {et~al.}(2025)SWGO, Abreu, Alfaro, Alfonso, Andrade, Angüner, Anita-Rangel, Aquines-Gutiérrez, Arcaro, Arceo, Arteaga-Velázquez, Assis, Solares, Bakalova, Bandeira, Bangale, de~Almeida, Batista, Batković, Bazo, Belmont, Bennemann, BenZvi, Bernal, Bian, Bigongiari, Bottacini, Branada, Brogueira, Brown, Bulik, Caballero-Mora, Camarri, Cao, Cao, Cao, Capistrán, Cardillo, Casentini, Castromonte, Chadwick, Chanamé, Chang, Chen, Chiavassa, Chytka, Colalillo, Conceição, Consolati, Cordero, Costa, Covarelli, Cui, Cui, Angelis, de~Gouveia Dal~Pino, de~Menezes, Desiati, Lalla, Pierro, Sciascio, Vélez, Dib, Dingus, Djuvsland, Dobrigkeit, Mendes, Dorigo, Doro, dos Reis, Vernois, Elsaesser, Engel, Ergin, Errando, Fang, Fazzi, Feng, Feroci, Ferreira, Fraija, Fraija, Franceschini, Franco, Funk, Galleguillos, Gao, Gao, Reyes, Garcia, Garfias, Giacinti, Gibilisco, Giovanni, Glombitza, Goksu, Gong, González, González, Goodman, Grieco, Gu, Guarino, Guedes, Gyeong, Haist, Han, Hansen, Harding, Cadena,
  Herzog, Hinton, Hofmann, Hou, C., Hu, Huang, Huentemeyer, Iriarte, Isaković, Jardin-Blicq, Junoy, Juryšek, Kaci, Khelifi, Kieda, Monaca, Mura, Lang, Lapington, Laspiur, Lavitola, Lee, Leitl, Lemoine-Goumard, Lessio, Lewis, Li, Li, Li, Li, Liberti, Lin, Lineros, Liu, Liu, Liu, Longo, Luo, Lv, Macerata, Magugliani, Malone, Mancilla, Mandat, Manganaro, Mariani, Mariazzi, Mariotti, Marrodan, Martínez-Huerta, Martins, Medina, Melo, Mendes, Meza, Micali, Miceli, Miozzi, Miranda, Enriquez, Mitchell, Molinario, Montero, Morales-Olivares, Morselli, Mossini, Mostafá, Muleri, Nardi, Nayak, Negro, Nellen, Nisa, Novotny, Olivera-Nieto, Omodei, Orlando, Ortolani, Osorio-Archila, Ota, Otiniano, Ou, Paoloni, Pepe, Perca, Peresano, Araujo, Petrosillo-Lago, Piano, Piccolo, Pichel, Pimenta, Pirke, Porcelli, Porter, Prandini, Pratts, Pretsch, Qi, Qin, Rainò, Recabarren, Regeard, Reisenegger, Remy, Ren, Rescic, Reville, Reyes, Rho, Riquelme, Rivadeneira, Fernandez, Rossi, Rovero, Ruina, Ruiz-Velasco, Salazar, Salotto,
  Sanchez, Sandoval, Sansone, Santander, Santonico, Santos, Santos-Guevara, Sartirana, Schneider, Schneider, Schoorlemmer, Schröder, Schussler, Schutte, Serna-Franco, Shoaib, Smith, Smith, Son, Soto, Spencer, Springer, Stewart, Stuani, Sun, Tambone, Tang, Tang, Tapia, Tavani, Terrier, Terzić, Tollefson, Tomé, Torres, Torres-Escobedo, Trinchero, Turner, Ulloa, Valore, van Eldik, Vega, Quispe, Viana, Vícha, Vigorito, Vittorini, Wang, Wang, Wang, Wang, Wang, Wang, Wang, Waqas, Watson, Werner, White, Wiebusch, Wohlleben, Xi, Xiao, Xiao, Xiao, Yang, Yang, Yang, Yang, Yanyachi, Yao, Zavrtanik, Zhang, Zhang, Zhang, Zhang, Zhang, Zhang, Zhang, Zhang, Zhang, Zhang, Zhao, Zhao, Zhou, Zhu, Zhu, Zhu, Zhu, Zuo, \& Zyla}]{SWGO2025}
SWGO, Abreu, P., Alfaro, R., {et~al.} 2025, Science Prospects for the Southern Wide-field Gamma-ray Observatory: SWGO.
\newblock \doarXiv{2506.01786}

\bibitem[{Vianello {et~al.}(2015)Vianello, Lauer, Younk, Tibaldo, Burgess, Ayala, Harding, Hui, Omodei, \& Zhou}]{3ml}
Vianello, G., Lauer, R.~J., Younk, P., {et~al.} 2015, The Multi-Mission Maximum Likelihood framework (3ML).
\newblock \doarXiv{1507.08343}

\bibitem[{{Wakely} \& {Horan}(2008)}]{TeVCat}
{Wakely}, S.~P., \& {Horan}, D. 2008, International Cosmic Ray Conference, 3, 1341

\bibitem[{Wilks(1938)}]{Wilks}
Wilks, S.~S. 1938, The Annals of Mathematical Statistics, 9, 60.
\newblock \url{http://www.jstor.org/stable/2957648}

\end{thebibliography}
\bibliographystyle{aasjournal}


\end{document}